\PassOptionsToPackage{numbers,sort&compress}{natbib}  
\documentclass[preprintnumbers,amsmath,amssymb,prd, notitlepage,nofootinbib,twocolumn, superscriptaddress]{revtex4}

\usepackage{graphicx}
\usepackage{dcolumn}
\usepackage{bm}
\usepackage{subfigure}
\usepackage{wasysym}
\usepackage{verbatim}
\usepackage{color}
\usepackage{amsmath}
\usepackage[utf8]{inputenc}

\usepackage[hidelinks]{hyperref}

\newcommand{\be}{\begin{equation}}                                  
\newcommand{\ee}{\end{equation}}                                    
\newcommand{\ba}{\begin{eqnarray}}                                  
\newcommand{\ea}{\end{eqnarray}}                                    
                                          
\newcommand{\barr}{\begin{array}}                                    
\newcommand{\earr}{\end{array}}

\def\ellb{{\boldsymbol{ \ell}}}

\usepackage{aas_macros}

\begin{document}

\title{The Atacama Cosmology Telescope: Component-separated maps of CMB temperature and the thermal Sunyaev-Zel'dovich effect}

\author{Mathew S.~Madhavacheril}
\email[Corresponding author: ]{mmadhavacheril@perimeterinstitute.ca}
\affiliation{Centre for the Universe, Perimeter Institute for Theoretical Physics, Waterloo, ON, Canada N2L 2Y5}
\affiliation{Department of Astrophysical Sciences, Princeton University, 4 Ivy Lane, Princeton, NJ, USA 08544}  
\author{J.~Colin Hill}
\affiliation{Department of Physics, Columbia University, 550 West 120th Street, New York, NY, USA 10027}
\affiliation{Center for Computational Astrophysics, Flatiron Institute, 162 5th Avenue, New York, NY, USA 10010}
\affiliation{School of Natural Sciences, Institute for Advanced Study, Princeton, NJ, USA 08540}
\author{Sigurd Næss}
\affiliation{Center for Computational Astrophysics, Flatiron Institute, 162 5th Avenue, New York, NY, USA 10010}
\author{Graeme E.~Addison}
\affiliation{Department of Physics and Astronomy, Johns Hopkins University, Baltimore, MD 21218, USA}
\author{Simone Aiola}
\affiliation{Center for Computational Astrophysics, Flatiron Institute, 162 5th Avenue, New York, NY, USA 10010}
\author{Taylor~Baildon}
\affiliation{Department of Physics, University of Michigan, Ann Arbor, USA 48103}
\author{Nicholas Battaglia}
\affiliation{Department of Astronomy, Cornell University, Ithaca, NY 14853, USA}
\author{Rachel Bean}
\affiliation{Department of Astronomy, Cornell University, Ithaca, NY 14853, USA}
\author{J.~Richard~Bond}
\affiliation{Canadian Institute for Theoretical Astrophysics, University of Toronto, 60 St.~George St., Toronto, ON M5S 3H8, Canada}
\author{Erminia Calabrese}
\affiliation{School of Physics and Astronomy, Cardiff University, The Parade, Cardiff, CF24 3AA, UK}
\author{Victoria Calafut}
\affiliation{Department of Astronomy, Cornell University, Ithaca, NY 14853, USA}
\author{Steve K.~Choi}
\affiliation{Department of Astronomy, Cornell University, Ithaca, NY 14853, USA}
\author{Omar Darwish}
\affiliation{Department of Applied Mathematics and Theoretical Physics, University of Cambridge, Wilberforce Road, Cambridge CB3 0WA, UK}
\author{Rahul Datta}
\affiliation{Department of Physics, University of Michigan, Ann Arbor, USA 48103}
\author{Mark J.~Devlin}
\affiliation{Department of Physics and Astronomy, University of Pennsylvania, 209 South 33rd Street, Philadelphia, PA, USA 19104}
\author{Joanna Dunkley}
\affiliation{Joseph Henry Laboratories of Physics, Jadwin Hall, Princeton University, Princeton, NJ 08544, USA}
\affiliation{Department of Astrophysical Sciences, Princeton University, 4 Ivy Lane, Princeton, NJ, USA 08544}
\author{Rolando D\"unner}
\affiliation{Instituto de Astrof\'isica and Centro de Astro-Ingenier\'ia, Facultad de F\'isica, Pontificia Universidad Cat\'olica de Chile, Av.  Vicu\~na Mackenna 4860, 7820436 Macul, Santiago, Chile}
\author{Simone Ferraro}
\affiliation{Physics Division, Lawrence Berkeley National Laboratory, Berkeley, CA 94720, USA}

\author{Patricio A. Gallardo}
\affiliation{Department of Physics, Cornell University, 109 Clark Hall, Ithaca, NY 14853, USA}

\author{Vera Gluscevic}
\affiliation{Department of Physics and Astronomy, University of Southern California, Los Angeles, CA 90089-0484, USA}

\author{Mark Halpern}
\affiliation{Department of Physics and Astronomy, University of British Columbia, Vancouver, BC, Canada V6T 1Z1}

\author{Dongwon Han}
\affiliation{Physics and Astronomy Department, Stony Brook University, Stony Brook, NY 11794, USA}
\affiliation{Center for Computational Astrophysics, Flatiron Institute, 162 5th Avenue, New York, NY, USA 10010}
\author{Matthew Hasselfield}
\affiliation{Center for Computational Astrophysics, Flatiron Institute, 162 5th Avenue, New York, NY, USA 10010}
\author{Matt~Hilton}
\affiliation{Astrophysics \& Cosmology Research Unit, School of
Mathematics, Statistics \& Computer Science, University of
KwaZulu-Natal, Westville Campus, Durban 4041, South Africa}
\author{Adam~D.~Hincks}
\affiliation{Canadian Institute for Theoretical Astrophysics, University of Toronto, 60 St.~George St., Toronto, ON M5S 3H8, Canada}
\author{Ren\'ee~Hlo\v{z}ek}
\affiliation{Department of Astronomy and Astrophysics, University of Toronto, 50 St.~George St., Toronto, ON M5S 3H4, Canada}
\affiliation{Dunlap Institute for Astronomy and Astrophysics, University of Toronto, 50 St.~George St., Toronto, ON M5S 3H4, Canada}
\author{Shuay-Pwu Patty Ho}
\affiliation{Joseph Henry Laboratories of Physics, Jadwin Hall, Princeton University, Princeton, NJ 08544, USA}

\author{Kevin M.~Huffenberger}
\affiliation{Department of Physics, Florida State University, Tallahassee, Florida 32306, USA}
\author{John P.~Hughes}
\affiliation{Department of Physics and Astronomy, Rutgers University, 136 Frelinghuysen Road, Piscataway, NJ 08854-8019 USA}
\author{Brian J. Koopman}
\affiliation{Department of Physics, Yale University, New Haven, CT 06520}
\author{Arthur Kosowsky}
\affiliation{Department of Physics and Astronomy, University of Pittsburgh, Pittsburgh, PA, USA 15260}
\author{Martine~Lokken}
\affiliation{Department of Astronomy and Astrophysics, University of Toronto, 50 St.~George St., Toronto, ON M5S 3H4, Canada}
\affiliation{Canadian Institute for Theoretical Astrophysics, University of Toronto, 60 St.~George St., Toronto, ON M5S 3H8, Canada}
\affiliation{Dunlap Institute for Astronomy and Astrophysics, University of Toronto, 50 St.~George St., Toronto, ON M5S 3H4, Canada}
\author{Thibaut~Louis}
\affiliation{Laboratoire de l'Acc\'el\'erateur Lin\'eaire, Univ. Paris-Sud, CNRS/IN2P3, Universit\'e Paris-Saclay, Orsay, France}
\author{Marius Lungu}
\affiliation{Joseph Henry Laboratories of Physics, Jadwin Hall, Princeton University, Princeton, NJ 08544, USA}

\author{Amanda MacInnis}
\affiliation{Physics and Astronomy Department, Stony Brook University, Stony Brook, NY 11794, USA}

\author{Lo\"ic Maurin}
\affiliation{Institut d’Astrophysique Spatiale, CNRS, Univ. Paris-Sud, Universit\'e Paris-Saclay, F-91400 Orsay, France}

\author{Jeffrey J.~McMahon}
\affiliation{Department of Physics, University of Michigan, Ann Arbor, USA 48103}

\author{Kavilan~Moodley}
\affiliation{Astrophysics \& Cosmology Research Unit, School of
Mathematics, Statistics \& Computer Science, University of
KwaZulu-Natal, Westville Campus, Durban 4041, South Africa}

\author{Federico Nati}
\affiliation{Department of Physics, University of Milano--Bicocca, Piazza della Scienza, 3, Milan, Italy, 20126
}

\author{Michael D. Niemack}
\affiliation{Department of Physics, Cornell University, 109 Clark Hall, Ithaca, NY 14853, USA}
\affiliation{Department of Astronomy, Cornell University, Ithaca, NY 14853, USA}
\author{Lyman A. Page}
\affiliation{Joseph Henry Laboratories of Physics, Jadwin Hall, Princeton University, Princeton, NJ 08544, USA}
\author{Bruce Partridge}
\affiliation{Department of Physics and Astronomy, Haverford College,Haverford, PA, USA 19041}
\author{Naomi Robertson}
\affiliation{Institute of Astronomy, Madingley Road, Cambridge CB3 0HA, UK}
\affiliation{Kavli Institute for Cosmology, University of Cambridge, Madingley Road, Cambridge CB3 OHA, UK}
\author{Neelima Sehgal}
\affiliation{Physics and Astronomy Department, Stony Brook University, Stony Brook, NY 11794, USA}
\affiliation{Center for Computational Astrophysics, Flatiron Institute, 162 5th Avenue, New York, NY, USA 10010}

\author{Emmanuel~Schaan}
\affiliation{Physics Division, Lawrence Berkeley National Laboratory, Berkeley, CA 94720, USA}
\author{Alessandro Schillaci}
\affiliation{Department of Physics, California Institute of Technology, Pasadena, California 91125, USA}
\author{Blake D. Sherwin}
\affiliation{Department of Applied Mathematics and Theoretical Physics, University of Cambridge, Wilberforce Road, Cambridge CB3 0WA, UK}
\affiliation{Kavli Institute for Cosmology, University of Cambridge, Madingley Road, Cambridge CB3 OHA, UK}
\author{Crist\'{o}bal Sif\'{o}n}
\affiliation{Instituto de F\'isica, Pontificia Universidad Cat\'olica de Valpara\'iso, Casilla 4059, Valpara\'iso, Chile}
\author{Sara M.~Simon}
\affiliation{Department of Physics, University of Michigan, Ann Arbor, USA 48103}
\author{David N.~Spergel}
\affiliation{Center for Computational Astrophysics, Flatiron Institute, 162 5th Avenue, New York, NY, USA 10010}
\affiliation{Department of Astrophysical Sciences, Princeton University, 4 Ivy Lane, Princeton, NJ, USA 08544}  
\author{Suzanne T.~Staggs}
\affiliation{Joseph Henry Laboratories of Physics, Jadwin Hall, Princeton University, Princeton, NJ 08544, USA}
\author{Emilie R.~Storer}
\affiliation{Joseph Henry Laboratories of Physics, Jadwin Hall, Princeton University, Princeton, NJ 08544, USA}
\author{Alexander van Engelen}
\affiliation{School of Earth and Space Exploration, Arizona State University, Tempe, AZ, 85287, USA}
\author{Eve M.~Vavagiakis}
\affiliation{Department of Physics, Cornell University, 109 Clark Hall, Ithaca, NY 14853, USA}

\author{Edward J.~Wollack}
\affiliation{NASA/Goddard Space Flight Center, Greenbelt, MD 20771, USA}

\author{Zhilei Xu}
\affiliation{Department of Physics and Astronomy, University of Pennsylvania, 209 South 33rd Street, Philadelphia, PA, USA 19104}

\date{\today}

\begin{abstract}
Optimal analyses of many signals in the cosmic microwave background (CMB) require map-level extraction of individual components in the microwave sky, rather than measurements at the power spectrum level alone.  To date, nearly all map-level component separation in CMB analyses has been performed exclusively using satellite data.  In this paper, we implement a component separation method based on the internal linear combination (ILC) approach which we have designed to optimally account for the anisotropic noise (in the 2D Fourier domain) often found in ground-based CMB experiments.  Using this method, we combine multi-frequency data from the \emph{Planck} satellite and the Atacama Cosmology Telescope Polarimeter (ACTPol) to construct the first wide-area ($\approx 2100$ sq.~deg.), arcminute-resolution component-separated maps of the CMB temperature anisotropy and the thermal Sunyaev-Zel'dovich (tSZ) effect sourced by the inverse-Compton scattering of CMB photons off hot, ionized gas.  Our ILC pipeline allows for explicit deprojection of various contaminating signals, including a modified blackbody approximation of the cosmic infrared background (CIB) spectral energy distribution. The cleaned CMB maps will be a useful resource for CMB lensing reconstruction, kinematic SZ cross-correlations, and primordial non-Gaussianity studies.  The tSZ maps will be used to study the pressure profiles of galaxies, groups, and clusters through cross-correlations with halo catalogs, with dust contamination controlled via CIB deprojection. The data products described in this paper are available on LAMBDA.
\end{abstract}

\maketitle

\section{Introduction}
Beginning with \emph{COBE}~\cite{COBECompSep1,COBECompSep2} and continuing with \emph{WMAP}~\cite{Bennett2013,Hinshaw2013,Tegmark2003} and \emph{Planck}~\cite{Planck2018legacy,Planck2015compsep,Planck2015compsepFG,Planck2018compsep}, the advent of multi-frequency, wide-area microwave sky surveys has allowed the extraction of maps of particular components from the total observed sky signal.  This extraction process, known as component separation, can be designed to use various characteristics of the signals, including their frequency, spatial, and angular scale dependences.  This procedure is complementary to the power spectrum-level foreground marginalization approach that has become standard for the inference of cosmological parameters, which does not use map-level foreground cleaning at multipoles $\ell \gtrsim 30$~\cite{Hinshaw2013,Sievers2013,Dunkley2013,Story2013,Planck2018likelihood}.  While all of the information in the primary cosmic microwave background (CMB) fluctuations is contained in their angular power spectrum (under the assumption of Gaussianity), and thus a map is not strictly required for cosmological parameter analysis, this is not the case for any other signals of interest in the microwave sky.  Thus, map-level extraction of these other signals is extremely useful for a wide variety of scientific applications.

Map-level extraction of the CMB itself is necessary for many important analyses, including CMB lensing reconstruction, searches for primordial non-Gaussianity, tests of isotropy and searches for anomalies, and defining masks for power spectrum-level analyses (by delineating regions where CMB component separation is effective).  In addition, important secondary anisotropy signals can be accessed via component separation.  Cleaned CMB maps contain the kinematic Sunyaev-Zel'dovich (kSZ) signal, as this effect preserves the blackbody spectrum of the CMB to lowest order in $(v/c)$, where $v$ is the line-of-sight electron velocity.  Thus, these maps can be used for kSZ measurements via cross-correlations~\cite{Hill2016,Ferraro2016,Planck2013kSZ,Planck2015kSZ}.  The thermal Sunyaev-Zel'dovich (tSZ) effect, on the other hand, generates a distinctive non-blackbody spectral distortion in the CMB, whose frequency dependence can be used to extract tSZ maps from multi-frequency data sets.  Given its high degree of non-Gaussianity, these tSZ maps contain much more information than the tSZ power spectrum alone~\cite{Wilson2012,Bhattacharya2012,Hill-Sherwin2013,Hill2014,Crawford2014,Planck2015ymap,Coulton2018,Thiele2019}.  In addition, they are valuable tools for cross-correlation analyses with a wide variety of large-scale structure data sets~\cite{Hill-Spergel2014,VanWaerbeke2014,BHM2015,Vikram2017,Hill2018,Alonso2018,Pandey2019a,Pandey2019b}.

Thus far, map-level component separation has primarily been performed using CMB satellite data because ground-based experiments lacked sufficiently sensitive multi-frequency data.  This situation has recently changed with the advent of high-sensitivity, multi-frequency receivers on ground-based experiments like the Atacama Cosmology Telescope (ACT)  and the South Pole Telescope (SPT) \cite{spt3g,spt3g2,spt3g3}.  In particular, the full Advanced ACTPol data set will include coverage at five frequencies from 28 to 230 GHz~\cite{Henderson2016}.  The upcoming Simons Observatory will extend this further with six frequencies from 27 to 280 GHz~\cite{SO2019}, while overlapping data from CCAT-prime will include coverage at higher frequencies as well~\cite{CCATp2018}.  Thus, it is an opportune time to build the necessary analysis infrastructure for multi-frequency component separation with these data sets.

Several previous works have explored co-addition of ground-based data with satellite data. Most notably, a previous work~\cite{PACT2019} considered a joint map co-add of \emph{Planck} and ACT data from the original ACT MBAC survey. Since that work used only one high-resolution ACT channel at 148 GHz, on small scales the resulting map is simply a rescaled version of the ACT 148 GHz map (see their Fig.~2).  Also, Ref.~\cite{ChownEtAl} constructed single-frequency co-additions of {\it Planck} and SPT data (e.g., SPT 90 GHz combined with \emph{Planck} 100 GHz). In contrast, in this work, we use high-resolution ground-based observations at multiple frequencies to produce arcminute-resolution maps of the CMB and tSZ signals.

In the tSZ context, parallel efforts in the past decade have used multi-frequency data in matched-filter-based analyses to detect individual tSZ clusters.  This multi-frequency matched filter (MMF) approach (e.g.,~\cite{JB2006}) has been applied to data from \emph{Planck}~\cite{PlanckClusters2015} and SPT~\cite{Williamson2011} to extract catalogs of tSZ-detected clusters (ACT tSZ cluster catalogs to date have relied on single-frequency matched filters~\cite{2011ApJ...737...61M,HiltonEtAl}).  However, these analyses do not produce large-area, unbiased tSZ maps comparable to those produced in this analysis or in Ref.~\cite{Planck2015ymap}.  While applying the MMF to multi-frequency maps yields a map that contains the tSZ cluster signal, it is not an unbiased map that faithfully contains all modes (up to the resolution limit): due to the application of the cluster-template filter, all of the large-scale modes are lost.  In addition, the foreground contribution to the frequency-frequency covariance matrix used in the MMF approach is generally not derived from the data itself, and thus is likely to be sub-optimal; however, detailed treatments of anisotropic noise have been implemented in MMF cluster-finding (as considered in detail below for our method).  We emphasize that the output from an MMF analysis cannot be used for unbiased wide-area cross-correlations or other map-based statistical analyses (unless such an analysis is verified after being mowdified to retain the large scales), as is intended for the data products constructed here.

The data used in this analysis include \emph{Planck} frequency maps from 30 to 545 GHz (eight channels) from the 2015 release and ACT data at 98 and 150 GHz collected by the ACTPol receiver~\cite{thornton/2016,Naess2014,Louis2017} during the 2014 and 2015 seasons. The total area of the component-separated maps produced in this analysis is $\approx 2100$ square degrees. Due to large-scale atmospheric noise in the ACT data, \emph{Planck} dominates the information content at low multipoles in our component-separated maps, while ACT dominates at high multipoles where the \emph{Planck} noise rises rapidly due to the instrument's coarser angular resolution.  On intermediate scales, both experiments contribute substantial information. 

To construct CMB and tSZ maps from these data, we implement an internal linear combination (ILC) algorithm~\cite{COBECompSep1,Bennett2003,Tegmark2003,Eriksen2004,Delabrouille2009,Remazeilles2011}.  The ILC approach aims to obtain a linear co-addition of the input maps that minimizes the variance of the final map while preserving the signal of interest in an unbiased way, relying solely on knowledge of the frequency dependence of this component.  This method has the advantages of computational efficiency and robustness to foregrounds with unexpected spectral properties due to its semi-blind nature.  On the other hand, its primary disadvantage is the difficulty of characterizing leakage of various foregrounds into the final map; no particular foreground is explicitly removed in the standard ILC analysis.  The latter drawback can be mitigated to a large extent by ``constrained'' ILC methods that explicitly null some foreground(s), as we discuss and implement below.  However, it is infeasible to fully deproject all foregrounds, and thus the auto-statistics (e.g., the auto-power spectrum) of the resulting maps must be interpreted with significant caution.\footnote{This issue is the reason that the cosmological parameter analysis of the primary CMB power spectrum in \emph{Planck} did not rely on component-separated maps~\cite{Planck2015compsep,Planck2018compsep,PlanckCosmology}.}  In particular, we do not attempt to interpret the auto-power spectra of the derived CMB or tSZ maps in this work.

Nevertheless, the maps constructed here will have a rich array of scientific applications.  The tSZ maps can be used for cross-correlations with galaxy, group, and cluster catalogs selected at many wavelengths (e.g., optical \cite{SDSS,DES,KIDS,HSC,LSST,DESI,EUCLID}, infrared \cite{WISE2010,WFIRST}, or X-ray \cite{ROSAT,EROSITA}).  These studies can probe the behavior of the gas pressure profile deep into the interiors of clusters and at the virial radii of galaxies and groups~\cite{BH2020SWP}, due to the unprecedented high resolution of our wide-area tSZ maps (FWHM = 1.6 arcminutes, in contrast to 10 arcminutes when using \emph{Planck} alone~\cite{Planck2015ymap}).  The CMB temperature maps can be used for a variety of kSZ analyses, including cross-correlations with spectroscopic and photometric galaxy samples.  They can also be used for CMB lensing reconstruction, particularly in novel estimators that are insensitive to tSZ foreground biases~\cite{MMHill}.  We expect that a diverse set of future analyses will follow from these data products, and from improved data products to be produced imminently from a broader set of Advanced ACTPol single-frequency maps over more area with additional frequency information.

The rest of this paper is organized as follows.  In Sec.~\ref{sec:theory}, we review the theory underlying our map co-addition procedure and describe our models of relevant sky components.  In Sec.~\ref{sec:data}, we describe the ACT and \emph{Planck} maps used in this analysis.  Sec.~\ref{sec:processing} presents the data processing steps applied to the maps, while Sec.~\ref{sec:algorithm} details the algorithm that we apply to co-add these data to produce maps of the CMB temperature and tSZ effect.  We then present the resulting maps in  Sec.~\ref{sec:results}, including several validation tests.  In Sec.~\ref{sec:sims}, we describe a set of simulations used for additional validation and for future use in covariance estimation.  We discuss future prospects and conclude in Sec.~\ref{sec:discussion}. The data products described in this paper are available on LAMBDA\footnote{\url{https://lambda.gsfc.nasa.gov/product/act/act_dr4_derived_maps_get.cfm}}.

\section{Component-Separated Maps from Multi-frequency Data}
\label{sec:theory}

\begin{figure}[t]
\includegraphics[width=\columnwidth]{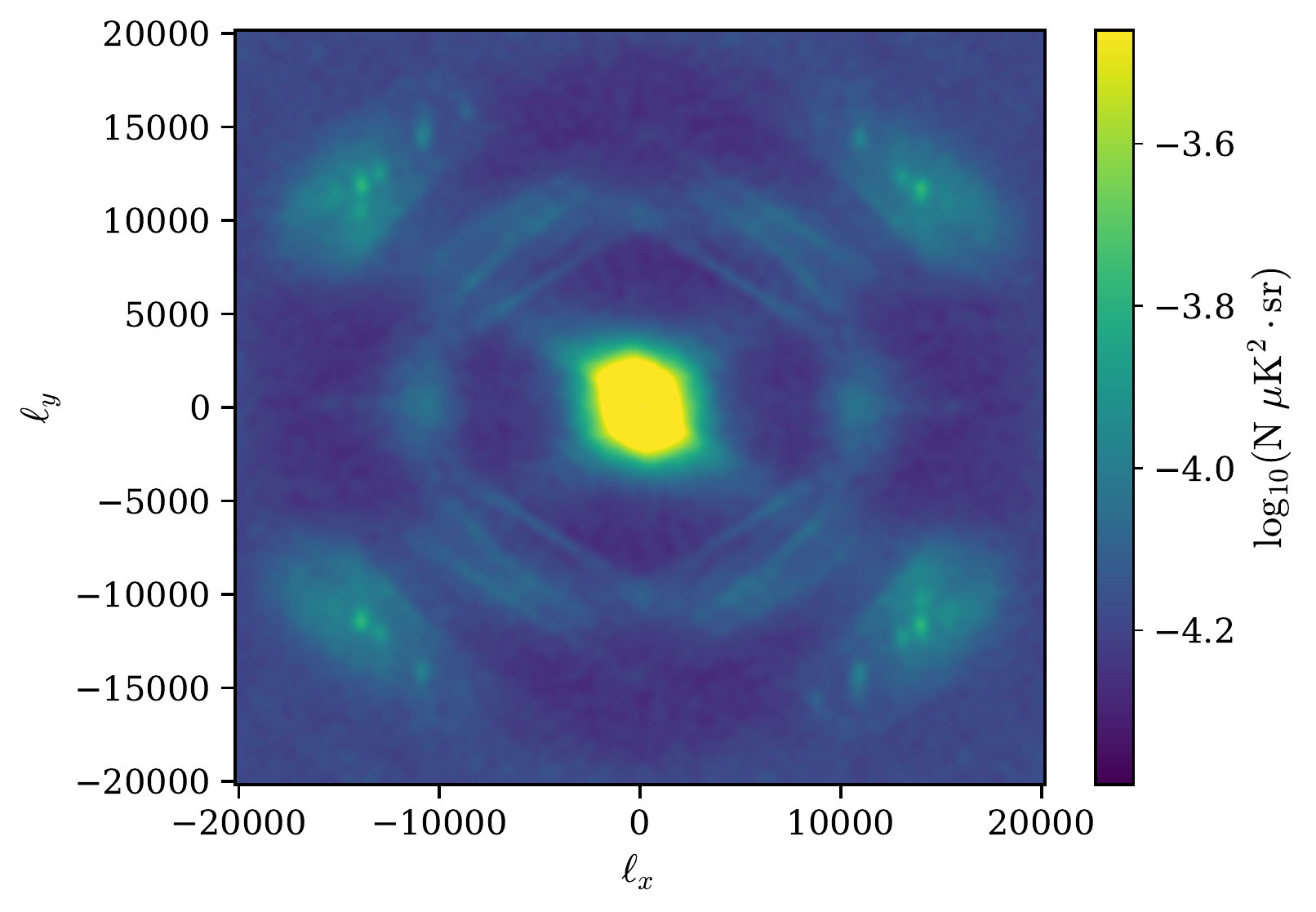}
\centering
\caption{Anisotropy of instrument noise in one of the ACT single-frequency maps entering our component separation analysis. The map is from the 150 GHz channel of the PA3 array from 2015 observations in the {\tt D56} region; see Sec. \ref{sec:data} for more details. We show the logarithm of the noise power in units of $\mu {\rm K}^2\cdot\rm{sr}$ in the 2D Fourier plane with angular wavenumbers shown on the axes. To make the anisotropy in the noise power visible, the color scale is saturated in the central region where atmospheric noise dominates. The noise power has been smoothed with an averaging block width of $\Delta \ell=400$. The anisotropy is primarily due to atmosphere-induced stripe-like noise in the scan directions (the X pattern) and detector correlations (the hexagonal pattern). While the instrument power shown here is an average over the entire sky region considered, the magnitude and directionality of anisotropy varies as a function of position in the map.}\label{fig:noise}
\end{figure}

We construct component-separated maps following the ILC approach~\cite{COBECompSep1,Bennett2003,Tegmark2003,Eriksen2004,Delabrouille2009,Remazeilles2011}, but applied in a novel domain in comparison to earlier works. In the ILC approach, one chooses a domain in which to represent the individual frequency maps and linearly co-add them with weights determined such that (a) the total power spectrum of the resulting map is minimized while (b) the resulting map has unit response to the component of interest, which has an assumed spectral energy distribution (SED), and optionally (c) the response to a contaminant with another assumed SED is nulled. Since the ACT maps we work with do not span a large area on the sphere ($\approx 2100$ sq.~deg.), we choose to work in the 2D Fourier pixel domain of rectangular maps in a cylindrical projection. We define a two-dimensional wavevector $\ellb$, rather than using spherical harmonics.  An advantage of working in the 2D Fourier domain is that it allows us to optimally downweight anisotropic noise in microwave sky maps from ground-based experiments (see Figure~\ref{fig:noise}). The anisotropic structure in the instrument noise power is primarily due to atmospheric noise, which correlates observations in the scanning directions. The stripe-like noise in the two primary scanning directions shows up as an ``X'' in the 2D noise power spectrum.  Nearby detectors also see nearby correlated parts of the atmosphere. The interplay between the atmosphere and detector layout in the focal plane also shows up in the 2D noise power spectrum as a hexagonal pattern.

In the 2D Fourier domain, the harmonic transform of any general linear map co-add that treats each 2D Fourier pixel at $\ellb$ independently can be written as
\be
\label{eq.coadd}
M(\ellb) = \sum_i w_i(\ellb)M_i(\ellb) \,,
\ee
where the index $i$ runs over any number of maps $M_i$, each of which has its own particular frequency bandpass and angular beam.  These maps are linearly combined with weights $w_i$ to satisfy desired properties of the co-add.

Throughout our analysis, we use maps $M_i$ that are calibrated in units of differential CMB blackbody temperature expressed in $\mu$K, with $T_{\rm CMB} = 2.726$ K assumed for the CMB monopole temperature~\cite{Fixsen2009}. Under this assumption, blackbody components like the CMB and kSZ signals are identical across all maps and have unit response, thus implying a constraint $\sum_i w_i=1$ in the ILC for these components. More generally, we may be interested in extracting a ``signal'' component $s(\ellb)$ whose SED yields a per-channel response $f_i$ in CMB temperature units.  Furthermore, we explicitly retain the beam (or point spread function) transmission in harmonic space $B_i(\ell)$, such that the total harmonic-space response factor for this component in each map is $a_i(\ell) = f_i B_i(\ell)$. This choice differs from some previous work (e.g.,~\cite{Remazeilles2011}), but it allows our pipeline to avoid explicit beam deconvolutions that can lead to numerical instabilities, particularly on small scales.  Furthermore, we assume that a second ``contaminant'' component $c(\ellb)$ may also be present with assumed SED $f'_i$ and overall response $a'_i(\ell) = f'_i B_i(\ell)$.  Any individual frequency map in the data set can then be modeled as
\be
\label{eq.twocompmodel}
M_i(\ellb) = a_i(\ell) s(\ellb) + a'_i(\ell) c(\ellb)  + n_i(\ellb) \,,
\ee
where each map has a noise contribution $n_i(\ellb)$ comprised of the sum of the instrument and atmospheric noise, as well as all other foreground contaminants that may be present.  In the ILC approach, we assume that the latter contaminants are uncorrelated with the signal of interest. This assumption is violated by the tSZ-CIB correlation~(e.g.,~\cite{Addison2012}); however, explicit deprojection of the CIB in the tSZ ILC map can partially mitigate this issue. In this work, we only consider explicit modeling and deprojection of a single contaminant $c$ in a given ILC map (though the choice of $c$ can vary), because of limited frequency coverage on small scales.  With future data at additional frequencies, we will explore explicit modeling of multiple components.

In the idealized case of a monochromatic ($\delta$-function) bandpass, the responses $f_i$ are scale-independent numbers that are directly related to the SED of the component of interest. With a bandpass of non-zero width, $f_i$ acquires ``color corrections'' that depend on the shape of the bandpass.  If the bandpass and beam were fully separable, these would still be scale-independent numbers.  However, in general, the responses $f_i$ become scale-dependent functions of $\ell$ (corresponding to different behavior in the diffuse and compact limits) when the beam changes slightly as a function of frequency within the bandpass. These scale-dependent color correction effects are accounted for as described in Appendix~\ref{app:color}, with amplitudes from $1\% - 10\%$ for the ACT maps in our analysis, motivating careful consideration of these effects in future high-precision analyses (see, e.g.,~\cite{WardBandpass}).      

\subsection{Covariance and Deprojection}

The optimal weights $w_i$ for use in Eq.~\ref{eq.coadd} can be derived by requiring that the ILC map co-add has minimum variance while satisfying additional constraints. Allowing for the presence of a second component allows us to produce two types of maps \cite{Remazeilles2011}: (1) ``Standard'' ILC (no contaminant deprojection) where the sole constraint is that the map has unit response to the component of interest $s(\ellb)$ (i.e., $\sum w_i a_i = 1$), and (2) ``Constrained'' ILC (with contaminant deprojection) where the map is also required to have null response to the second component $c(\ellb)$ (i.e., $\sum w_i a'_i = 0$).

This results in the following expression for the weights (e.g.,~\cite{Remazeilles2011}):
\newcommand{\cinv}{C^{-1}}
\be
\label{eq.ILCweights}
w_i = \frac{\left(a'_j\cinv_{jk}a'_k\right)a_j\cinv_{ji} - \left(a_j\cinv_{jk}a'_k\right)a'_j\cinv_{ji}}{\left(a_j\cinv_{jk}a_k\right)\left(a'_j\cinv_{jk}a'_k\right)-\left(a_j\cinv_{jk}a'_k\right)^2} \,,
\ee
where we have suppressed the $\ellb$-dependence for clarity, and the covariance matrix $C$ is the expectation value of the cross-map power-spectrum $\langle M_i(\ellb) M^*_j(\ellb) \rangle$. In the ``no deprojection'' (Standard ILC) case, $w_i = a_j\cinv_{ji} / (a_j\cinv_{jk}a_k)$ can be obtained as a special case in the limit $a'_i\rightarrow 0$.  The expectation value of the covariance matrix $C$ is formally over realizations of the underlying statistical fields, and thus it contains contributions from sample variance of the astrophysical quantities as well as the power spectrum of the experimental noise. For example, if the CMB were the only component in the sky and had known power spectrum $C_{\ell}$, one would have $C_{ij} = B_i(\ell)B_j(\ell) C_{\ell} + N^{ij}_{\ell}$ where $N^{ij}$ are the noise cross-power spectra of the maps.

In practice, the map power spectra receive contributions from astrophysical foregrounds whose relative amplitude can depend in an unpredictable way on sky position,
angular scale, and frequency. For these reasons, a covariance matrix based on theoretical predictions can result in a sub-optimal co-add, leaving large foreground residuals in the result. Thus, an empirical estimate of the power spectrum of the maps is preferred; this is a primary motivation of the ILC method. This empirical estimate can be obtained by smoothing or averaging $M_i(\ellb) M^*_j(\ellb)$ over domains of similar $\ellb$ (e.g., in annular bins centered on $\ell$ with width $\Delta \ell$).  Such empirical estimates are prone to problems of signal loss, sometimes referred to as ``ILC bias'' (see~\cite{Delabrouille2009} for a detailed discussion), since the dependence of the weights on the data causes the estimator for the map co-add to be non-linear in the data.  The ILC bias primarily affects large scales due to the small number of modes available at low $\ell$. As described in Sec.~\ref{sec:algorithm}, we employ a strategy to preserve the information content of the 2D power spectra while minimizing ILC bias. We verify this strategy on simulations (see Sec.~\ref{sec:sims}).

As a point of comparison, the methodology employed here has strong similarities to the semi-blind component separation algorithms employed by \emph{Planck}, particularly the needlet ILC (NILC) algorithm~\cite{Delabrouille2009}.  One distinction is that our ILC weights include anisotropic dependence in the Fourier domain, which is not considered in NILC. Unlike NILC, our ILC weights do not vary with spatial position, but this is not a significant disadvantage as we only analyze relatively small regions of sky far from the Galactic plane, with observed sky fraction $f_{\rm sky} \lesssim 0.05$. We anticipate that our results could be moderately improved on large angular scales by including spatial dependence in the ILC weights in a needlet-like manner; the needlet approach was demonstrated on ACT and \emph{Planck} simulations in~\cite{HeteroILC}.  The other semi-blind methods used by \emph{Planck} include spectral-matching independent component analysis (SMICA) and pixel-space linear template fitting methods, which both produced results similar to NILC~\cite{Planck2015compsep}, as did an independent analysis using the semi-blind Local-Generalized Morphological Component Analysis method~\cite{Bobin2016}.  On large scales where the number of modes available for covariance estimation is limited, parametric fitting component separation methods, such as the Commander code used by \emph{Planck}, are more well-suited than ILC or ICA methods.  With sufficient frequency coverage, parametric methods can also perform well on small scales.  Given the moderate number of frequency channels available on small scales here, we focus only on the semi-blind ILC approach.

\subsection{Frequency dependence of components}

\begin{figure}[t]
\includegraphics[width=\columnwidth]{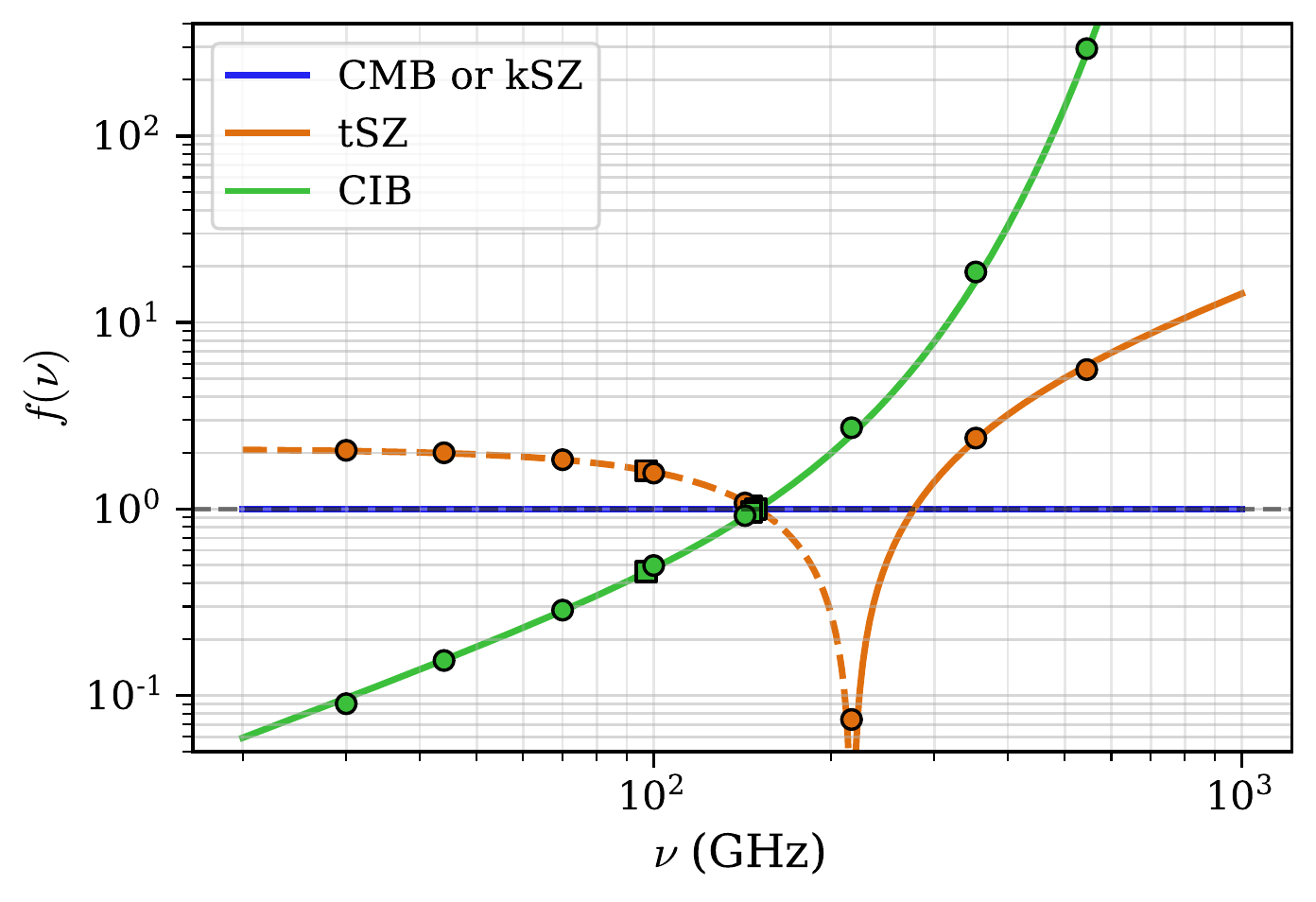}
\centering
\caption{Relative frequency dependences of the CMB+kSZ, tSZ, and CIB components in units
of differential CMB temperature, normalized at 150 GHz. The tSZ spectrum Eq.~\ref{eq:tszresponse}
is negative (orange dashed line) below 218 GHz. The CIB is the fiducial
modified blackbody SED Eq.~\ref{eq:fcib} (green line). The CMB+kSZ (blue line) is constant
in these units.  The circles and squares show the color-corrected response factors for each frequency band used in this work for {\emph{Planck}} and ACT, respectively, positioned horizontally at the approximate central frequencies of the bands.}\label{fig:sed}
\end{figure}

\begin{figure*}[t]
\includegraphics[width=\textwidth]{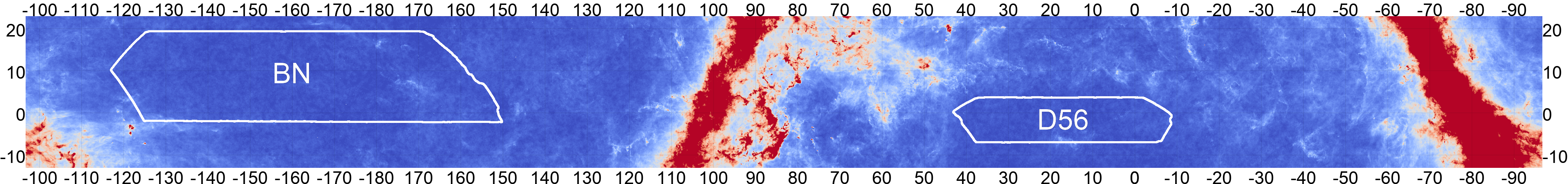}
\centering
\caption{Sky regions analyzed in this work. The vertical axis (declination) and horizontal axis (right ascension) are labeled in degrees. The {\tt BN} region (1633 sq.~deg) and {\tt D56} region (456 sq.~deg.) that we analyze are labeled.  The background shows the \emph{Planck} 353 GHz temperature map in this sky region.}
\label{fig:footprint}
\end{figure*}

In this work, we focus on reconstructing ILC maps of two components: (1) the blackbody component consisting primarily of the lensed CMB and kSZ signals, hereafter referred to as the ``CMB+kSZ'' component (subscripted with just ``CMB'' in formulae for brevity), and (2) the Compton-$y$ fluctuations arising from the tSZ effect, which probes the line-of-sight-integrated pressure of ionized gas. Each of these components is reconstructed with various ILC approaches: (1) CMB+kSZ with no deprojection (standard ILC), tSZ deprojected, or CIB deprojected; and (2) tSZ with no deprojection (standard ILC), CMB deprojected, or CIB deprojected. Here, CIB refers to the cosmic infrared background, whose SED we take to be a modified blackbody spectrum with associated dust temperature, as described below.  This treatment of the CIB is only approximate; in reality since it constitutes the sum of emission from dusty galaxies across a range of redshifts it is not actually a fully coherent field whose signal is simply rescaled across frequencies. Instead, different frequency maps contain intrinsically different CIB fluctuations \cite{CIBDecorr1,CIBDecorr2,CIBDecorr3,CIBDecorr4,CIBDecorr5}.  Nevertheless, maps that are relatively close in frequency contain highly coherent CIB fluctuations, and thus deprojecting a CIB-like spectrum can remove significant contamination in the ILC.

Producing these maps requires us to assume a frequency response $f_i$ in the response factors $a_i=B_i(\ell)f_i$ for each component that is either preserved ($a_i$ for CMB+kSZ and tSZ) or deprojected ($a'_i$ for CMB, tSZ, and CIB). Eq.~\ref{eq.ILCweights} shows that the overall normalization of the frequency dependence of a deprojected component does not matter, as it cancels in the weights. As mentioned above, for the CMB+kSZ component, the frequency dependence for our calibrated maps in differential CMB units is simple: $f_i=1$.

For the tSZ effect, the map we wish to obtain is a map of the dimensionless Compton-$y$ parameter, which is related to the tSZ temperature anisotropy contribution via~\cite{ZS1969,SZ1970}:
\be
\Delta T_{\rm{tSZ}}(\nu) = y T_{\rm{CMB}} f_{\rm{tSZ}}(\nu) \,,
\ee
where
\be
\label{eq:tszresponse}
f_{\rm{tSZ}}(\nu) = x\frac{e^x+1}{e^x-1} - 4
\ee
with $x=h\nu/k_BT_{\rm{CMB}}$. This implies that for a monochromatic bandpass, $f_i = T_{\rm{CMB}} f_{\rm{tSZ}}(\nu_i)$ for the tSZ effect (see Appendix~\ref{app:color} for color corrections in the realistic case). Eq.~\ref{eq:tszresponse} is the non-relativistic tSZ spectral dependence; we neglect relativistic corrections to the tSZ effect~(e.g.,~\cite{Nozawa2006}) as these are only relevant for the most massive clusters in the Universe.  If necessary for a particular analysis, relativistic corrections can be accounted for by modeling the SED with a moment expansion and deprojecting this component in the ILC~\cite{Chluba2017,Remazeilles2019}.

Finally, for the CIB SED, we use a modified blackbody:
\be
\label{eq:fcib}
f_{\rm{CIB}}(\nu) \propto \frac{\nu^{3+\beta}}{e^{h\nu/(k_B T_{\rm CIB})} - 1} \left( \left. \frac{dB(\nu, T)}{dT} \right|_{T=T_{\rm CMB}} \right)^{-1} \,, 
\ee
where $\beta = 1.2$, $T_{\rm CIB} = 24$ K, and the final factor is the conversion from specific intensity units to differential CMB temperature units ($B$ is the Planck function).  The overall normalization of this SED is not relevant, as we only consider it as a contaminant to be deprojected.  The SED parameters are consistent with predictions of the sky-average CIB SED for the CIB halo model fit to the \emph{Planck} CIB power spectra measurements ~\cite{Planck2013CIB}.\footnote{The SED parameters here do not correspond to physical SED parameters of an actual infrared source.}   The frequency dependence of the CMB+kSZ, tSZ, and CIB components, normalized to their values at 150 GHz, is shown in Figure~\ref{fig:sed}.  The figure shows the color-corrected responses in the scale-independent limit, i.e., at $\ell=0$.  In all analyses performed in this paper, we account for the full scale-dependent responses, as detailed in Appendix~\ref{app:color}.

\section{Data}
\label{sec:data}


\begin{table*}[ht]
\label{tab:maps}
\centering
\caption{Maps used for component separation. The central frequencies are not intended to be precise; we use the full bandpass in our analysis. While the beam FWHM for \emph{Planck} reflects what we assume in our analysis, for ACT, only rough estimates are shown in this table; we use the appropriate harmonic transfer function in our analysis. The noise sensitivities are those reported by \emph{Planck} for LFI~\cite{LFIMaps} and HFI~\cite{Planck2015HFIcal}, and those for ACT estimated from the multipole region $5000 < \ell < 5500$.}
\begin{tabular}[t]{l|l|l|l|l|l|l} 
\textbf{Name} & \textbf{Season} & \textbf{Array} & \textbf{Region} & \textbf{Freq.} & \textbf{FWHM} & \textbf{Noise} \\ 
 &  &  &  & (GHz) & (arcmin) & ($\mu {\rm K} \cdot$arcmin) \\ \hline 
P030 & Planck PR2 & - & - & 30 & 32.408 & 195.1 \\ 
P044 & Planck PR2 & - & - & 44 & 27.100 & 226.1 \\ 
P070 & Planck PR2 & - & - & 70 & 13.315 & 199.1 \\ 
P100 & Planck PR2 & - & - & 100 & 9.69 & 77.4 \\ 
P143 & Planck PR2 & - & - & 143 & 7.30 & 33.0 \\ 
P217 & Planck PR2 & - & - & 217 & 5.02 & 46.8 \\ 
P353 & Planck PR2 & - & - & 353 & 4.94 & 153.6 \\ 
P545 & Planck PR2 & - & - & 545 & 4.83 & 818.2 \\ 
BN\_1\_150 & ACTPol 2015 & PA1 & {\tt BN} & 150 & 1.4 & 66.8 \\ 
BN\_2\_150 & ACTPol 2015 & PA2 & {\tt BN} & 150 & 1.4 & 35.7 \\ 
BN\_3\_098 & ACTPol 2015 & PA3 & {\tt BN} & 98 & 2.2 & 31.9 \\ 
BN\_4\_150 & ACTPol 2015 & PA3 & {\tt BN} & 150 & 1.4 & 47.1 \\ 
D56\_1\_150 & ACTPol 2014 & PA1 & {\tt D56} & 150 & 1.4 & 26.6 \\ 
D56\_2\_150 & ACTPol 2014 & PA2 & {\tt D56} & 150 & 1.4 & 18.6 \\ 
D56\_3\_150 & ACTPol 2015 & PA1 & {\tt D56} & 150 & 1.4 & 26.9 \\ 
D56\_4\_150 & ACTPol 2015 & PA2 & {\tt D56} & 150 & 1.4 & 18.0 \\ 
D56\_5\_098 & ACTPol 2015 & PA3 & {\tt D56} & 98 & 2.2 & 17.3 \\ 
D56\_6\_150 & ACTPol 2015 & PA3 & {\tt D56} & 150 & 1.4 & 26.9 
\end{tabular}
\end{table*}

We now apply the above formalism to data from the \emph{Planck} satellite and ACT to construct co-added ILC maps of the CMB+kSZ and tSZ signals.  We work with maps in two distinct, non-overlapping regions of the sky, labeled {\tt BN} and {\tt D56}. Roughly, the {\tt BN} region spans RA of 117 degrees to 150 degrees and declination of -2 degrees to 19 degrees with a total effective area in our analysis mask of 1633 sq.~deg, while the {\tt D56}
region spans RA of -9 degrees to 43 degrees and declination of -7 degrees to 4 degrees with a total effective area of 456
sq.~deg. We show these sky regions in Figure~\ref{fig:footprint}. These regions correspond to areas of the sky that have deep observations from ACT in 2014 and 2015.

\subsection{ACT}
\label{sec:ACTdata}

We use the intensity (temperature) data taken with the ACTPol receiver at 98 GHz and 150 GHz, using only the night-time data taken during 2014 and 2015, where night-time data refers to observations between 23:00 and 11:00 UTC. This data set consists of maps made from detector arrays PA1 and PA2 both observing at 150 GHz in 2014 (for {\tt D56}) and 2015 (for {\tt BN} and {\tt D56}), and a dichroic detector array PA3 observing at 98 GHz and 150 GHz in 2015 (for {\tt BN} and {\tt D56}).  Details of the ACTPol instrument, including the detector arrays, can be found in~\cite{thornton/2016}.  A more detailed description of the full data set will appear in \cite{ChoiEtAl,AiolaEtAl}.  The Advanced ACTPol instrument has collected more data from 2016 on and the source subtraction procedure used in this analysis uses information on the location of compact sources inferred from co-add maps that include data up to the 2017 observing season for greater precision in source parameters. Similarly, in the stacking analysis on tSZ-selected clusters used for validation of the Compton-$y$ maps (see Sec.~\ref{sec:stacks}), we include confirmed cluster locations inferred from co-add maps that include data up to the 2018 season.  The ILC maps, however, do not contain any post-2015 ACT data.

When including ACT data in the ILC co-addition procedure, we exclude all Fourier modes with $\ell<500$, since these modes are measured well by \emph{Planck}, while the ACT data in this region of Fourier space can have low signal-to-noise due to atmospheric noise. At $\ell>500$, the ACT transfer function is unity to within 0.2\% (see \cite{ChoiEtAl,AiolaEtAl} for details).

\begin{figure}
\includegraphics[width=\columnwidth]{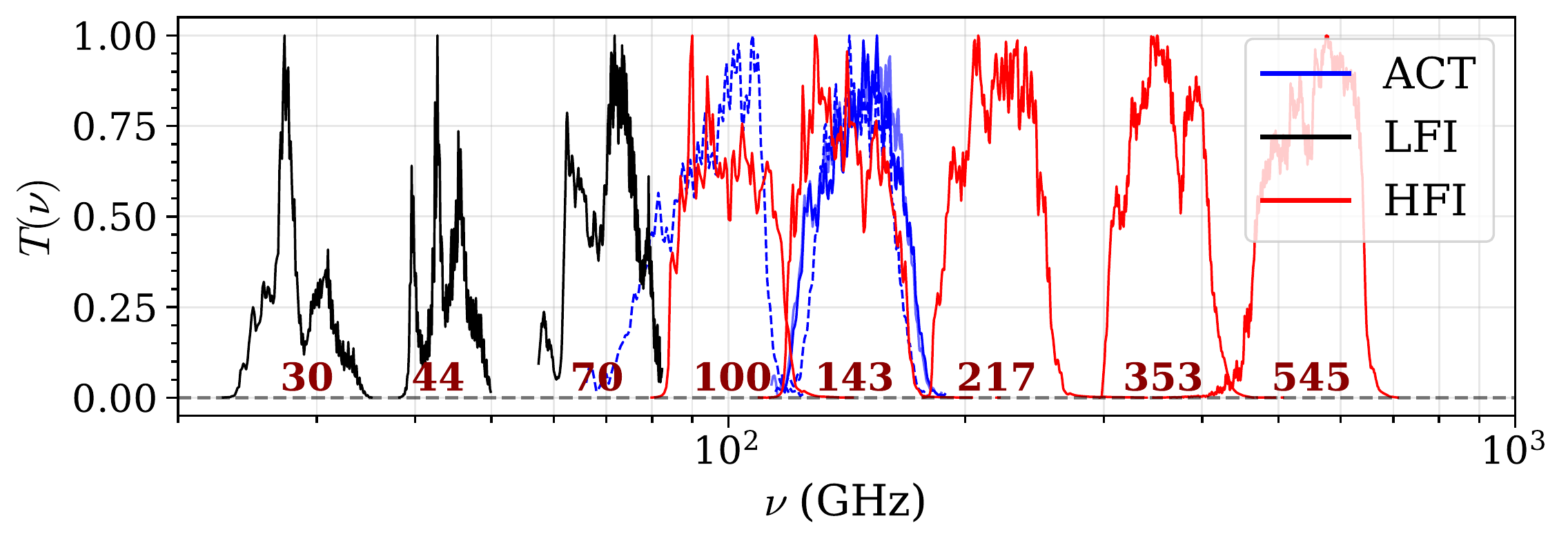}
\centering
\caption{Bandpass transmission from \emph{Planck} and ACT for maps used in
component separation, normalized to the maximum in each. The \emph{Planck} LFI bands are shown in black, the HFI bands in red, and the ACTPol bands in blue (dichroic PA3 array in dashed).  Three ACTPol bandpasses lie very close to one another at 150 GHz, namely, the PA1 and PA2 arrays (solid blue) and the PA3 array (dashed blue). We show in dark red text in units of GHz the rough central values of ``frequency groups'', where the grouping is done only for the purpose of obtaining co-added spectra for covariance estimation (see Sec.~\ref{sec:covest}).}\label{fig:bandpass}
\end{figure}

The ACTPol bandpasses for the three arrays used in this work (PA1, PA2, PA3) are shown in blue in Figure~\ref{fig:bandpass}.  The 150 GHz bandpasses are quite similar and lie nearly on top of one another in the figure. The bandpasses were measured using a Fourier Transform Spectrometer (FTS) at the ACT site after the receiver was installed in the telescope \cite{thornton/2016}. The ACT beams were measured from observations of Uranus (see \cite{ChoiEtAl,AiolaEtAl}) and subsequently corrected for the fact that the Rayleigh-Jeans spectrum of Uranus is different from the CMB blackbody spectrum. Appendix \ref{app:color} describes color corrections accounting for the measured bandpasses. 

\subsection{\emph{Planck}}
\label{sec:Planckdata}

The \emph{Planck} data set that we use consists of maps from both the Low
Frequency Instrument (LFI) and High Frequency Instrument (HFI).  These maps come
from the PR2 (2015) data release. We refrain from using the PR3 (2018) release
maps since the effective intensity bandpasses become component-dependent due to
the polarization systematics-cleaning procedures~\cite{Planck2018compsep}; this
issue would make our analysis much less straightforward. To avoid numerical
issues due to noise in cross-correlations that enter the covariance matrix
calculation, we exclude low signal-to-noise ratio modes of the \emph{Planck}
maps by using modes with $20 < \ell < 300$ for LFI 30 and 44 GHz, $20 < \ell <
2000$ for LFI 70 GHz, and $20 < \ell < 5800$ for HFI 100, 143, 217, 353, and 545
GHz. We exclude $\ell<20$ from the \emph{Planck} maps since this is close to the
largest scale supported in our sky regions. We do not include the 857 GHz maps
due to their more uncertain calibration~\cite{Planck2015HFIcal}. We use
the \emph{Planck} full-mission maps for the co-addition process, and the HFI
half-mission maps and LFI half-ring maps for covariance calculations from data
splits.  We use bandpass transmission data provided by \emph{Planck} for the
LFI~\cite{Planck2013LFIdataproc} and HFI~\cite{Planck2013HFIspectralresp}
channels. We verify that our implementation of these bandpasses reproduces the
unit conversions and color corrections in Table 6
of~\cite{Planck2013HFIspectralresp} to high precision. The LFI and HFI
bandpasses are shown in Figure~\ref{fig:bandpass} in black and red,
respectively.  We treat the \emph{Planck} LFI and HFI beams as Gaussian, with
FWHM values given in
Table~\ref{tab:maps}~\cite{Planck2015LFIbeams,Planck2015HFIbeams}. We use the
FWHM of the Gaussian whose solid angle is equivalent to that of the effective
beams, as provided by \emph{Planck} (see Table 3 of~\cite{Planck2015LFIbeams}
and Table 3 of~\cite{Planck2015HFIbeams}), following the \emph{Planck} $y$-map
construction~\cite{Planck2015ymap}. Any small deviations from Gaussianity in the
tails of the \emph{Planck} beams lie in high-multipole regions where the ACT
channels dominate in our ILC (see Figure~\ref{fig:weights}).

\section{Data Processing}
\label{sec:processing}

\subsection{Source detection and subtraction}
\label{sec:srcsub}

Since we will be calculating Fourier transforms in the ILC analysis, we wish to remove any bright compact objects that can cause ringing. Removing compact sources has the additional advantage of reducing the overall level of contamination from radio sources in lensing maps and cross-correlations derived from these products. We perform  compact source subtraction in three steps as done in \cite{AiolaEtAl,ChoiEtAl}, prior to running the component separation algorithm. First, a fiducial catalog is built for each frequency using a matched-filter source-finder on co-added maps using all night-time ACT data from 2013 to 2017, with sources detected at 5$\sigma$ or more being kept. Then per-season, per-array catalogs are built by using this catalog as a template and fitting only the source amplitudes on the individual season maps. This fit is performed jointly for sources that overlap spatially in the maps, and the amplitudes in the original cross-season catalog are used as weak ($10^{-3}$ relative weight) priors to break any degeneracies. Sources in these per-season, per-array catalogs are subtracted from each map, with the exception of the \emph{Planck} 545 GHz map; see Sec.~\ref{sec:Planckproc}. The subtraction is performed by modeling the point source to appear in the map with a profile identical to the beam. Since the subtraction threshold is set based on the signal-to-noise ratio (SNR) in the total map, but amplitudes are then fit in individual season maps, it is not straightforward to express the source subtraction level in terms of a specific flux. However, the typical detection noise in the co-added 2013-2017 map is 1.4 mJy at 98 GHz and 0.9 mJy at 150 GHz for {\tt D56} and 2.1 mJy at 98 GHz and 1.8 mJy at 150 GHz for {\tt BN}. The detection threshold for source subtraction is then roughly 5 times these numbers. In total, this subtraction was performed on about 700/2800 point sources for {\tt D56}/{\tt BN} at 98 GHz and 900/2200 at 150 GHz.

\subsection{Processing of \emph{Planck} data}
\label{sec:Planckproc}

Each \emph{Planck} map is reprojected from HEALPix~\cite{HEALPix}\footnote{http://healpix.sourceforge.net} to the pixelization corresponding to a Plate Carr\'{e}e (CAR) projection (i.e., equirectangular conformal on the equator, see e.g. \cite{WCS1}) with a 0.5 arcmin pixel size used in the ACT data. This reprojection is done by calculating the spherical harmonic transform of the HEALPix map (band-limited to $\ell<3072$ for LFI maps and $\ell<6144$ for HFI maps) and then calculating the inverse spherical harmonic transform on to the CAR pixelization of the ACT maps after applying a rotation from Galactic to Equatorial coordinates. Following this reprojection, the \emph{Planck} maps are source-subtracted using the method described in Sec.~\ref{sec:srcsub} for all frequencies from 30 GHz to 353 GHz. For the 545 GHz map, we do not follow the procedure in Sec.~\ref{sec:srcsub}, because many new sources are present at this frequency that do not appear in the catalog made from co-added 98 GHz and 150 GHz ACT data. Instead, we use the publicly available \emph{Planck} Catalog of Compact Sources~\cite{Planck2015PCCS} for sources detected at \emph{Planck} 545 GHz (PR2 2015 PCCS2), and subtract compact sources based on the reported fluxes in that catalog.

\subsection{Processing of ACT data}
\label{sec:ACTproc}

While the source subtraction described in Sec. \ref{sec:srcsub} significantly reduces compact source contamination, we must account for two classes of compact contamination in the ACT maps. We categorize the residuals in the ACT data from all arrays and patches from 2013-2016, although only a fraction of them appear in the 2014--2015 {\tt BN} and {\tt D56} patches considered in this work. The classes are:

(a) Residuals at the locations of bright sources that received special treatment in the map-maker (see \cite{NaessSources} for details). There are at most 132 of these locations across all arrays and patches from 2013--2016.

(b) Residuals at the locations of bright sources that did not receive special treatment in the map-maker, but appear in external catalogs and may be extended. The source subtraction procedure used in \cite{ChoiEtAl,AiolaEtAl} handles extended sources well, but the brightest of these leave visible residuals. A cut at SNR $= 90$ is used to identify 23 such sources across all arrays and patches from 2013--2016 that are not already in the previous list.

To handle contamination from the above classes of objects, we ``inpaint''
circular holes centered on them. We fill circular holes around these objects by
finding the maximum-likelihood solution for pixels within the hole constrained
by the pixels in a context region around the hole, and subsequently adding a
realization of ``noise'' to the pixels inside the hole by sampling the
covariance matrix for the maximum-likelihood solution. This corresponds to the
brute-force solution in \cite{Bucher:2012}, rather than the conjugate gradient
algorithm presented there involving all pixels in the map. When only a small
number of objects need to be inpainted, the former is much faster than the
latter. All holes are chosen to be 6 arcminutes in radius, embedded in a square
context region of width 20 arcminutes. The noise model used to build the
covariance matrix assumes (a) a CMB spectrum diagonal in Fourier space and
corresponding to a fiducial lensed theory spectrum obtained with
CAMB \cite{CAMB},\footnote{\url{http://camb.info}} which is convolved with the
one-dimensional beam corresponding to the array under consideration, (b)
inhomogeneous white noise diagonal in pixel space obtained from the inverse
variance per pixel output from the map-maker, and (c) that the solutions in each
array are independent of each other despite sharing common CMB. The solution is
performed jointly in Stokes $(I, Q, U)$-space even though some of the selected
sources are unpolarized. While the inpainted $Q$ and $U$ maps are not used in
the products described here, they are used in the CMB lensing reconstruction
presented in \cite{DarwishEtAl}. Within these holes, the inpainting
procedure described above ignores correlated noise, lensing, foregrounds, and
correlations due to common signal and atmosphere between arrays. However, since
these effects on the covariance matrix (used only for inpainting) are at most at
the tens of percent level, and since only a total fraction of the map of less
than 0.3\% is inpainted this way, we expect negligible changes to the final
component separated maps themselves due to these approximations.

\subsection{Preparation}
\label{sec:dprep}

After compact source subtraction and inpainting, we apply an apodized mask to each map. The apodized mask restricts our analysis to the well-crosslinked region used for power spectrum measurements in \cite{ChoiEtAl,AiolaEtAl}. In addition, we deconvolve the pixel window function of the ACT maps in 2D Fourier space, but we leave the HEALPix pixel window function of the \emph{Planck} maps untouched, treating it as entering the assumed \emph{Planck} beam transfer function.

\section{Algorithm}
\label{sec:algorithm}
\begin{figure*}[t]
\includegraphics[width=0.425\textwidth]{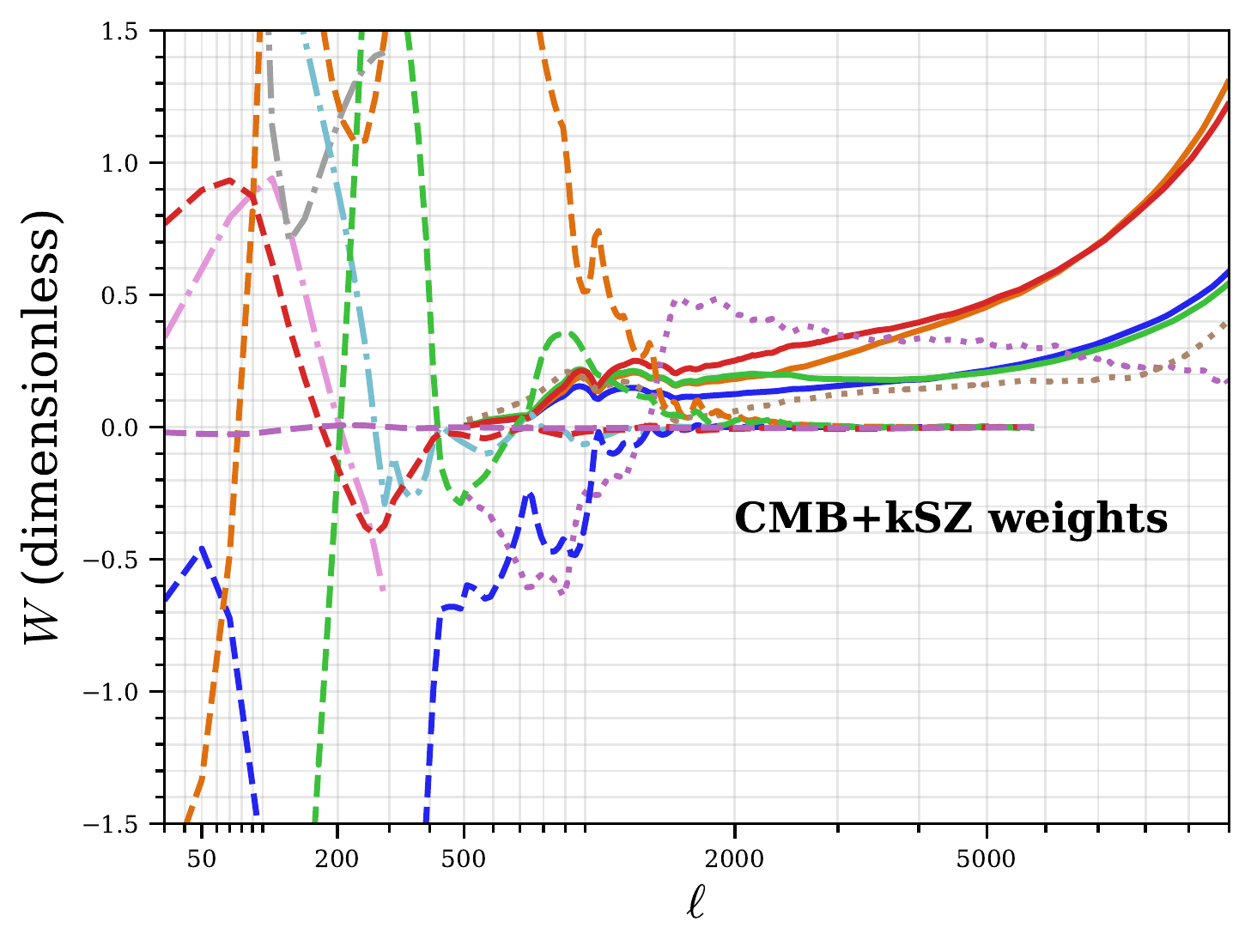} 
\includegraphics[width=0.56\textwidth]{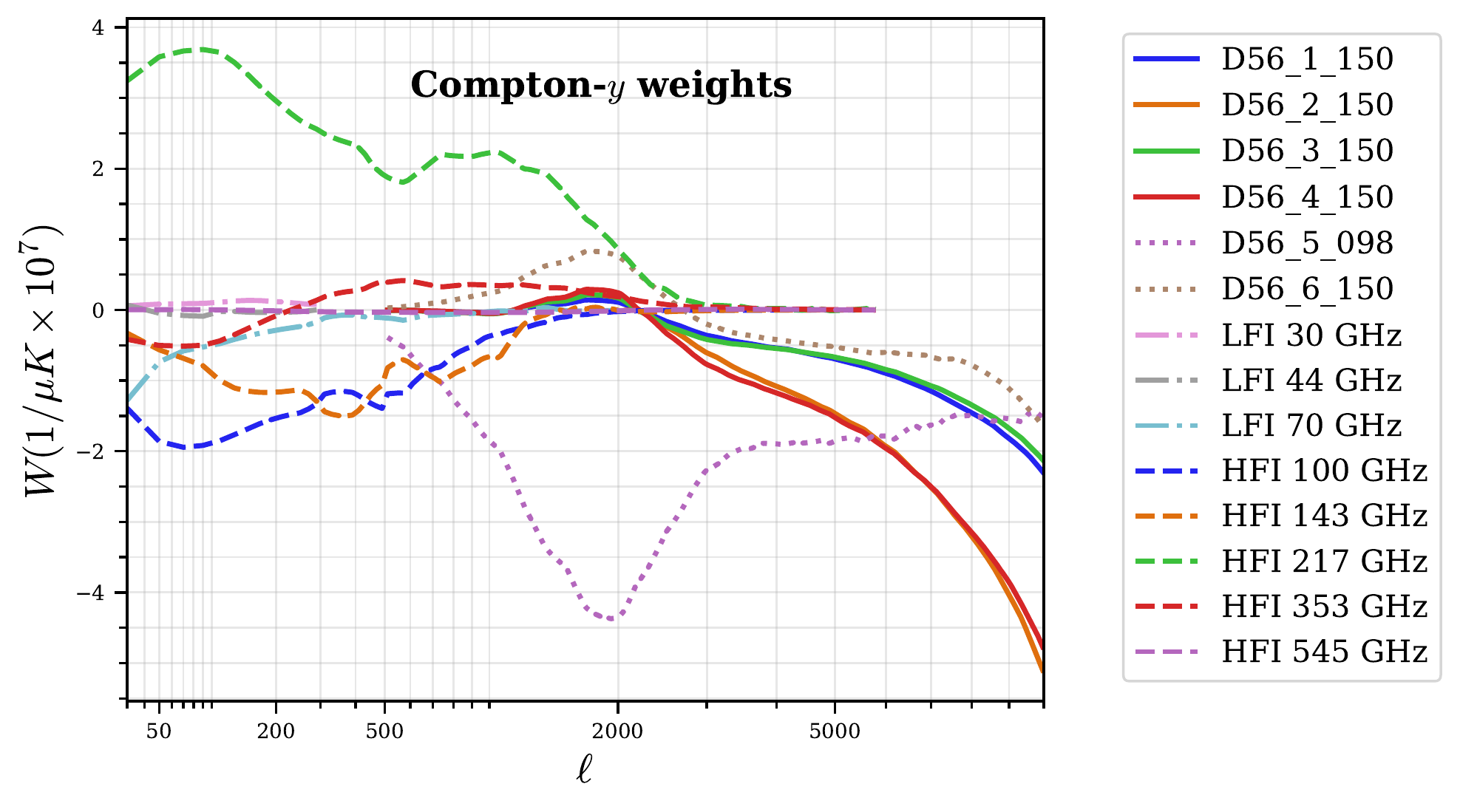}
\centering
\caption{ILC weights for CMB+kSZ (left) and Compton-$y$ (right) maps with no
contaminants deprojected in the {\tt D56} region.  While our ILC method operates
in 2D Fourier space, we show here 1D binned weights as a function of multipole
for each of the maps entering into the ILC reconstruction. \emph{Planck} LFI
weights are shown with dash-dotted lines, HFI weights with dashed lines, the ACT dichroic array weights with dotted and the rest of the ACT array weights with solid lines. The small scales are clearly dominated by ACT, due to
its high resolution, while the large scales are dominated by \emph{Planck}, due
to the significant large-scale atmospheric noise in ACT, as well
as \emph{Planck}'s wider frequency coverage. The ACT 150 and 98 GHz channel
weights approximately overlap with the HFI 143 and 100 GHz channel weights,
respectively, in the region where both experiments have non-negligible
signal-to-noise. The array names for the ACT maps are the same as those in Table~\ref{tab:maps}. The horizontal axis has been transformed by $x^{0.3}$ to highlight information from both large and small scales.}
\label{fig:weights}
\end{figure*}

Since we use the CAR projection, each map is on a 2D grid for which a 2D Fourier space can be constructed. As described in Sec.~\ref{sec:data}, we construct maps in two distinct regions of the sky where deep observations of ACTPol data are available, {\tt D56} and {\tt BN}. For each of these regions, we use maps of the microwave sky measured by ACTPol (98, 150 GHz) as well as by the \emph{Planck} LFI (30, 44, 70 GHz) and HFI (100, 143, 217, 353, 545 GHz) instruments. At each of the aforementioned frequencies, one ``array'' is available from \emph{Planck} (corresponding to the full \emph{Planck} mission), and multiple arrays are available for ACTPol (corresponding to multiple observation seasons and detector arrays). For each \emph{Planck} array, two splits of the data are available (corresponding to half-mission splits for HFI and half-ring splits for LFI) and four splits are available for ACTPol (corresponding to the time-ordered-data splitting scheme described in \cite{ChoiEtAl,AiolaEtAl}). The splits have independent instrument noise and thus can be used to build instrument noise estimates.

After defining a domain consisting of pixels in the 2D Fourier space of the map geometry in each region ({\tt BN} or {\tt D56}) under consideration, our algorithm for constructing component maps consists of the following steps:
\begin{enumerate}
    \item For each pair of arrays, estimate the instrument noise covariance from the power spectra of difference maps (see Sec.~\ref{sec:covest} for details on Steps 1-6);
    \item Estimate the sky covariance by subtracting the noise estimate from the total spectrum;
    \item Where multiple arrays are available for nearby frequencies, co-add the sky covariance with optimal weights; 
    \item Smooth the co-added signal spectra calculated for each frequency pair in radial annuli in 2D Fourier space and assign to each array pair, since the sky signal is expected to be fairly isotropic;
    \item Smooth the noise covariance in Cartesian sub-blocks of 2D Fourier space to preserve information about the noise anisotropy and assign the result to the relevant array pair;
    \item Use the smoothed signal and noise covariances to get an estimate of the total covariance for each array pair;
    \item Solve for the component of interest in each Fourier pixel using the estimated array-array covariance, known response of each array to each component, deprojecting components with known spectra as needed, and inverse-Fourier transform to obtain the corresponding component-separated maps (see Sec.~\ref{sec:coadd} for details).
\end{enumerate}
We describe these steps in detail in the following subsections. In Sec.~\ref{sec:sims}, we describe how we apply the entire algorithm described here to simulations, in order to verify our pipeline and inform analysis choices made here.

\subsection{Covariance estimation}
\label{sec:covest}

Steps 1--6 are described here. In a given patch (either {\tt D56} or {\tt BN}), we construct an $N_a \times N_a$ covariance matrix for each 2D Fourier-space pixel, from each of the $N_a$ frequency maps. For {\tt D56}, $N_a=14$, corresponding to eight \emph{Planck} frequencies, two detector sets from ACT Season 2014 at 150 GHz (PA1 and PA2), three detector sets from ACT Season 2015 at 150 GHz (PA1, PA2, and PA3 at 150 GHz), and one detector set from ACT Season 2015 at 98 GHz (PA3 at 98 GHz). For {\tt BN}, $N_a=12$, corresponding to the eight \emph{Planck} frequencies, three detector sets from ACT Season 2015 at 150 GHz (PA1, PA2, and PA3 at 150 GHz), and one detector set from ACT Season 2015 at 98 GHz (PA3 at 98 GHz).

We split the calculation of the covariance $C$ into a sum of signal $S$ and noise $N$ parts. The signal part includes the contribution from all components that are constant in time, including the lensed CMB and astrophysical foregrounds. The noise part is calculated from splits of the data interleaved in time.

For Step 1, we estimate the noise power needed for the diagonal of the covariance matrix and for the pair corresponding to ACT PA3 at 150 GHz and ACT PA3 at 98 GHz, which have up to 40\% correlated atmospheric noise since they come from the same dichroic detector array. For all other pairs (including ACT-\emph{Planck} pairs), the instrument noise correlation is expected to be zero.

To get a noise estimate for a given array pair, we do the following. For each array in the pair, we obtain a difference map $d_i = s_i - c$ by subtracting the ``co-add'' map $c$ from each split map $s_i$ indexed by $0<i<k$, where ``co-add'' here refers to the map containing all of the observations in all of the splits. This removes all signal including CMB and foregrounds, but also any potential common systematic like ground pickup. The co-add map is $c = \sum s_i h_i / \sum h_i$ where $h_i$ is the inverse variance in each pixel. The number of splits is $k=4$ for ACT and $k=2$ for \emph{Planck}.

We then calculate the 2D FFT of the above difference map $\tilde{d}_i = \mathrm{FFT}(d_i~\tilde{h}_i~m_a)$ after weighting with the normalized inverse variance $\tilde{h}_i = h_i/\sum_i h_i$,  together with the apodized mask $m_a$ (see Sec.~\ref{sec:dprep}). With the Fourier transforms, we build an estimate for the noise power of the co-add in 2D Fourier-space by averaging over the noise power spectra obtained from each split:
\be
\label{eq.noisepower}
\mathcal{N}_{ab} = \frac{1}{k (k - 1) } \sum_i \frac{1}{w^i_2} \tilde{d}_i^a \tilde{d}_i^{b*} \,,
\ee
where $a$ and $b$ may index two different arrays as in the case of the PA3 98 GHz-150 GHz noise cross-power, and $w^i_2 = \langle (\tilde{h}_i~m_a(\mathbf{\hat{n}}))^2 \rangle$ accounts for the loss in power due to the apodized mask. This completes Step 1.

For Step 2, to calculate the signal power spectrum, we first consider the case where the array pairs have no correlated instrument noise (which includes all off-diagonals of the covariance matrix, except for the ACT PA3 98-150 GHz pair). For this case, we simply calculate the 2D cross-power of the total co-add maps of each array. For the case of diagonal elements and the 98-150 GHz pair, we subtract the unsmoothed noise estimate $\mathcal{N}_{ab}$ from the 2D cross-power of the total co-add maps. 

To improve the SNR of this signal covariance estimate, we co-add similar
estimates of signal covariance that belong to the same ``frequency group'' (Step 3, and see Figure \ref{fig:bandpass}). For example, despite the slightly different bandpasses, the \emph{Planck} 143 GHz signal power should be nearly identical to the signal power of any of the ACT 150 GHz arrays, so we co-add all of these into a single estimate which is used for each of those arrays.  The weights in the power spectrum co-add account for the differences in beams and noise in the different arrays. 

For Step 4, we next smooth the signal power estimates by averaging the 2D power spectra in annular bins of width $\Delta\ell=160$ and linearly interpolating the result back on to the 2D Fourier grid, thus providing us an estimate of the signal covariance $S$.

For Step 5, the final instrument noise power estimate $N_{ab}$ is obtained by block-averaging $\mathcal{N}_{ab}$, i.e. averaging the 2D power within blocks of size $\Delta\ell \times \Delta\ell$; since the noise power spectrum is estimated from the data itself, it must be smoothed to avoid biases due to chance fluctuations. We smooth the noise power estimate by treating the 2D noise power as a 2D image. We then smooth this ``image'' by applying a low-pass filter that removes high-frequency modes in the Fourier transform of the 2D noise power spectrum. We choose the low-pass filter so that the 2D noise power is effectively averaged in blocks of width $\Delta\ell=400$. This procedure smooths the 2D Fourier power while preserving anisotropy. Before smoothing, the 2D power is either (1) transformed into its logarithm (for the diagonal auto-noise-power case, with a numerical correction applied for the fact that the logarithm changes the distribution of averages), or (2) whitened, in the case of the cross-noise-power calculated between PA3 150 GHz and 98 GHz. The whitening is done by fitting (in $\ell>500$) the radially binned power spectrum to the functional form $[(\frac{\ell_{\rm{knee}}}{\ell})^{-\alpha} + 1]  w^2$ and dividing out that fit, where $w$ is the white noise floor determined at large $\ell$. The corresponding inverse transform for each case is applied after smoothing.

\begin{figure*}
\includegraphics[width=0.95\textwidth]{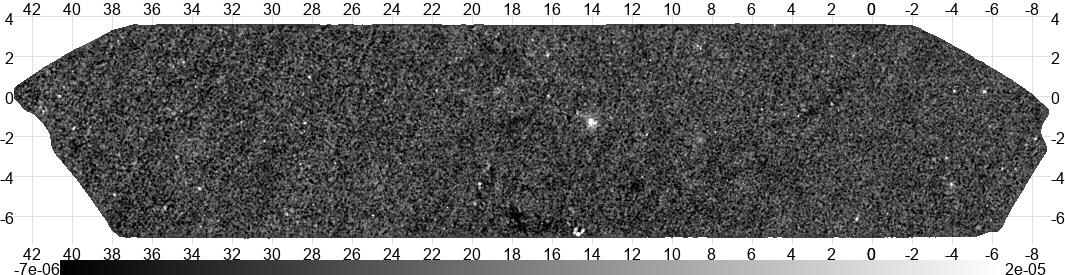} \\
\includegraphics[width=0.95\textwidth]{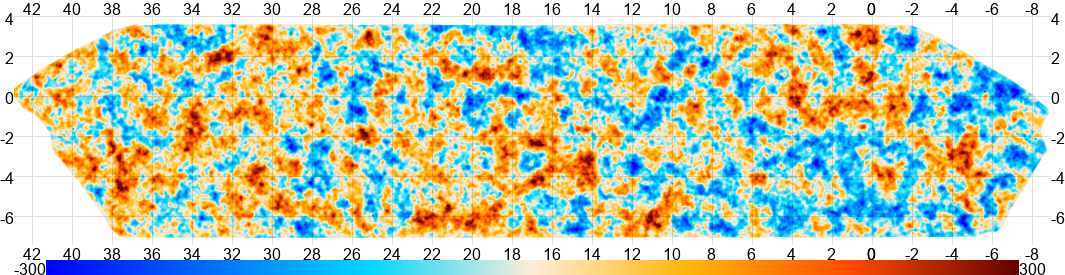} \\ 
\includegraphics[width=0.95\textwidth]{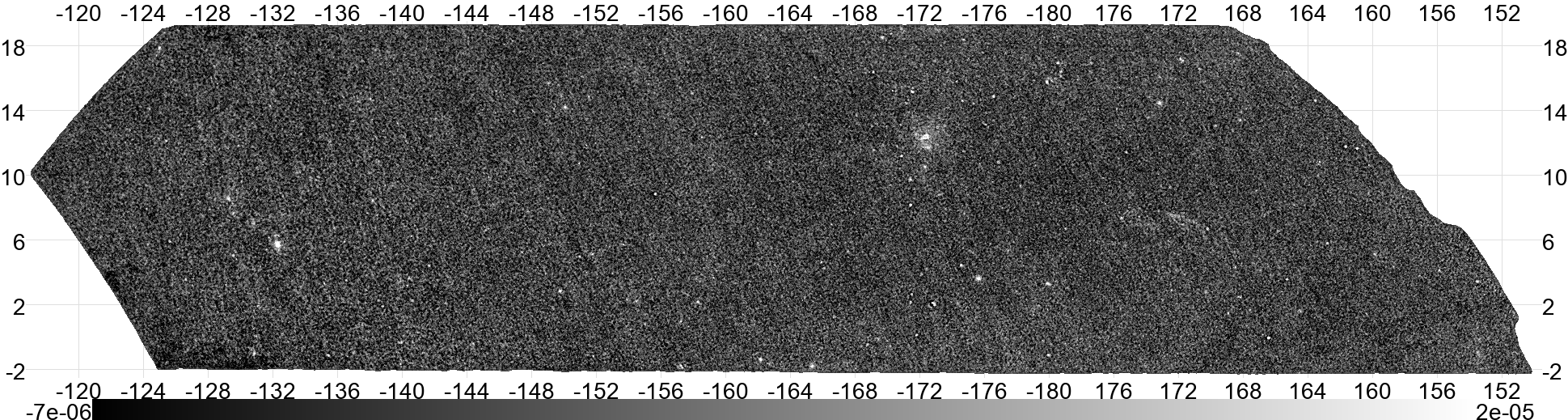} \\
\includegraphics[width=0.95\textwidth]{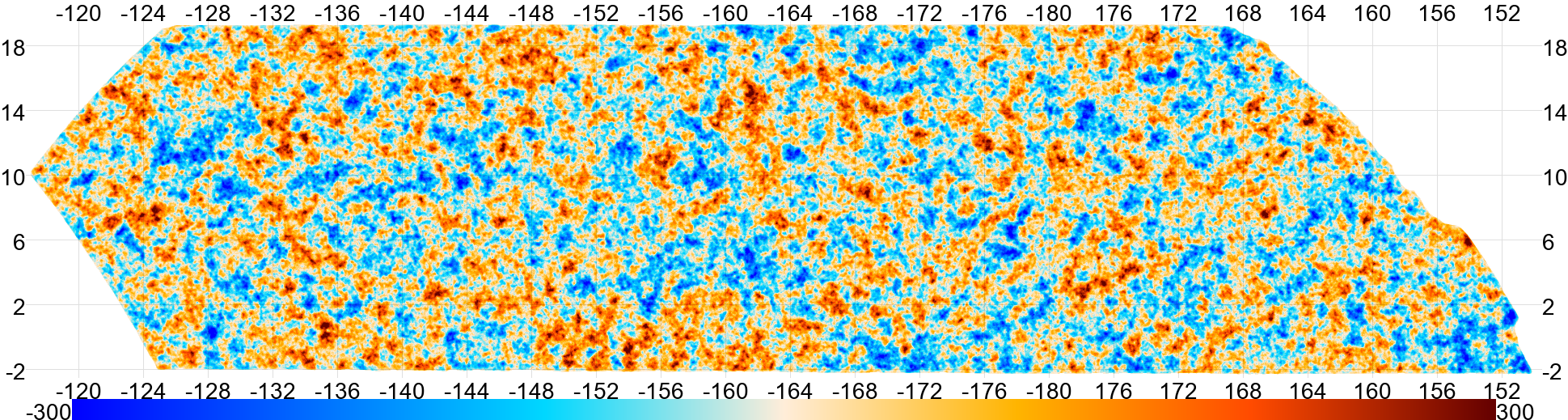} \\ 
\centering
\caption{Component-separated maps of the Compton-$y$ (gray-scale, first and third panels) and CMB+kSZ (false color, second and fourth panels) temperature anisotropies from our joint ACTPol + \emph{Planck} analysis. The vertical axis (declination) and horizontal axis (right ascension) are labeled in degrees. These maps have an effective beam corresponding to a 1.6 arcmin FWHM Gaussian. No additional filtering is applied in these images. The top two panels show the standard ILC Compton-$y$ map (dimensionless units) and the CMB+kSZ map (units of $\mu \rm{K}$) in the smaller {\tt D56} region ($\approx 456$ sq.~deg.), respectively, and the bottom two panels show the same for the larger {\tt BN} region ($\approx 1633$ sq.~deg.). The information visible by eye in the CMB+kSZ maps here is dominated by \emph{Planck} since we do not include ACT information in the co-add for $\ell<500$, but see Figure~\ref{fig:zoom} for a zoomed-in image where ACT also contributes substantial information. Because the Compton-$y$ maps are noise-dominated, inhomogeneities from the \emph{Planck} scanning pattern can be seen as diagonal stripes. However, since the tSZ distribution is highly non-Gaussian, many galaxy clusters can be seen by eye as saturated (white) points in the Compton-$y$ maps. The core of the Virgo cluster can be seen through its diffuse emission centered on \{RA=$-172.3^{\circ}$,DEC=$12.3^{\circ}$\}. Evidence of Galactic contamination can also be seen in the Compton-$y$ maps (as inferred from comparisons with \emph{Planck} 545 GHz maps where dust emission from the Galaxy dominates), e.g., at approximately  \{RA=$171^{\circ}$,DEC=$7^{\circ}$\}, \{RA=$14.5^{\circ}$,DEC=$-7^{\circ}$\} and \{RA=$17.8^{\circ}$,DEC=$-6.5^{\circ}$\}.}
\label{fig:fullmaps}
\end{figure*}

\begin{figure*}
\includegraphics[width=0.5\textwidth]{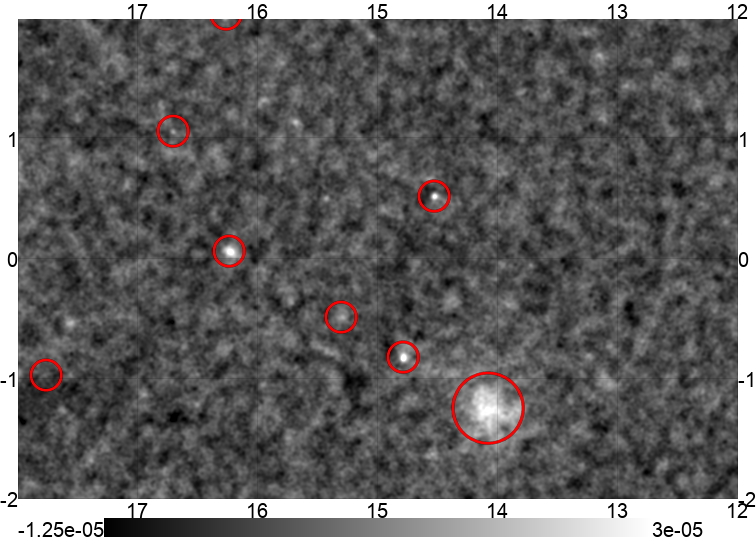} \\
\includegraphics[width=0.45\textwidth]{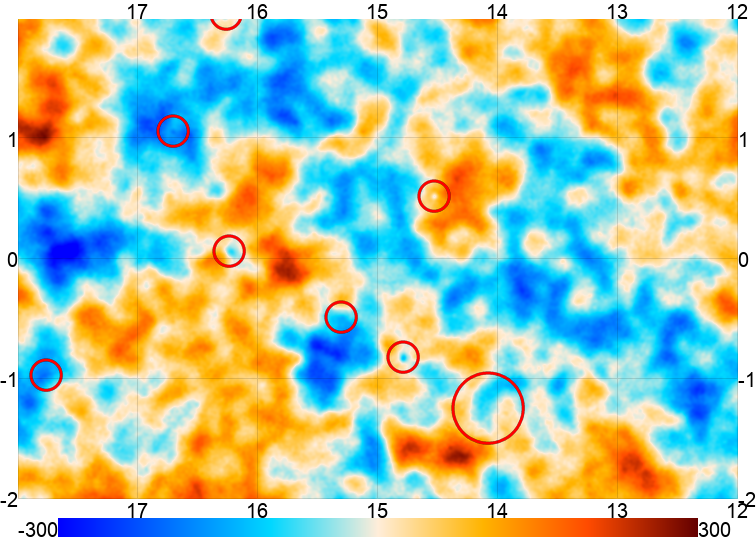} 
\includegraphics[width=0.45\textwidth]{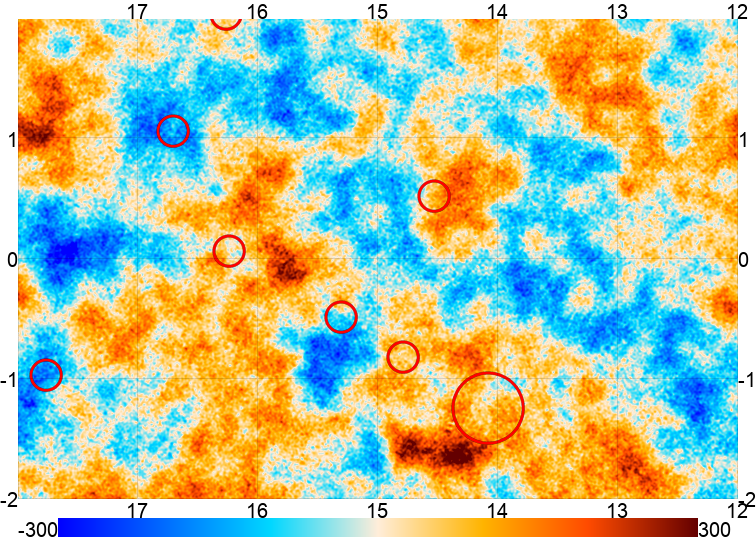}
\centering
\caption{Zoomed-in view of component-separated maps in a deep patch ($\approx 24$ sq.~deg.~out of the available 2100 sq.~deg.) near the
  equatorial region. The vertical axis (declination) and horizontal axis (right
  ascension) are labeled in degrees. The top panel shows the minimum-variance (standard ILC) Compton-$y$ map in
  this region (dimensionless color scale). Bright spots corresponding to galaxy clusters can be seen, in particular a large diffuse region in
the bottom right corresponding to Abell 119, a nearby cluster at $z=0.044$.  Known ACT tSZ clusters from \cite{HiltonEtAl} corresponding to an ACT detection signal-to-noise ratio $>4$ are circled in red, with Abell 119 (not detected previously in ACT due to its low redshift) located in the largest circle. The
bottom left panel shows the minimum-variance (standard ILC) CMB+kSZ map in this region.  Clear residual contamination at the locations of bright tSZ clusters can be seen as negative decrements due to
the map receiving large weights from the ACT 150 GHz and 98 GHz channels on small scales. The
bottom right panel shows the constrained ILC CMB+kSZ map with the tSZ signal deprojected, i.e., with the contaminant field $c$ in Eq.~\ref{eq.twocompmodel} corresponding to tSZ. The color scale for both CMB+kSZ maps is in $\mu$K. No residuals can be seen at the locations of clusters in the tSZ-deprojected map, at the cost of higher noise (see Figure \ref{fig:ypower}). All maps have been re-convolved to an effective Gaussian beam of FWHM
2.2 arcmin for display here (the native resolution of the standard ILC maps is 1.6 arcmin and that of the constrained ILC maps is 2.4 arcmin).}
\label{fig:zoom}
\end{figure*}

\begin{figure}
\includegraphics[width=\columnwidth]{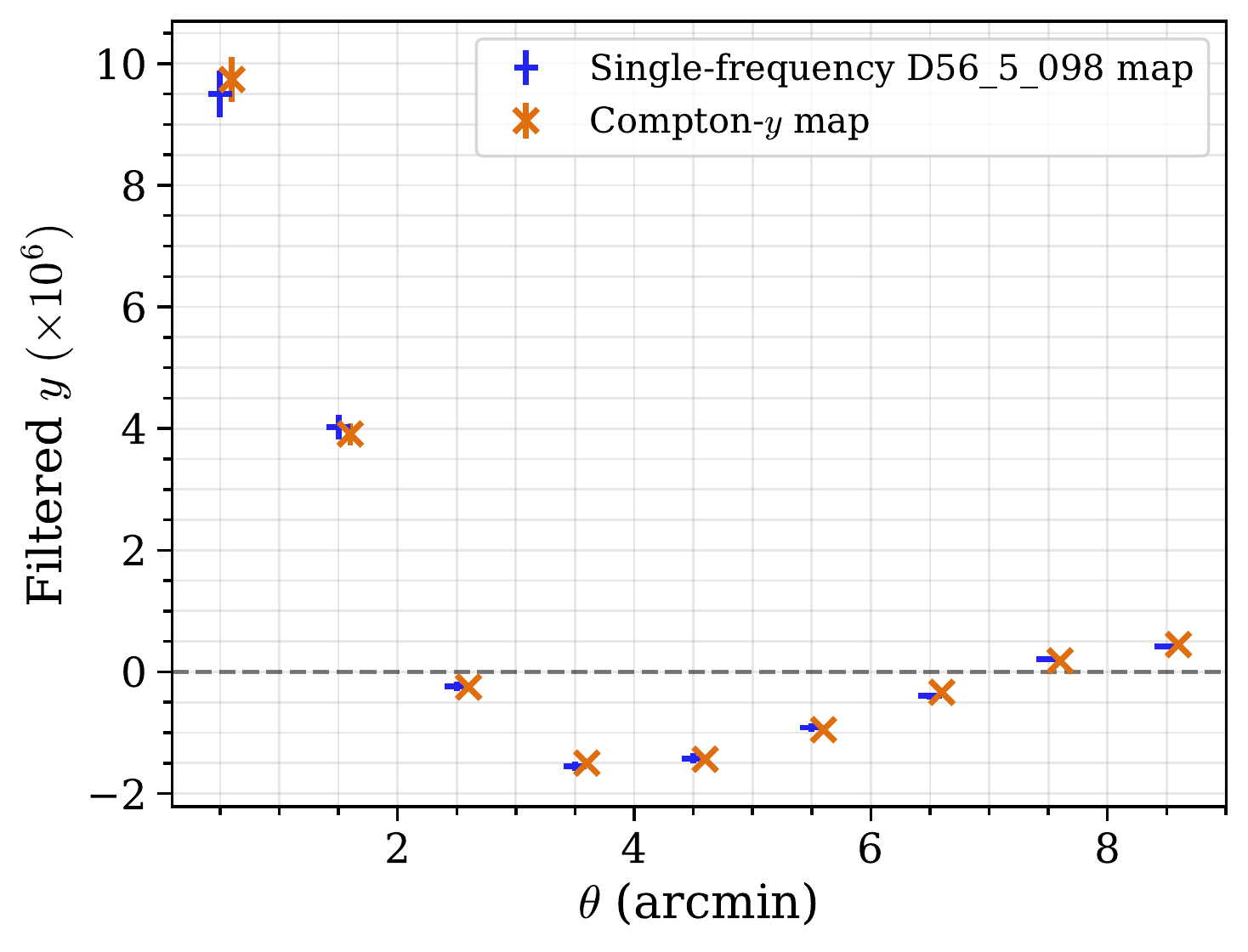}
\centering
\caption{Comparison of the profile of a stack on the locations of 179 ACT confirmed clusters in the joint
  \emph{Planck} + ACTPol standard ILC Compton-$y$ map in the {\tt D56} region with the same stack on a rescaled version of
  the single-frequency map from ACTPol array PA3 at 98 GHz. The rescaling factors
  apply corrections that include calibration, beam, and scale-dependent color
  corrections that have already been accounted for in the Compton-$y$ map. Both
  maps have also been high-pass filtered to $\ell>2000$ to reduce large
correlated scatter in the single-frequency maps (from atmospheric noise on large angular scales), which
would make comparison difficult. (Horizontal positions of the points have been
offset slightly for clarity. The errors on the profile from the Compton-$y$ map are smaller than the plot symbols.)}\label{fig:yprofile}
\end{figure}

\begin{figure}
\includegraphics[width=0.99\columnwidth]{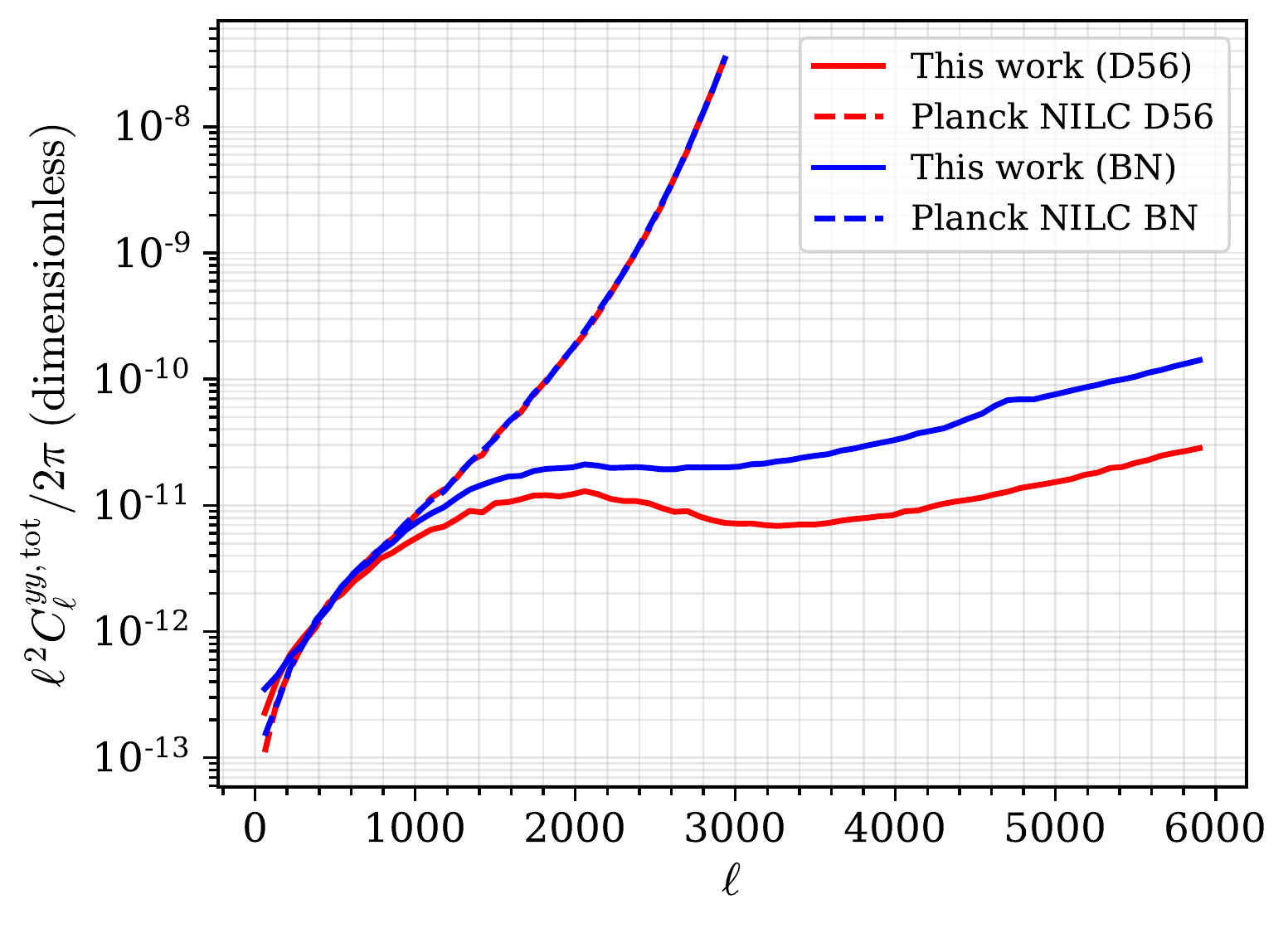} \\
\includegraphics[width=0.99\columnwidth]{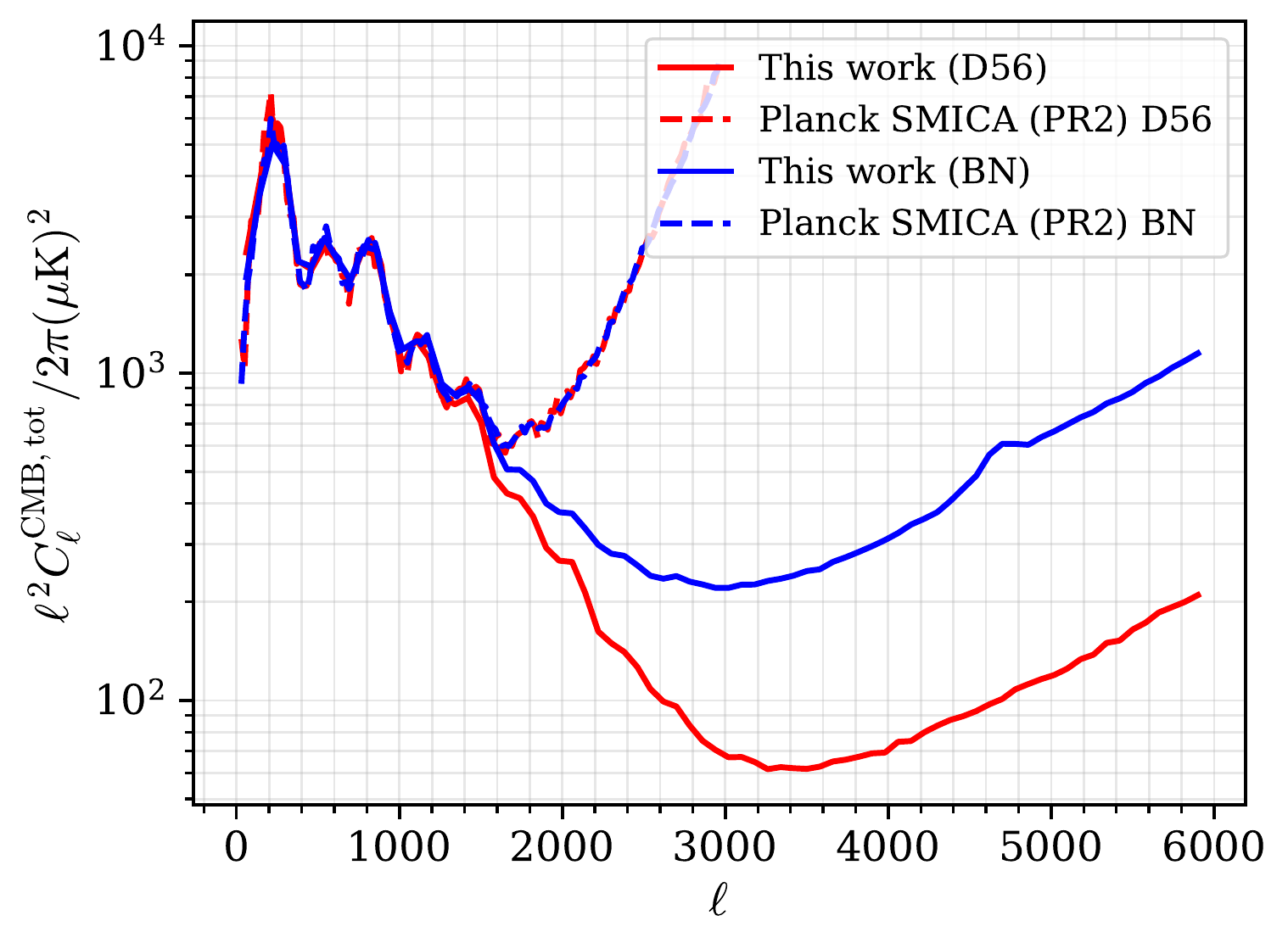} \\
\includegraphics[width=0.99\columnwidth]{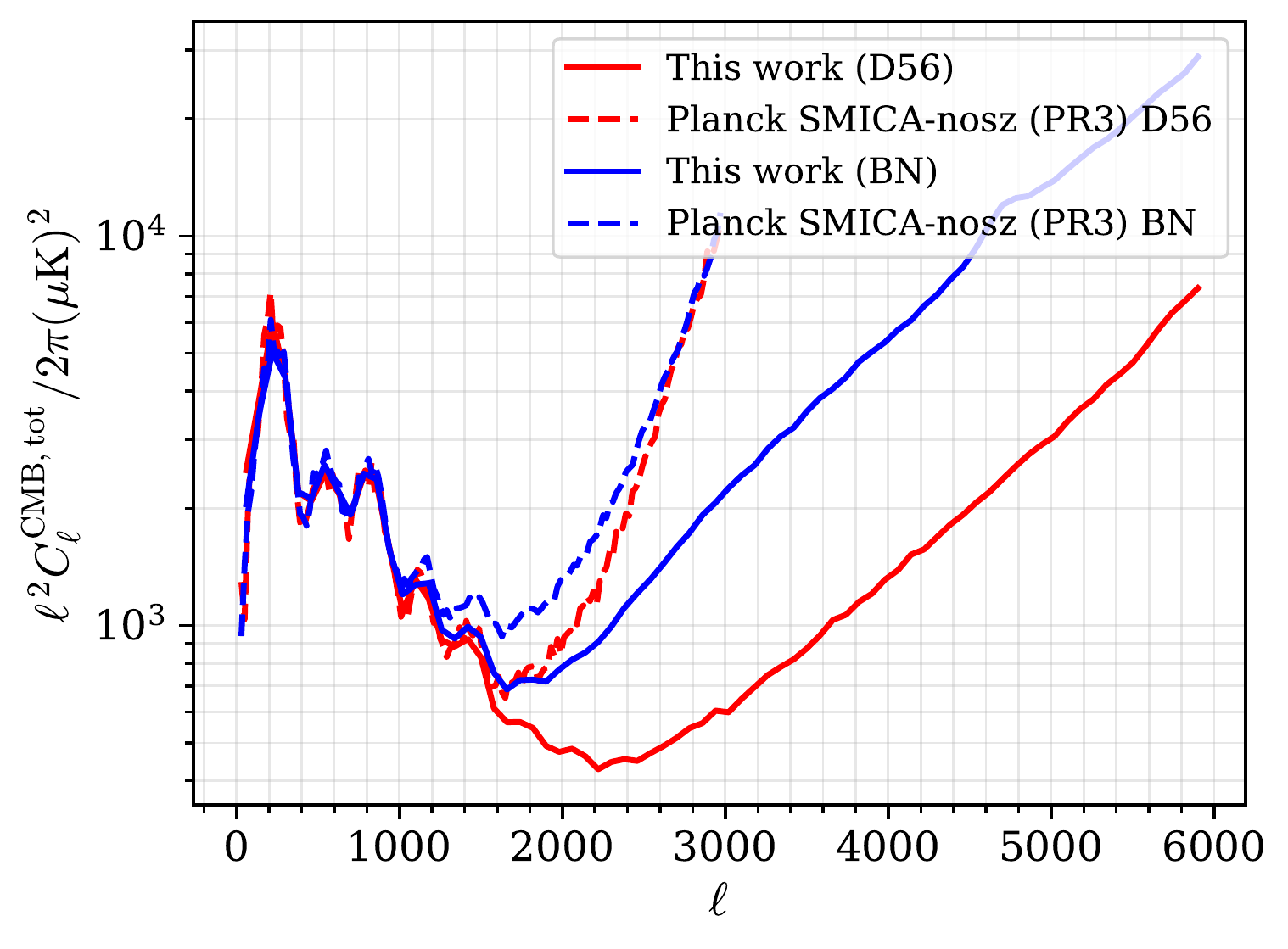}

\centering
\caption{Total beam-deconvolved auto-power spectra (including noise bias and residual foregrounds) of our ACTPol + \emph{Planck} ILC maps compared to similar maps from {\it Planck}. The top panel shows the Compton-$y$ power in the {\tt D56} (red) and {\tt BN} (blue) regions for this work (solid) and the {\it Planck} needlet ILC Compton-$y$ map (dashed). The middle panel shows similar power spectra for the CMB+kSZ maps, but compared to the {\it Planck} PR2 SMICA map. The bottom panel shows corresponding spectra for tSZ-deprojected versions of the CMB+kSZ maps. In all cases, orders of magnitude improvement in the signal-to-noise per mode can be seen for scales $\ell>2000$ where the \emph{Planck} spectra become completely noise dominated.}
\label{fig:ypower}
\end{figure}

\begin{figure}
\includegraphics[width=0.6\columnwidth]{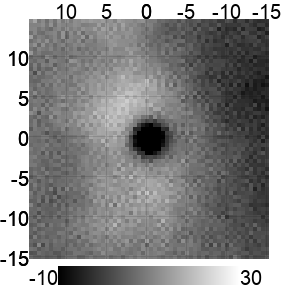} \\
\includegraphics[width=0.6\columnwidth]{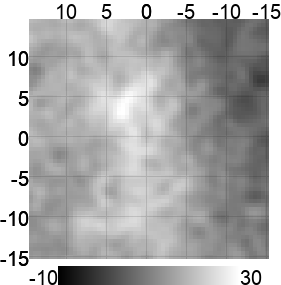}
\centering
\caption{Stacks on component-separated ILC CMB+kSZ maps at the locations of ACT tSZ-selected
  clusters with no foregrounds deprojected (top panel) and with the tSZ effect deprojected
  (bottom panel). The horizontal and vertical axes are in units of arcminutes and the color-scale units are $\mu$K.  As discussed for the \emph{Planck} cleaned CMB+kSZ maps in~\cite{MMHill}, it
  can be seen here that the
  standard ILC procedure leaves behind large tSZ residuals, whereas explicitly
  deprojecting the tSZ frequency dependence eliminates this residual, at the cost of higher noise. The fuzzy vertical stripe seen here is likely a chance alignment of correlated noise from the CMB, as a differently shaped feature is seen in an analogous stack in {\tt BN}. The
  maps used here are the joint \emph{Planck} + ACTPol ILC results in the {\tt D56} region, with the minimum variance map re-convolved to the same beam as the tSZ-deprojected map. This plot shows that our tSZ deprojection leaves no obvious residuals in the CMB+kSZ maps even at the locations of massive galaxy clusters.}
  \label{fig:ystack}
\end{figure}

Summing the signal and noise contributions (Step 6) completes the calculation of the total inter-array covariance $C=S+N$ for the ILC analysis. The above smoothing procedures are important for controlling the amplitude of the ILC bias that arises due to the finite number of modes used in estimating the covariance from the data. The choice of $\Delta\ell=160$ in the signal smoothing in particular is the smallest bin-width that reduces the ILC bias to an acceptable level (i.e., either well within the statistical error or at the $<0.5\%$ level), as determined from the pipeline simulations described in Sec.~\ref{sec:sims}.

\subsection{Co-addition and deprojection}
\label{sec:coadd}

Step 7 is described here. Given the covariance matrix estimated above, we calculate the weights (Eq.~\ref{eq.ILCweights}) for the CMB+kSZ and tSZ ILC constructions with and without deprojection of various contaminating components, as described in Sec.~\ref{sec:theory}. 

The azimuthally averaged, radially binned weights are shown in Figure~\ref{fig:weights} for the {\tt D56} region to aid in interpretation of the relevant contributions from various arrays to the component-separated maps (weights for the {\tt BN} region are similar and omitted for brevity). On large scales, \emph{Planck} dominates the information content, as these modes are lost to atmospheric noise in ACT.  The wide frequency coverage of \emph{Planck} is useful for mitigating the large number of foregrounds present on large scales.  On small scales, the ACT 150 and 98 GHz channels dominate the information content, due to their high resolution and low noise levels compared to \emph{Planck}.  Interestingly, both the 98 and 150 GHz channels contribute to the Compton-$y$ reconstruction out to very high $\ell$, whereas 150 GHz mostly dominates the very high $\ell$ CMB+kSZ reconstruction.  Also, the ACT 150 (98) GHz channel weights approximately overlap with those of the \emph{Planck} 143 (100) GHz channels in the range of angular scales where both arrays have good signal-to-noise, as expected due to their similar bandpasses.

The relative weights of the different ACT arrays in Figure~\ref{fig:weights} may be surprising upon initial inspection.  Considering the tSZ ILC weights in the bottom panel, the Compton-$y$ response is only 60\% greater at 98 GHz than at 150 GHz (see Figure~\ref{fig:sed}), yet in the range $1000 < \ell < 3000$ the weight for the 98 GHz dataset D56\_5\_098 (PA3 98 GHz) is about $-4 \times 10^{-7} \, \mu{\rm K}^{-1}$ while the 150 GHz data sets only have weights around $0.5 \times 10^{-7} \, \mu{\rm K}^{-1}$. This is not because D56\_5\_098 is more sensitive than the other arrays --- in fact, its sensitivity is lower than the average. Instead, this is because the dominating contaminant in this multipole range is not noise, but the CMB. In the regime where a number of arrays have much lower noise power than that contributed by the CMB itself, the actual sensitivities of those arrays do not matter. Moreover, having multiple arrays with the same properties does not help more than just having one. Instead, what drives the behavior of the weights are the different linear combinations of Compton-$y$ and CMB in each channel, which depend on the array bandpasses. For ACT, the data sets can be divided into three bandpass groups: e.g. for {\tt D56} into D56\_5\_098 (PA3 98 GHz), D56\_1-4\_150 (PA1 and PA2 150 GHz) and D56\_6\_150 (PA3 150 GHz), with the latter differing from the other 150 GHz datasets by having a 20\% narrower bandpass (see Figure~\ref{fig:bandpass}). In the CMB-dominated regime, the data sets in each group do not contribute independent information, and so have to share weight, making the weight per member inversely proportional to the number of members. For $\ell > 3000$ this ceases to be the case as the ILC becomes noise-dominated, and each array's weight becomes proportional to the tSZ response (Eq.~\ref{eq:tszresponse}) times the array inverse noise variance.

We apply the ILC weights to construct a linear combination of the 2D Fourier transforms of the ACT and \emph{Planck} arrays, as in Eq.~\ref{eq.coadd}. The resulting co-add is then multiplied by a beam transfer function and inverse-Fourier transformed to provide the corresponding ILC map in real space. The beam transfer functions of the final maps are chosen so as to prevent a rising (blue) spectrum on small scales. We choose a 1.6 arcmin FWHM Gaussian for maps with no deprojection and a 2.4 arcmin FWHM Gaussian for maps with deprojection to achieve this property.

\section{Results and Validation}
\label{sec:results}

Figure~\ref{fig:fullmaps} shows the component-separated maps of the  Compton-$y$ and CMB+kSZ fields produced by our pipeline, for the standard ILC (no deprojection) analysis in the {\tt BN} and {\tt D56} regions. Figure~\ref{fig:zoom} shows a zoomed-in view of a small patch of sky in {\tt D56} ($\approx 24$ sq.~deg.~out of the available 2100 sq.~deg.) for three of our component-separated maps: the standard ILC Compton-$y$ map (no contaminant deprojection), the standard ILC CMB+kSZ map, and the constrained ILC CMB+kSZ map with the tSZ signal deprojected.  Many galaxy clusters can be seen by eye in the Compton-$y$ map, including Abell 119 in {\tt D56} (Figures~\ref{fig:fullmaps} and \ref{fig:zoom}) and the Virgo cluster in {\tt BN} (Figure~\ref{fig:fullmaps}). Comparison of the standard ILC and constrained ILC CMB+kSZ maps in Figure~\ref{fig:zoom} clearly shows the effect of tSZ residuals in the former map, thus demonstrating the need for the latter in CMB lensing and kSZ analyses.

We now validate the reconstructed ILC maps by measuring their power spectra and cross-correlating them with other data sets.

\subsection{Cluster stacks}
\label{sec:stacks}

An important validation of the Compton-$y$ map is comparison of the stacked
profile of massive, high-SNR galaxy clusters with that from a single-frequency map. This ensures that
the application of the component separation pipeline has not introduced an
effective harmonic space transfer function or calibration offset. For this
comparison, we pick as our reference map the ACT Season 2015 PA3 98 GHz map in the {\tt BN} and {\tt D56}
regions, due to its relatively high weight in the Compton-$y$ map.  We stack on the locations of tSZ-selected confirmed clusters from the 2013-2018 maps with $\rm{SNR}>5$  (179 locations in {\tt D56} and 425 locations in {\tt BN}) after applying the appropriate analytic transfer function
that allows direct comparison with the stack on the standard ILC Compton-$y$ map
(which is reconvolved to the same beam as the reference map). The 
transfer function applied to the reference map is
\be
T(\ellb) =
  \frac{1}{f(\ell)P(\ellb)} \,,
\ee
where $P(\ellb)$ is the pixel window function to be deconvolved in 2D Fourier space and $f(\ell)$
is the tSZ response in Eq.~\ref{eq:tszresponse}, but with scale-dependent corrections for the bandpass and small changes of the beam within the bandpass (see Appendix~\ref{app:color}). Before stacking, both the reference map and the Compton-$y$ map are
additionally high-pass filtered to $\ell>2000$ to suppress scatter due to correlated noise. The
resulting comparison in the {\tt D56} region is shown in Figure~\ref{fig:yprofile}, showing excellent agreement
between the Compton-$y$ map and the reference map.  Error bars are estimated from the scatter within the stacking sample, and so do not include large-scale correlated noise.  Note that errors in the ILC map are smaller. In Sec.~\ref{sec:fgres}, we discuss stacking on cluster locations in the ILC CMB+kSZ maps.

\subsection{Total power spectra and comparison with \emph{Planck} maps}

In Figure~\ref{fig:ypower}, we show the power spectra of a selection of the ILC maps. These
power spectra are calculated under the flat-sky approximation, do not account for
mode coupling or noisy directions in 2D Fourier space, and receive
contributions from all components in the map including instrument noise and
foregrounds. However, the improvement over \emph{Planck} in resolution and signal-to-noise per mode
can be clearly seen at high multipoles. At low multipoles (large angular scales) where
the component-separated maps effectively only include \emph{Planck} data, our ILC maps have
higher noise than those from the \emph{Planck} component separation pipelines, likely
due to insufficient accounting for varying foreground levels in our constant-sky-covariance approximation (in contrast to, e.g., needlet ILC).

In order to interpret the CMB or Compton-$y$ power spectrum, it is important to estimate
its signal power spectrum without noise bias and accounting for foreground
residuals. These steps are beyond the scope of this
work since they require (1) constructing splits of the component-separated maps that have
independent instrument noise and (2) marginalizing over the expected
contribution of foreground residuals to the power spectrum. We emphasize that
the released ILC maps are not suitable for measurements of the power spectra
of either the CMB or tSZ signals due to these complications. The maps should only be used in
applications where noise bias does not appear (e.g., lensing maps for
cross-correlations, and tSZ maps for cross-correlations). The tSZ maps in
particular are noise-dominated at the power spectrum level, although the very large non-Gaussianity of the
tSZ distribution allows one to see clusters by eye without any filtering.

\subsection{Foreground residuals and dust contamination}
\label{sec:fgres}

When interpreting cross-correlations of the ILC maps, it is important to
keep in mind that standard ILC component separation with no deprojection only attempts to minimize variance due to
foreground residuals; it does not fully eliminate the foreground residuals in general. In the case of maps that
deproject some component, there can still be residuals due to inaccuracy of the assumed
SED (though not for the CMB, kSZ or tSZ) or due to decorrelation across frequencies. These issues are especially of concern
when working with CIB/dust-deprojected maps.

The effect of dust contamination in particular can be seen when stacking the tSZ
map at the locations of sources like quasars. The manifestation of dust
contamination in a stack on Compton-$y$ maps is not immediately obvious. For example, if the tSZ map with no
deprojection is stacked at the location of SDSS quasars \cite{SDSSQuasars}, a high signal-to-noise
feature can be detected, with a relative deficit in the center surrounded by
a ring. Using simulations that use a fiducial dust SED, we
have confirmed that such a ring is the response of the Compton-$y$ map weights
(Figure~\ref{fig:weights}) to a dusty point source. This can be understood as arising due to the
fact that on small scales (where instrument noise dominates) the $y$-map is predominantly constructed from $y \sim -A_{98}-A_{150}$, which is negative for a dusty region that is positive at these frequencies. On large
scales (where CMB sample variance dominates), the weights force the map to be
$y \sim -A_{98}+A_{150}$, which is net positive for a dusty region that is brighter at 150 GHz than 98 GHz.
Here $A_{98}$ and $A_{150}$ are the dominant ACT arrays at 98 GHz and 150 GHz,
respectively. We also confirm that when instead stacking these quasars on
dust-deprojected $y$-maps, the ring signal disappears leaving behind a small
residual positive signal that may be due to tSZ from AGN feedback in quasars
(see, e.g., \cite{CrichtonQuasars} and \cite{HallQuasars}). When working with stacks on such
sources, we recommend applying the weights we provide in the data release to
realistic simulations to aid interpretation. Both the level of dust residuals in no-deprojection (standard ILC) maps and the noise in the dust-deprojected maps should be significantly reduced with high-resolution 230 GHz data collected with the Advanced ACTPol instrument in 2016 and onward.

Residual foregrounds in the ILC CMB+kSZ maps are also of concern, especially for lensing reconstruction. Contamination by the tSZ effect can be assessed by stacking on the locations of ACT tSZ clusters in the CMB+kSZ maps. As was seen for \emph{Planck} maps in~\cite{MMHill}, a large negative decrement can be observed in such a stack since the majority of the weight in the ILC CMB+kSZ map comes from the 98 and 150 GHz maps, where the tSZ signal is negative. This can be seen in the top panel of Figure \ref{fig:ystack}. When the tSZ signal is deprojected, the identical stack (bottom panel) shows no detectable tSZ residual, suggesting these maps are suitable for the tSZ-cleaned estimator derived in \cite{MMHill}.  These maps can also be used for applying kSZ estimators to high-mass cluster stacks.

\section{Simulations}
\label{sec:sims}

\begin{figure*}
\includegraphics[width=0.45\textwidth]{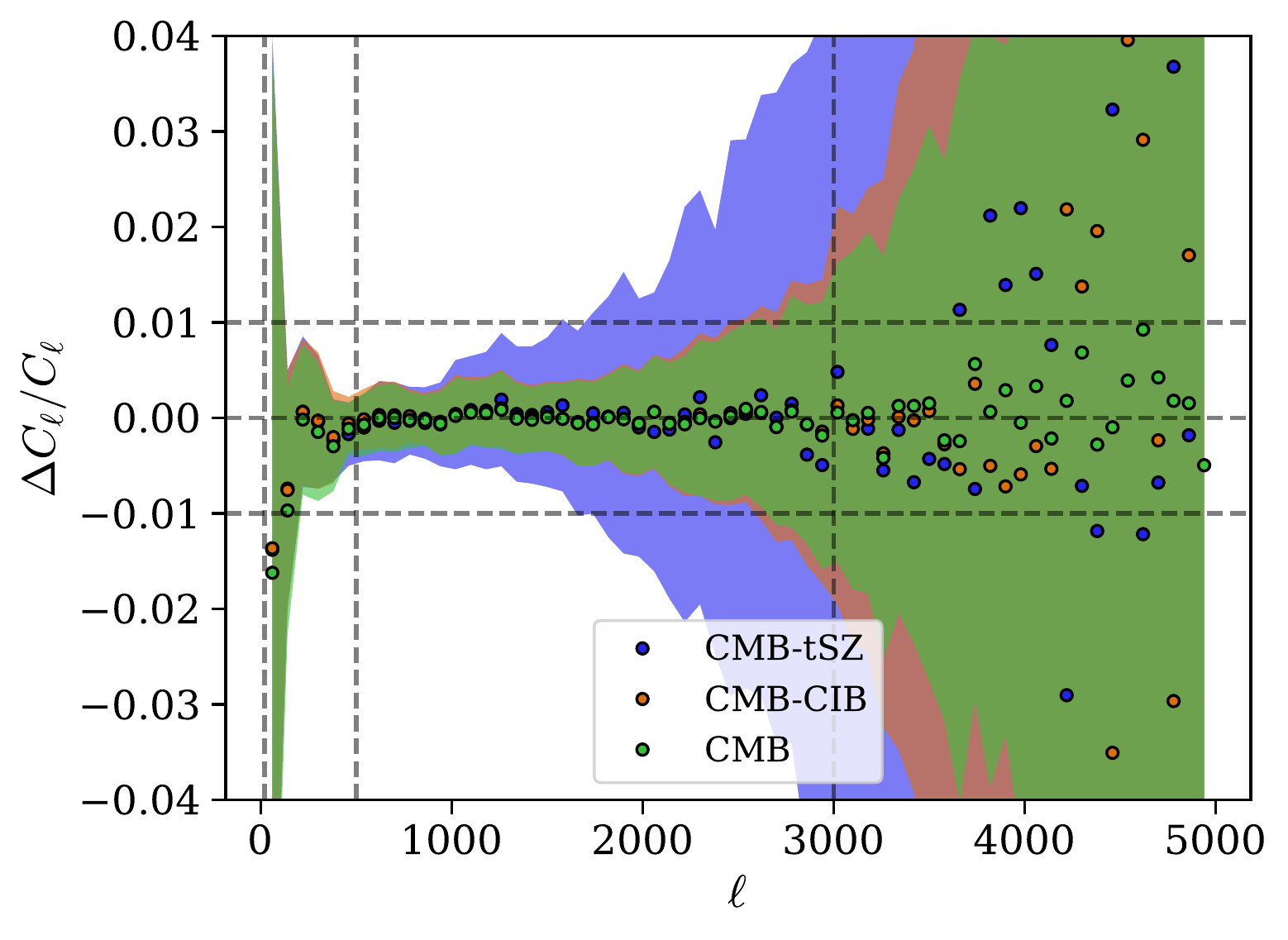}
\includegraphics[width=0.45\textwidth]{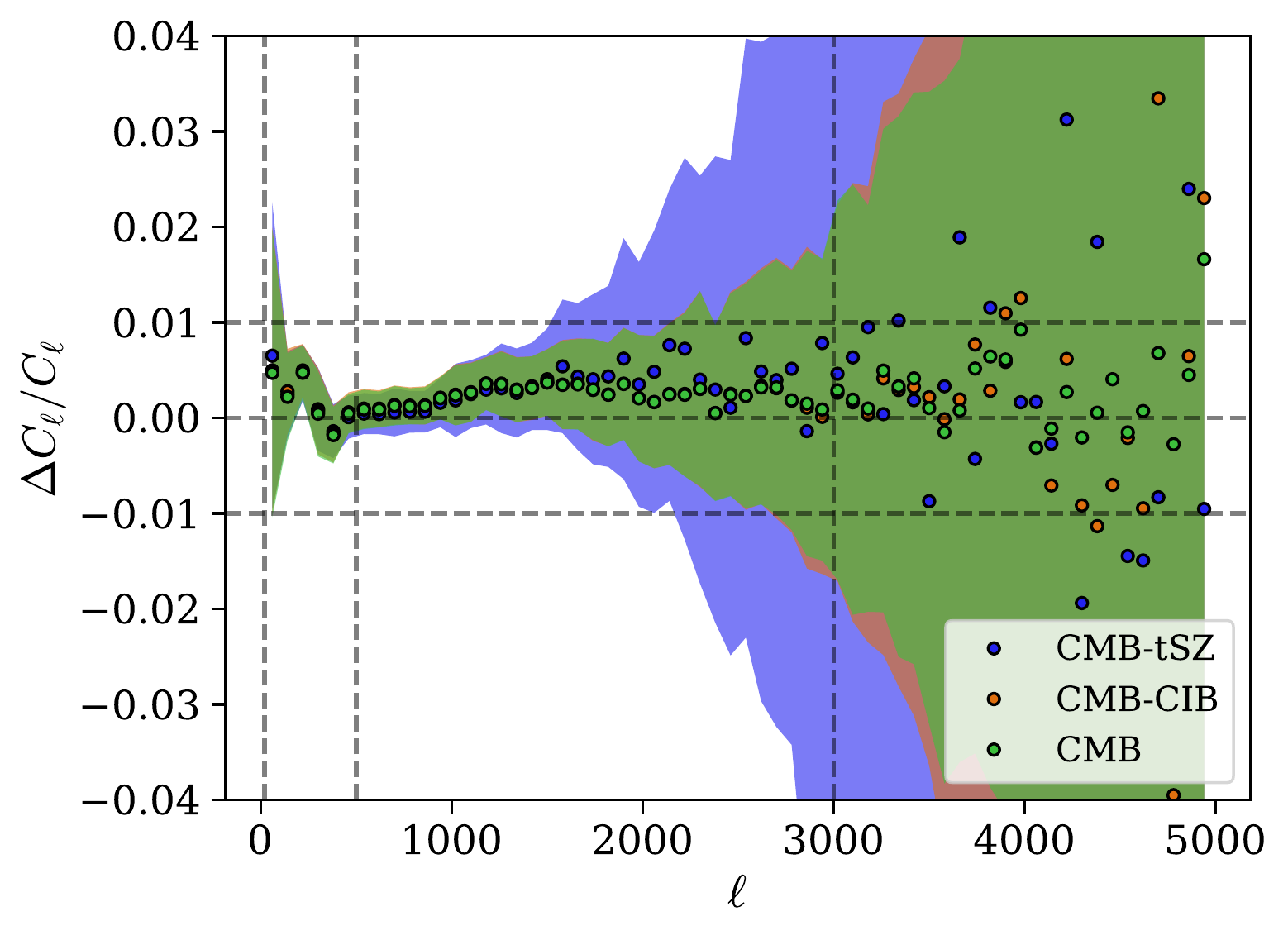} \\
\includegraphics[width=0.45\textwidth]{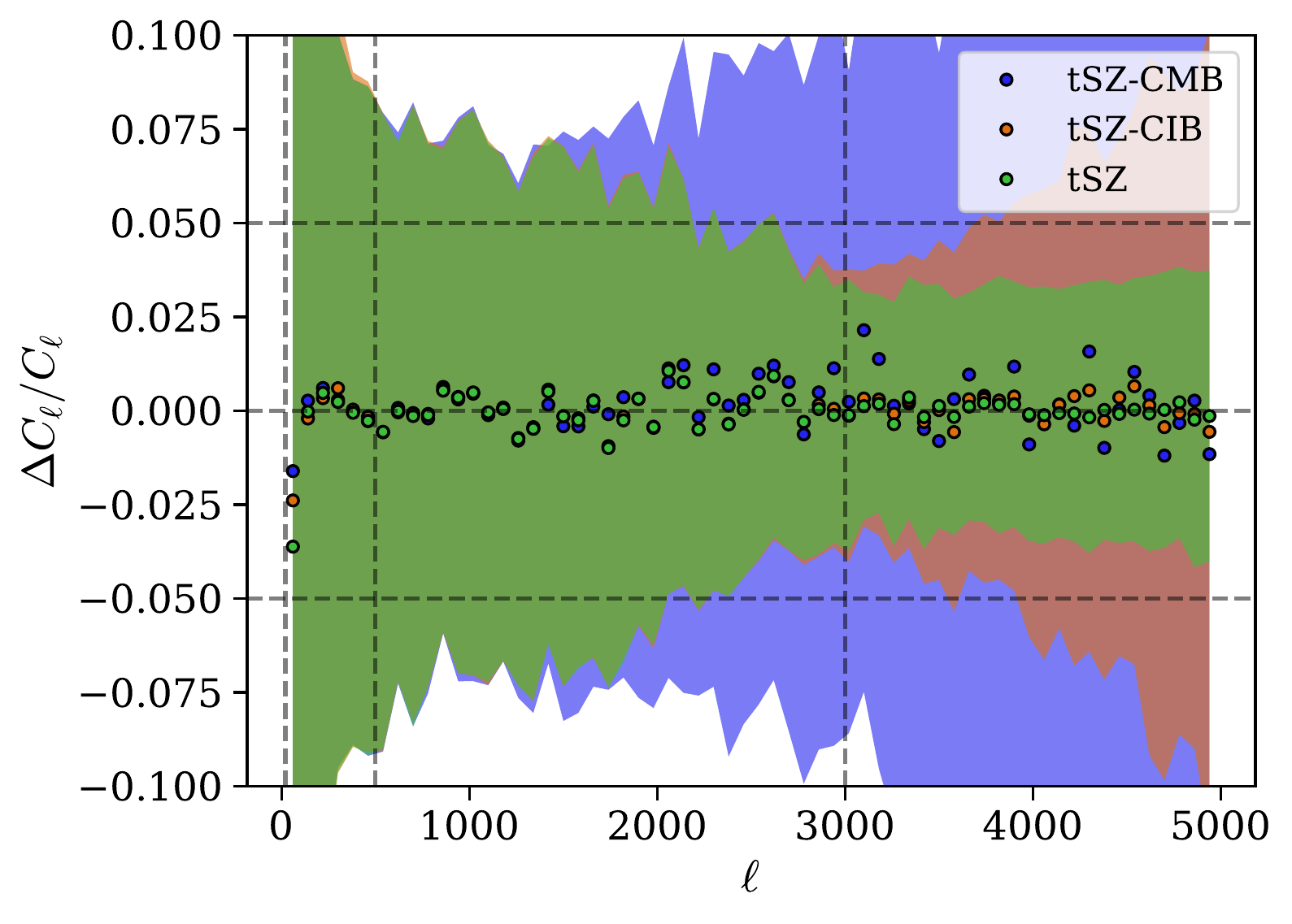}
\includegraphics[width=0.45\textwidth]{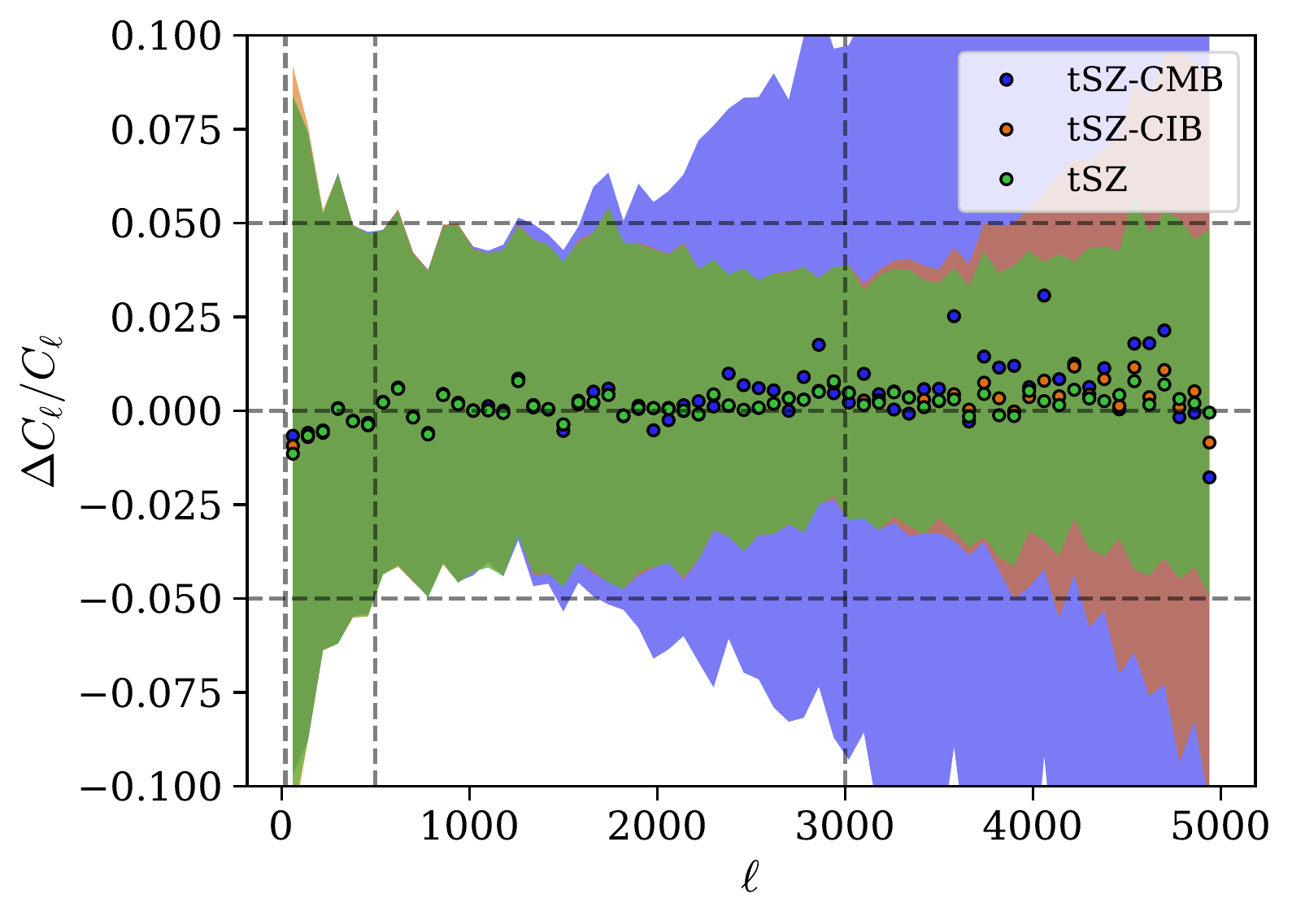}
\centering
\caption{The difference of the cross-spectrum of the reconstructed component and the input component and the auto-spectrum of the input component, normalized to the latter. Symbols are the difference averaged over 160
simulations. The shaded bands show 68\% uncertainties for a single realization. In these simulations the unlensed CMB and foregrounds
(including tSZ) are Gaussian, and the CMB is lensed with a Gaussian
convergence field. The left and right panels show results for the {\tt D56} and {\tt BN} region, respectively.  The top and bottom panels show results for CMB+kSZ and tSZ reconstructions, respectively, each for various constrained ILC options as labeled in the legends (the second component in each label refers to the deprojected field).  The dashed vertical lines indicate $\ell=20,500,3000$. This plot shows that our pipeline is able to recover the underlying CMB+kSZ and tSZ signals with sufficient accuracy for cross-correlation and lensing analyses with these maps.}\label{fig:verify}
\end{figure*}

We use simulations to validate our component separation pipeline and inform analysis choices that control potential sources of bias due to the use of an empirical covariance matrix (so-called ``ILC bias'' in this context). These simulations can also be used for covariance estimation for cross-correlation analyses. Our simulation suite includes contributions from the lensed CMB, Gaussian instrument noise under a constant covariance noise model, Gaussian realizations of the tSZ field, and Gaussian realizations of ``residual'' foregrounds. Details of the lensed CMB and instrument noise simulation can be found in \cite{ChoiEtAl,AiolaEtAl}, but we summarize the salient points here.  Gaussian realizations of the unlensed CMB are drawn on the full sky using a model fit to the \emph{Planck} data~\cite{PlanckCosmology}.  These simulations are then lensed by a Gaussian realization of the lensing field, using the \texttt{pixell} software package.\footnote{\url{https://github.com/simonsobs/pixell}} Noise realizations are generated from the smoothed noise power spectrum of the map measured from splits and subsequently modulated by the detector hit counts in the region.

Since our verification procedure requires checking the fidelity of reconstructing the lensed CMB and the Compton-$y$ distribution, we simulate the tSZ signal by generating a Gaussian random Compton-$y$ field from a template tSZ power spectrum~\cite{Battaglia2010,Battaglia2012tSZPS} for each realization and scaling it by the color-corrected frequency response factors. Finally, we add Gaussian realizations of ``residual foregrounds'' corresponding to the contributions from all other foregrounds by drawing from a covariance matrix that is constructed from empirically fitting measured spectra of the arrays to a power-law model. For each map array pair in the covariance matrix, we calculate the signal power spectrum (using independent noise splits if necessary, as is the case when calculating the diagonals of the covariance matrix for example) and subtract a fiducial lensed CMB and tSZ contribution. The residual measured power spectrum is then fit to
\be
R(\ell) = w + a_1 \left(\frac{\ell}{3000}\right)^{e_1}  \,,
\ee
where $a_1>0$, $w>0$, and $0<e_1<2$ are free parameters. Physically, the first term is expected to capture a Poisson contribution and the second a residual extragalactic contribution important on small scales. We only fit array auto-spectra to the data. The cross-spectra in the covariance matrix used for generating simulations use correlation coefficients from a theoretical model \cite{Dunkley2013} scaled by the fit auto-spectra on the diagonal.  For all the fits, the minimum multipole is $\ell>20$ for \emph{Planck} and $\ell>1000$ for ACT. For the maximum multipole of the fits, we use $\ell<5000$ for ACT, $\ell<2000$ for \emph{Planck} HFI, $\ell<1000$ for LFI 70 GHz, and $\ell<300$ for all other LFI. Realizations drawn from the covariance matrix constructed this way when added to the lensed CMB, tSZ, and instrument noise should result in simulations with identical total power spectra between all pairs of arrays. This allows our simulations of component-separated maps to have nearly identical statistics as the data maps at the power spectrum level.

We generate 160 realizations of the simulations described above by combining the
lensed CMB, tSZ, residual foregrounds and instrument noise in the same model as
the data (e.g., two splits of each \emph{Planck} map and four splits of each ACT map)
and pass these through the same pipeline as the data, applying the entire algorithm described in Sec.~\ref{sec:algorithm} to each of the simulation realizations. Importantly, the
covariance estimation is repeated for each realization since only such a
procedure can detect any ILC bias due to correlations between the estimated
covariance and the data (or simulation realization) itself. The final products of the pipeline consist of (1) maps of the CMB+kSZ with no deprojection, tSZ deprojection, and CIB deprojection, and (2) maps of Compton-$y$ with no deprojection, CMB deprojection, and CIB deprojection. Each of (1) and (2) are then cross-correlated with the input lensed CMB and input Compton-$y$ map, respectively. The resulting cross-spectrum is then binned and averaged over the 160 simulations and compared to the auto-spectrum of the input components. We show the resulting comparison in Figure~\ref{fig:verify}. We confirm that with our analysis choices we control any bias in the resulting maps to be either within the statistical error or at the  $<0.5\%$ level, sufficient for cross-correlation and lensing analyses with these maps.

\section{Discussion}
\label{sec:discussion}

We have presented component-separated maps of the CMB+kSZ and Compton-$y$ signals from the joint analysis of \emph{Planck} and ACTPol temperature data from the 2014 and 2015 observing seasons. While these maps span a sky fraction of 5\% compared to the 65\% spanned by \emph{Planck}, they have orders of magnitude higher signal-to-noise per mode on small scales. In particular, they improve the resolution of the tSZ effect from 10 arcminutes to 1.6 arcminutes. In addition, we provide versions of the maps that explicitly null the contributions from components with an assumed SED (CMB/kSZ, tSZ, or fiducial CIB model). We have shown that in the case of a CMB+kSZ map with tSZ deprojected, no residual tSZ contamination can be seen when the map is stacked at the locations of massive ($\approx 10^{14-15} M_{\odot}$) tSZ-selected clusters. 

A number of science applications are opened up by these new data products. Although explicit deprojection of a specific contaminant increases the final noise level in the ILC map, it is useful in applications where significant reduction of bias at the cost of a small increase in noise is desirable. For example, bias due to tSZ contamination has been shown to be the largest foreground contaminant in temperature-based CMB lensing reconstruction, for both the auto-power spectrum~\cite{vanEngelen2014,Osborne} and cross-correlations with $z \lesssim 1$ tracers~\cite{MMHill,Omori2019,Baxter2019}.  This bias can be mitigated by using tSZ-deprojected CMB maps in an ``asymmetric'' lensing estimator~\cite{MMHill}, which allows for the elimination of tSZ-induced bias in CMB lensing cross-correlations with little penalty in signal-to-noise.\footnote{See \cite{DarwishEtAl} for a demonstration of a tSZ-free lensing cross-correlation.} The loss of signal-to-noise approaches zero in the limit of ``CMB halo lensing'' reconstruction on the smallest scales.  Similarly, a CMB lensing cross-correlation with CIB maps or with high-redshift galaxies (e.g., selected from \emph{Wide-field Infrared Survey Explorer} data~\cite{WISE2010}) can benefit from asymmetric lensing estimators that use our CIB-deprojected CMB+kSZ maps. Since the CIB and tSZ fields are correlated, high-precision cross-correlations will likely require deprojecting both components (which the ILC formalism naturally allows for) or combining component separation with other bias-avoidance methods like shear reconstruction \cite{Prince2018,SchaanFerraro} or bias hardening \cite{Osborne}.

In addition to CMB lensing reconstruction, component-separated CMB maps are also crucial for primordial non-Gaussianity measurements.  The \emph{WMAP} and \emph{Planck} non-Gaussianity analyses relied on such foreground-cleaned maps for analyses of the primordial bispectrum and trispectrum~\cite{Bennett2013,Planck2015NG,Planck2018NG}.  The influence of Galactic and extragalactic foregrounds on these measurements has recently received attention~\cite{Coulton2018,Hill2018NG,Jung2018}.  It was argued in~\cite{Hill2018NG} that explicit deprojection of tSZ and/or CIB contamination would be necessary for unbiased non-Gaussianity constraints from high-resolution, post-\emph{Planck} CMB temperature maps.  We thus expect that the maps produced here will be helpful for such analyses.

Foreground-cleaned CMB+kSZ maps are also useful for kSZ measurements. The projected-field kSZ estimator~\cite{Dore2004,Hill2016,Ferraro2016}, which involves a bispectrum of the form $\langle TTg \rangle$ where $T$ is the CMB+kSZ temperature map and $g$ is a large-scale structure tracer, requires exquisite suppression of foreground residuals.  We anticipate that variations of the estimator that mix minimum-variance CMB+kSZ maps and foreground-deprojected CMB+kSZ maps will likely be valuable. For ``kSZ tomography''-type estimators that involve a bispectrum of the form $\langle Tgg \rangle$~\cite{Smith2018} (e.g., the velocity-field template estimator~\cite{HoTemplate,SchaanTemplate} and pairwise-momentum estimator~\cite{HandPairwise,deBernardisPairwise}), formally, symmetry considerations show that the presence of foreground residuals does not introduce a bias at leading order. However, in practice when performing a kSZ-tomography stack on a CMB+kSZ map with foreground residuals, the steep scaling of the tSZ contamination with halo mass ($y \sim M^{5/3}$) introduces a large amount of scatter in the highest mass bin (which is also where the kSZ signal is largest). Instead of discarding the most massive clusters, as is commonly done, the kSZ signal from these clusters can simply be measured using the (noisier) tSZ-deprojected CMB+kSZ map.

Maps of the Compton-$y$ signal enable a variety of cosmological and astrophysical analyses.  Recent years have seen an explosion in tSZ-based studies of galaxy formation and the physics of the intra-cluster medium (ICM), driven by the \emph{Planck} data~(e.g.,~\cite{Planck2013stack,Hill-Spergel2014,VanWaerbeke2014,Greco2015,BHM2015,Vikram2017,Hill2018,Alonso2018,Pandey2019a,Pandey2019b}).  Most of these analyses use cross-correlations of large-area Compton-$y$ maps with other tracers of large-scale structure (e.g., galaxies, clusters, or weak lensing maps).  The dramatic improvement in angular resolution afforded by the tSZ maps produced in this work (from 10 arcmin with \emph{Planck} to 1.6 arcmin here) will enable transformative gains in these studies, yielding tight constraints on models of baryonic feedback in structure formation~(e.g.,~\cite{LeBrun2015,Hill2018,CMBS4DSR}).  Beyond cross-correlations, the auto-statistics of Compton-$y$ maps are powerful probes of cosmology and the ICM.  However, such analyses require precise and sufficiently flexible modeling of residual foreground contributions to the component-separated map auto-statistics, as well as the removal or modeling of noise contributions.  We defer these steps to future work, and emphasize that the Compton-$y$ maps released here are intended solely for use in cross-correlation analyses, where only a small number of potential foreground contaminants are relevant (generally CIB or radio sources).

Our pipeline has been designed such that the slow covariance matrix calculation is separate from a fast ($\approx$ few seconds) co-addition and deprojection step. Only the fast step depends on assumptions about the SEDs of the components. This will allow us to produce ILC maps as part of a Markov chain calculation involving marginalization over parameters of the SEDs. Such marginalization is especially useful for any analysis that involves assumptions regarding dust or CIB contamination, for which the SED parameters should be varied over some range. This can be incorporated into any cross-correlation analysis where such contamination is of concern. Furthermore, the ability to marginalize over foreground residuals is a necessary condition to extend our pipeline to unbiased measurements of the auto-correlation of the CMB and Compton-$y$ signals. Another necessary condition is the capability to remove the contribution of instrument noise to the auto-spectrum, e.g., through cross-correlation of ILC maps that are constructed from independent splits of the data. We thus stress that direct interpretation of the auto-power spectra of the maps released here is not recommended.

We have worked in the 2D Fourier domain to optimally account for the anisotropy of instrument noise present in ground-based experiments like ACT. However, our analysis is applied using power spectra calculated as an average within each of the two regions considered here. This averaging procedure washes out instrument noise anisotropy to some extent. Combined with the fact that in our deep patches, the CMB temperature dominates over noise and the Compton-$y$ signal in regions of harmonic space where the noise anisotropy is large, we find only percent-level improvements in the total power of our maps at $\ell>500$ compared to an analysis that isotropizes the information in the instrument noise. Nonetheless, this pilot study sets the formalism and pipeline in place for an improved analysis that accounts for spatial variations of both the amplitude of the instrument noise and its 2D Fourier anisotropy.

This paper represents the first step on an exciting path toward
component-separated ACT + \emph{Planck} maps covering significantly larger sky
fractions and using additional frequency coverage, as compared to the data used
in the analysis reported here. We will also extend the method developed here to
produce component-separated CMB polarization maps from these data.  Starting in
the 2016 observing season and continuing until now, the Advanced ACTPol
instrument has undertaken wide scans that altogether cover roughly half of the
sky, i.e., an order of magnitude more area than the regions analyzed here.  This
coverage includes complete overlap with the $\approx 5000$ sq.~deg.~covered by
the Dark Energy Survey (DES)~\cite{DES} and the planned $\approx 20000$
sq.~deg.~footprint of the Vera Rubin Observatory's Legacy Survey of Space and Time (LSST)~\cite{LSST},
and partial overlap of $\approx 8000$ sq.~deg.~with the Dark Energy
Spectroscopic Instrument (DESI)~\cite{DESI}, amongst other experiments (e.g.,
KiDS \cite{KIDS} and HSC \cite{HSC}).  In addition, the Advanced ACTPol receiver
includes detectors operating at 230 GHz, with arcminute resolution.  The 230 GHz data will be included in imminent component separation analyses, allowing for significantly improved dust and CIB removal, as well as simultaneous deprojection of multiple contaminants on small scales. A low-frequency array with bands centered at 28 and 41 GHz has been tested in the laboratory \cite{Koopman2018,Simon2018} and will be installed on the telescope in the next year, yielding nearly a full decade in frequency coverage with ACT.  These data will provide leverage in removing Galactic synchrotron and extragalactic radio source contamination.  Starting in the early 2020s, the Simons Observatory Large Aperture Telescope (LAT) will include significantly more sensitive coverage at all five ACT frequencies, as well as an additional 280 GHz channel.  Looking further ahead, the current design of the CMB-S4 LATs also includes channels at all six of these frequencies~\cite{CMBS4DSR}.  Future prospects are thus extremely bright for the production of high-sensitivity, high-resolution component-separated maps covering half of the sky, which will enable a vast range of exciting CMB science in the coming decade.

\acknowledgments

We are grateful to Hans Kristian Eriksen, Reijo Keskitalo, and Mathieu Remazeilles for informative discussions related to \emph{Planck} products and analysis. Some of the results in this paper have been derived using the healpy~\cite{Healpix1} and HEALPix~\cite{Healpix2} packages. This research made use of Astropy,\footnote{http://www.astropy.org} a community-developed core Python package for Astronomy \citep{astropy:2013, astropy:2018}. We also acknowledge use of the matplotlib~\cite{Hunter:2007} package and the Python Image Library for producing plots in this paper, and use of the Boltzmann code CAMB~\cite{CAMB} for calculating theory spectra. 

This work was supported by the U.S. National Science Foundation through awards AST-1440226, AST0965625 and AST-0408698 for the ACT project, as well as awards PHY-1214379 and PHY-0855887. Funding was also provided by Princeton University, the University of Pennsylvania, and a Canada Foundation for Innovation (CFI) award to UBC. ACT operates in the Parque Astron\'{o}mico Atacama in northern Chile under the auspices of the Comisi\'{o}n Nacional de Investigaci\'{o}n Cient\'{i}fica y Tecnol\'{o}gica de Chile (CONICYT). Computations were performed on the GPC and \emph{Niagara} supercomputers at the SciNet HPC Consortium. SciNet is funded
by the CFI under the auspices of Compute Canada, the Government of Ontario, the Ontario Research Fund
-- Research Excellence; and the University of Toronto. The development of multichroic detectors and lenses was supported by NASA grants NNX13AE56G and NNX14AB58G.   Colleagues at AstroNorte and RadioSky provide logistical support and keep operations in Chile running smoothly. We also thank the Mishrahi Fund and the Wilkinson Fund for their generous support of the project.

MSM acknowledges support from NSF grant AST-1814971. JCH acknowledges support from the Simons Foundation and the W.~M.~Keck Foundation Fund at the Institute for Advanced Study. Flatiron Institute is supported by the Simons Foundation. RB and VC acknowledge DoE grant DE-SC0011838, NASA ATP grants NNX14AH53G and 80NSSC18K0695, NASA ROSES grant 12-EUCLID12-0004 and funding related to the WFIRST Science Investigation Team. EC is supported by a STFC Ernest Rutherford Fellowship ST/M004856/2. SKC acknowledges support from the Cornell Presidential Postdoctoral Fellowship. RD thanks CONICYT for grant BASAL CATA AFB-170002. MHi
acknowledges funding support from the National Research Foundation, the
South African Radio Astronomy Observatory, and the University of
KwaZulu-Natal. LM received funding from CONICYT FONDECYT grant 3170846. KM acknowledges support from the National Research Foundation of South Africa. NS acknowledges support from NSF grant number 1513618. 

\appendix

\section{Color Corrections}
\label{app:color}

Since the input maps are in differential CMB temperature units and assumed to be calibrated (with beams $B_{\rm CMB}(\ell)$ that are spectrally matched to the CMB), the frequency response to the CMB blackbody SED is always unity. To compute responses for the tSZ and CIB (modified blackbody) SEDs, we must account for the width of the bandpass and for small changes of the beam within the bandpass. For a bandpass transmission function $W(\nu)$ and for a frequency response $f(\nu)$ in differential CMB temperature units (i.e., Eqs.~\ref{eq:tszresponse} and~\ref{eq:fcib}), the bandpass color-corrected response factor in the limit that the beam does not change with frequency is,
\be
\label{Eq:fi}
f_i = \frac{\int d\nu\,f(\nu)\,W(\nu)\,D_{\rm{CMB}}(\nu)}{\int d\nu\,W(\nu)\,D_{\rm{CMB}}(\nu)} \,,
\ee
where $D_{\rm CMB}(\nu) \equiv \frac{dB(\nu, T)}{dT} \rvert_{T=T_{\rm CMB}}$ converts $f(\nu)$ to units of specific intensity. Incorporating the effect of variation of the beam within the bandpass changes this to
\be
\label{Eq:fiell}
f_i(\ell) = \frac{1}{B_{\rm CMB}(\ell)}\frac{\int d\nu\,f(\nu)\,W(\nu)\,D_{\rm{CMB}}(\nu)\,B(\ell,\nu)}{\int d\nu\,W(\nu)\,D_{\rm{CMB}}(\nu)} \,.
\ee
In our baseline analysis, we use the above expression, Eq.~\ref{Eq:fiell}, to calculate scale-dependent color corrections for the tSZ and CIB responses of ACT arrays. Assuming the optics are diffraction-limited, we approximate $B(\ell,\nu) = B_{\rm CMB}( \ell(\nu_0/\nu) ) $, where $\nu_0$ is the effective frequency \cite{thornton/2016} of the CMB assumed in the spectrally matched beams \citep[]{AiolaEtAl,ChoiEtAl}. Using simulations of the optical system, we have confirmed that the diffraction-limited approximation is accurate at the sub-percent level in the response factors. The corrections in Eq.~\ref{Eq:fiell} yield variations at the 5-10\% level as a function of multipole when compared to the scale-independent expression given in Eq.~\ref{Eq:fi}.  For the \emph{Planck} maps, we do not incorporate the effects of beam variation, and rather use the scale-independent color corrections in Eq.~\ref{Eq:fi}.

When interpreting results based on the component-separated maps produced here, note that the bandpasses measured for the ACT arrays have overall frequency shift uncertainties of $\approx 2.4$ GHz at 150 GHz and $\approx 1.5$ GHz at 98 GHz.  These uncertainties are due to systematics in the Fourier Transform Spectrometer used for the measurements~\cite{thornton/2016,Datta2019}. These shifts correspond to uncertainties in the tSZ color corrections of 3\% and 0.8\% for the 150 and 98 GHz channels, respectively. A high-precision analysis may require the production of a set of ILC maps with color corrections calculated from bandpasses whose frequencies have been shifted in accordance with these uncertainties.  For \emph{Planck}, the overall uncertainties from pre-launch bandpass measurements are smaller than these shifts~\cite{Planck2013LFIdataproc,Planck2013HFIspectralresp}.

Looking ahead to Advanced ACTPol and Simons Observatory~\cite{SO2019}, reductions in these systematic uncertainties will be necessary in order to avoid biasing measurements; e.g., for a Simons Observatory tSZ cross-correlation with LSST galaxies, absolute calibration of the bandpasses approaching 0.1\% precision will be necessary (see, e.g., \cite{WardBandpass} for similar considerations). However, the bandpass calibration requirements are less stringent for CMB power spectrum measurements (e.g., as used to constrain $N_{\rm eff}$ and other cosmological parameters), because calibration errors are absorbed in free parameters associated with the SEDs of astrophysical foregrounds (the tSZ contribution being a notable exception, as its SED has no free parameters).  The situation is also helped by the fact that the $TE$ power spectrum drives the constraining power on parameters~\cite{SO2019}, for which the only foregrounds are Galactic dust and synchrotron, whose free SED parameters can absorb bandpass uncertainties (at the cost of biasing the physical interpretation of those foreground parameters).

Finally, the ACT bandpass transmission curves used here do not include the effect of atmospheric opacity, since they are measured in the vicinity of the ACT receivers. In order to quantify this effect, we generate a set of atmospheric transmission curves for various precipitable water vapor (PWV) values ranging from 0 mm to 3 mm using the ALMA Atmospheric Modeling (AATM) code \cite{AATM} and propagate these to the color correction factors. We find only sub-percent differences compared to the fiducial results, so we do not include these effects in our baseline analysis.

\bibliography{msm}

\begin{thebibliography}{132}
\expandafter\ifx\csname natexlab\endcsname\relax\def\natexlab#1{#1}\fi
\expandafter\ifx\csname bibnamefont\endcsname\relax
  \def\bibnamefont#1{#1}\fi
\expandafter\ifx\csname bibfnamefont\endcsname\relax
  \def\bibfnamefont#1{#1}\fi
\expandafter\ifx\csname citenamefont\endcsname\relax
  \def\citenamefont#1{#1}\fi
\expandafter\ifx\csname url\endcsname\relax
  \def\url#1{\texttt{#1}}\fi
\expandafter\ifx\csname urlprefix\endcsname\relax\def\urlprefix{URL }\fi
\providecommand{\bibinfo}[2]{#2}
\providecommand{\eprint}[2][]{\url{#2}}

\bibitem[{\citenamefont{{Bennett} et~al.}(1992)\citenamefont{{Bennett},
  {Smoot}, {Hinshaw}, {Wright}, {Kogut}, {de Amici}, {Meyer}, {Weiss},
  {Wilkinson}, {Gulkis} et~al.}}]{COBECompSep1}
\bibinfo{author}{\bibfnamefont{C.~L.} \bibnamefont{{Bennett}}},
  \bibinfo{author}{\bibfnamefont{G.~F.} \bibnamefont{{Smoot}}},
  \bibinfo{author}{\bibfnamefont{G.}~\bibnamefont{{Hinshaw}}},
  \bibinfo{author}{\bibfnamefont{E.~L.} \bibnamefont{{Wright}}},
  \bibinfo{author}{\bibfnamefont{A.}~\bibnamefont{{Kogut}}},
  \bibinfo{author}{\bibfnamefont{G.}~\bibnamefont{{de Amici}}},
  \bibinfo{author}{\bibfnamefont{S.~S.} \bibnamefont{{Meyer}}},
  \bibinfo{author}{\bibfnamefont{R.}~\bibnamefont{{Weiss}}},
  \bibinfo{author}{\bibfnamefont{D.~T.} \bibnamefont{{Wilkinson}}},
  \bibinfo{author}{\bibfnamefont{S.}~\bibnamefont{{Gulkis}}},
  \bibnamefont{et~al.}, \bibinfo{journal}{\apjl}
  \textbf{\bibinfo{volume}{396}}, \bibinfo{pages}{L7} (\bibinfo{year}{1992}).

\bibitem[{\citenamefont{{Maino} et~al.}(2003)\citenamefont{{Maino}, {Banday},
  {Baccigalupi}, {Perrotta}, and {G{\'o}rski}}}]{COBECompSep2}
\bibinfo{author}{\bibfnamefont{D.}~\bibnamefont{{Maino}}},
  \bibinfo{author}{\bibfnamefont{A.~J.} \bibnamefont{{Banday}}},
  \bibinfo{author}{\bibfnamefont{C.}~\bibnamefont{{Baccigalupi}}},
  \bibinfo{author}{\bibfnamefont{F.}~\bibnamefont{{Perrotta}}},
  \bibnamefont{and} \bibinfo{author}{\bibfnamefont{K.~M.}
  \bibnamefont{{G{\'o}rski}}}, \bibinfo{journal}{\mnras}
  \textbf{\bibinfo{volume}{344}}, \bibinfo{pages}{544} (\bibinfo{year}{2003}),
  \eprint{astro-ph/0303657}.

\bibitem[{\citenamefont{{Bennett} et~al.}(2013)\citenamefont{{Bennett},
  {Larson}, {Weiland}, {Jarosik}, {Hinshaw}, {Odegard}, {Smith}, {Hill},
  {Gold}, {Halpern} et~al.}}]{Bennett2013}
\bibinfo{author}{\bibfnamefont{C.~L.} \bibnamefont{{Bennett}}},
  \bibinfo{author}{\bibfnamefont{D.}~\bibnamefont{{Larson}}},
  \bibinfo{author}{\bibfnamefont{J.~L.} \bibnamefont{{Weiland}}},
  \bibinfo{author}{\bibfnamefont{N.}~\bibnamefont{{Jarosik}}},
  \bibinfo{author}{\bibfnamefont{G.}~\bibnamefont{{Hinshaw}}},
  \bibinfo{author}{\bibfnamefont{N.}~\bibnamefont{{Odegard}}},
  \bibinfo{author}{\bibfnamefont{K.~M.} \bibnamefont{{Smith}}},
  \bibinfo{author}{\bibfnamefont{R.~S.} \bibnamefont{{Hill}}},
  \bibinfo{author}{\bibfnamefont{B.}~\bibnamefont{{Gold}}},
  \bibinfo{author}{\bibfnamefont{M.}~\bibnamefont{{Halpern}}},
  \bibnamefont{et~al.}, \bibinfo{journal}{\apjs}
  \textbf{\bibinfo{volume}{208}}, \bibinfo{eid}{20} (\bibinfo{year}{2013}),
  \eprint{1212.5225}.

\bibitem[{\citenamefont{{Hinshaw} et~al.}(2013)\citenamefont{{Hinshaw},
  {Larson}, {Komatsu}, {Spergel}, {Bennett}, {Dunkley}, {Nolta}, {Halpern},
  {Hill}, {Odegard} et~al.}}]{Hinshaw2013}
\bibinfo{author}{\bibfnamefont{G.}~\bibnamefont{{Hinshaw}}},
  \bibinfo{author}{\bibfnamefont{D.}~\bibnamefont{{Larson}}},
  \bibinfo{author}{\bibfnamefont{E.}~\bibnamefont{{Komatsu}}},
  \bibinfo{author}{\bibfnamefont{D.~N.} \bibnamefont{{Spergel}}},
  \bibinfo{author}{\bibfnamefont{C.~L.} \bibnamefont{{Bennett}}},
  \bibinfo{author}{\bibfnamefont{J.}~\bibnamefont{{Dunkley}}},
  \bibinfo{author}{\bibfnamefont{M.~R.} \bibnamefont{{Nolta}}},
  \bibinfo{author}{\bibfnamefont{M.}~\bibnamefont{{Halpern}}},
  \bibinfo{author}{\bibfnamefont{R.~S.} \bibnamefont{{Hill}}},
  \bibinfo{author}{\bibfnamefont{N.}~\bibnamefont{{Odegard}}},
  \bibnamefont{et~al.}, \bibinfo{journal}{\apjs}
  \textbf{\bibinfo{volume}{208}}, \bibinfo{eid}{19} (\bibinfo{year}{2013}),
  \eprint{1212.5226}.

\bibitem[{\citenamefont{{Tegmark} et~al.}(2003)\citenamefont{{Tegmark}, {de
  Oliveira-Costa}, and {Hamilton}}}]{Tegmark2003}
\bibinfo{author}{\bibfnamefont{M.}~\bibnamefont{{Tegmark}}},
  \bibinfo{author}{\bibfnamefont{A.}~\bibnamefont{{de Oliveira-Costa}}},
  \bibnamefont{and} \bibinfo{author}{\bibfnamefont{A.~J.}
  \bibnamefont{{Hamilton}}}, \bibinfo{journal}{\prd}
  \textbf{\bibinfo{volume}{68}}, \bibinfo{eid}{123523} (\bibinfo{year}{2003}),
  \eprint{astro-ph/0302496}.

\bibitem[{\citenamefont{{Planck Collaboration}
  et~al.}(2016{\natexlab{a}})\citenamefont{{Planck Collaboration}, {Adam},
  {Ade}, {Aghanim}, {Alves}, {Arnaud}, {Ashdown}, {Aumont}, {Baccigalupi},
  {Banday} et~al.}}]{Planck2015compsepFG}
\bibinfo{author}{\bibnamefont{{Planck Collaboration}}},
  \bibinfo{author}{\bibfnamefont{R.}~\bibnamefont{{Adam}}},
  \bibinfo{author}{\bibfnamefont{P.~A.~R.} \bibnamefont{{Ade}}},
  \bibinfo{author}{\bibfnamefont{N.}~\bibnamefont{{Aghanim}}},
  \bibinfo{author}{\bibfnamefont{M.~I.~R.} \bibnamefont{{Alves}}},
  \bibinfo{author}{\bibfnamefont{M.}~\bibnamefont{{Arnaud}}},
  \bibinfo{author}{\bibfnamefont{M.}~\bibnamefont{{Ashdown}}},
  \bibinfo{author}{\bibfnamefont{J.}~\bibnamefont{{Aumont}}},
  \bibinfo{author}{\bibfnamefont{C.}~\bibnamefont{{Baccigalupi}}},
  \bibinfo{author}{\bibfnamefont{A.~J.} \bibnamefont{{Banday}}},
  \bibnamefont{et~al.}, \bibinfo{journal}{\aap} \textbf{\bibinfo{volume}{594}},
  \bibinfo{eid}{A10} (\bibinfo{year}{2016}{\natexlab{a}}), \eprint{1502.01588}.

\bibitem[{\citenamefont{{Planck Collaboration}
  et~al.}(2016{\natexlab{b}})\citenamefont{{Planck Collaboration}, {Adam},
  {Ade}, {Aghanim}, {Arnaud}, {Ashdown}, {Aumont}, {Baccigalupi}, {Banday},
  {Barreiro} et~al.}}]{Planck2015compsep}
\bibinfo{author}{\bibnamefont{{Planck Collaboration}}},
  \bibinfo{author}{\bibfnamefont{R.}~\bibnamefont{{Adam}}},
  \bibinfo{author}{\bibfnamefont{P.~A.~R.} \bibnamefont{{Ade}}},
  \bibinfo{author}{\bibfnamefont{N.}~\bibnamefont{{Aghanim}}},
  \bibinfo{author}{\bibfnamefont{M.}~\bibnamefont{{Arnaud}}},
  \bibinfo{author}{\bibfnamefont{M.}~\bibnamefont{{Ashdown}}},
  \bibinfo{author}{\bibfnamefont{J.}~\bibnamefont{{Aumont}}},
  \bibinfo{author}{\bibfnamefont{C.}~\bibnamefont{{Baccigalupi}}},
  \bibinfo{author}{\bibfnamefont{A.~J.} \bibnamefont{{Banday}}},
  \bibinfo{author}{\bibfnamefont{R.~B.} \bibnamefont{{Barreiro}}},
  \bibnamefont{et~al.}, \bibinfo{journal}{\aap} \textbf{\bibinfo{volume}{594}},
  \bibinfo{eid}{A9} (\bibinfo{year}{2016}{\natexlab{b}}), \eprint{1502.05956}.

\bibitem[{\citenamefont{{Planck Collaboration}
  et~al.}(2018{\natexlab{a}})\citenamefont{{Planck Collaboration}, {Akrami},
  {Arroja}, {Ashdown}, {Aumont}, {Baccigalupi}, {Ballardini}, {Banday},
  {Barreiro}, {Bartolo} et~al.}}]{Planck2018legacy}
\bibinfo{author}{\bibnamefont{{Planck Collaboration}}},
  \bibinfo{author}{\bibfnamefont{Y.}~\bibnamefont{{Akrami}}},
  \bibinfo{author}{\bibfnamefont{F.}~\bibnamefont{{Arroja}}},
  \bibinfo{author}{\bibfnamefont{M.}~\bibnamefont{{Ashdown}}},
  \bibinfo{author}{\bibfnamefont{J.}~\bibnamefont{{Aumont}}},
  \bibinfo{author}{\bibfnamefont{C.}~\bibnamefont{{Baccigalupi}}},
  \bibinfo{author}{\bibfnamefont{M.}~\bibnamefont{{Ballardini}}},
  \bibinfo{author}{\bibfnamefont{A.~J.} \bibnamefont{{Banday}}},
  \bibinfo{author}{\bibfnamefont{R.~B.} \bibnamefont{{Barreiro}}},
  \bibinfo{author}{\bibfnamefont{N.}~\bibnamefont{{Bartolo}}},
  \bibnamefont{et~al.}, \bibinfo{journal}{arXiv e-prints}
  \bibinfo{eid}{arXiv:1807.06205} (\bibinfo{year}{2018}{\natexlab{a}}),
  \eprint{1807.06205}.

\bibitem[{\citenamefont{{Planck Collaboration}
  et~al.}(2018{\natexlab{b}})\citenamefont{{Planck Collaboration}, {Akrami},
  {Ashdown}, {Aumont}, {Baccigalupi}, {Ballardini}, {Band ay}, {Barreiro},
  {Bartolo}, {Basak} et~al.}}]{Planck2018compsep}
\bibinfo{author}{\bibnamefont{{Planck Collaboration}}},
  \bibinfo{author}{\bibfnamefont{Y.}~\bibnamefont{{Akrami}}},
  \bibinfo{author}{\bibfnamefont{M.}~\bibnamefont{{Ashdown}}},
  \bibinfo{author}{\bibfnamefont{J.}~\bibnamefont{{Aumont}}},
  \bibinfo{author}{\bibfnamefont{C.}~\bibnamefont{{Baccigalupi}}},
  \bibinfo{author}{\bibfnamefont{M.}~\bibnamefont{{Ballardini}}},
  \bibinfo{author}{\bibfnamefont{A.~J.} \bibnamefont{{Band ay}}},
  \bibinfo{author}{\bibfnamefont{R.~B.} \bibnamefont{{Barreiro}}},
  \bibinfo{author}{\bibfnamefont{N.}~\bibnamefont{{Bartolo}}},
  \bibinfo{author}{\bibfnamefont{S.}~\bibnamefont{{Basak}}},
  \bibnamefont{et~al.}, \bibinfo{journal}{arXiv e-prints}
  \bibinfo{eid}{arXiv:1807.06208} (\bibinfo{year}{2018}{\natexlab{b}}),
  \eprint{1807.06208}.

\bibitem[{\citenamefont{{Dunkley} et~al.}(2013)\citenamefont{{Dunkley},
  {Calabrese}, {Sievers}, {Addison}, {Battaglia}, {Battistelli}, {Bond}, {Das},
  {Devlin}, {D{\"u}nner} et~al.}}]{Dunkley2013}
\bibinfo{author}{\bibfnamefont{J.}~\bibnamefont{{Dunkley}}},
  \bibinfo{author}{\bibfnamefont{E.}~\bibnamefont{{Calabrese}}},
  \bibinfo{author}{\bibfnamefont{J.}~\bibnamefont{{Sievers}}},
  \bibinfo{author}{\bibfnamefont{G.~E.} \bibnamefont{{Addison}}},
  \bibinfo{author}{\bibfnamefont{N.}~\bibnamefont{{Battaglia}}},
  \bibinfo{author}{\bibfnamefont{E.~S.} \bibnamefont{{Battistelli}}},
  \bibinfo{author}{\bibfnamefont{J.~R.} \bibnamefont{{Bond}}},
  \bibinfo{author}{\bibfnamefont{S.}~\bibnamefont{{Das}}},
  \bibinfo{author}{\bibfnamefont{M.~J.} \bibnamefont{{Devlin}}},
  \bibinfo{author}{\bibfnamefont{R.}~\bibnamefont{{D{\"u}nner}}},
  \bibnamefont{et~al.}, \bibinfo{journal}{\jcap}
  \textbf{\bibinfo{volume}{2013}}, \bibinfo{eid}{025} (\bibinfo{year}{2013}),
  \eprint{1301.0776}.

\bibitem[{\citenamefont{{Planck Collaboration}
  et~al.}(2019{\natexlab{a}})\citenamefont{{Planck Collaboration}, {Aghanim},
  {Akrami}, {Ashdown}, {Aumont}, {Baccigalupi}, {Ballardini}, {Banday},
  {Barreiro}, {Bartolo} et~al.}}]{Planck2018likelihood}
\bibinfo{author}{\bibnamefont{{Planck Collaboration}}},
  \bibinfo{author}{\bibfnamefont{N.}~\bibnamefont{{Aghanim}}},
  \bibinfo{author}{\bibfnamefont{Y.}~\bibnamefont{{Akrami}}},
  \bibinfo{author}{\bibfnamefont{M.}~\bibnamefont{{Ashdown}}},
  \bibinfo{author}{\bibfnamefont{J.}~\bibnamefont{{Aumont}}},
  \bibinfo{author}{\bibfnamefont{C.}~\bibnamefont{{Baccigalupi}}},
  \bibinfo{author}{\bibfnamefont{M.}~\bibnamefont{{Ballardini}}},
  \bibinfo{author}{\bibfnamefont{A.~J.} \bibnamefont{{Banday}}},
  \bibinfo{author}{\bibfnamefont{R.~B.} \bibnamefont{{Barreiro}}},
  \bibinfo{author}{\bibfnamefont{N.}~\bibnamefont{{Bartolo}}},
  \bibnamefont{et~al.}, \bibinfo{journal}{arXiv e-prints}
  \bibinfo{eid}{arXiv:1907.12875} (\bibinfo{year}{2019}{\natexlab{a}}),
  \eprint{1907.12875}.

\bibitem[{\citenamefont{{Sievers} et~al.}(2013)\citenamefont{{Sievers},
  {Hlozek}, {Nolta}, {Acquaviva}, {Addison}, {Ade}, {Aguirre}, {Amiri},
  {Appel}, {Barrientos} et~al.}}]{Sievers2013}
\bibinfo{author}{\bibfnamefont{J.~L.} \bibnamefont{{Sievers}}},
  \bibinfo{author}{\bibfnamefont{R.~A.} \bibnamefont{{Hlozek}}},
  \bibinfo{author}{\bibfnamefont{M.~R.} \bibnamefont{{Nolta}}},
  \bibinfo{author}{\bibfnamefont{V.}~\bibnamefont{{Acquaviva}}},
  \bibinfo{author}{\bibfnamefont{G.~E.} \bibnamefont{{Addison}}},
  \bibinfo{author}{\bibfnamefont{P.~A.~R.} \bibnamefont{{Ade}}},
  \bibinfo{author}{\bibfnamefont{P.}~\bibnamefont{{Aguirre}}},
  \bibinfo{author}{\bibfnamefont{M.}~\bibnamefont{{Amiri}}},
  \bibinfo{author}{\bibfnamefont{J.~W.} \bibnamefont{{Appel}}},
  \bibinfo{author}{\bibfnamefont{L.~F.} \bibnamefont{{Barrientos}}},
  \bibnamefont{et~al.}, \bibinfo{journal}{\jcap}
  \textbf{\bibinfo{volume}{2013}}, \bibinfo{eid}{060} (\bibinfo{year}{2013}),
  \eprint{1301.0824}.

\bibitem[{\citenamefont{{Story} et~al.}(2013)\citenamefont{{Story},
  {Reichardt}, {Hou}, {Keisler}, {Aird}, {Benson}, {Bleem}, {Carlstrom},
  {Chang}, {Cho} et~al.}}]{Story2013}
\bibinfo{author}{\bibfnamefont{K.~T.} \bibnamefont{{Story}}},
  \bibinfo{author}{\bibfnamefont{C.~L.} \bibnamefont{{Reichardt}}},
  \bibinfo{author}{\bibfnamefont{Z.}~\bibnamefont{{Hou}}},
  \bibinfo{author}{\bibfnamefont{R.}~\bibnamefont{{Keisler}}},
  \bibinfo{author}{\bibfnamefont{K.~A.} \bibnamefont{{Aird}}},
  \bibinfo{author}{\bibfnamefont{B.~A.} \bibnamefont{{Benson}}},
  \bibinfo{author}{\bibfnamefont{L.~E.} \bibnamefont{{Bleem}}},
  \bibinfo{author}{\bibfnamefont{J.~E.} \bibnamefont{{Carlstrom}}},
  \bibinfo{author}{\bibfnamefont{C.~L.} \bibnamefont{{Chang}}},
  \bibinfo{author}{\bibfnamefont{H.~M.} \bibnamefont{{Cho}}},
  \bibnamefont{et~al.}, \bibinfo{journal}{\apj} \textbf{\bibinfo{volume}{779}},
  \bibinfo{eid}{86} (\bibinfo{year}{2013}), \eprint{1210.7231}.

\bibitem[{\citenamefont{{Ferraro} et~al.}(2016)\citenamefont{{Ferraro}, {Hill},
  {Battaglia}, {Liu}, and {Spergel}}}]{Ferraro2016}
\bibinfo{author}{\bibfnamefont{S.}~\bibnamefont{{Ferraro}}},
  \bibinfo{author}{\bibfnamefont{J.~C.} \bibnamefont{{Hill}}},
  \bibinfo{author}{\bibfnamefont{N.}~\bibnamefont{{Battaglia}}},
  \bibinfo{author}{\bibfnamefont{J.}~\bibnamefont{{Liu}}}, \bibnamefont{and}
  \bibinfo{author}{\bibfnamefont{D.~N.} \bibnamefont{{Spergel}}},
  \bibinfo{journal}{\prd} \textbf{\bibinfo{volume}{94}}, \bibinfo{eid}{123526}
  (\bibinfo{year}{2016}), \eprint{1605.02722}.

\bibitem[{\citenamefont{{Hill} et~al.}(2016)\citenamefont{{Hill}, {Ferraro},
  {Battaglia}, {Liu}, and {Spergel}}}]{Hill2016}
\bibinfo{author}{\bibfnamefont{J.~C.} \bibnamefont{{Hill}}},
  \bibinfo{author}{\bibfnamefont{S.}~\bibnamefont{{Ferraro}}},
  \bibinfo{author}{\bibfnamefont{N.}~\bibnamefont{{Battaglia}}},
  \bibinfo{author}{\bibfnamefont{J.}~\bibnamefont{{Liu}}}, \bibnamefont{and}
  \bibinfo{author}{\bibfnamefont{D.~N.} \bibnamefont{{Spergel}}},
  \bibinfo{journal}{\prl} \textbf{\bibinfo{volume}{117}}, \bibinfo{eid}{051301}
  (\bibinfo{year}{2016}), \eprint{1603.01608}.

\bibitem[{\citenamefont{{Planck Collaboration}
  et~al.}(2016{\natexlab{c}})\citenamefont{{Planck Collaboration}, {Ade},
  {Aghanim}, {Arnaud}, {Ashdown}, {Aubourg}, {Aumont}, {Baccigalupi}, {Banday},
  {Barreiro} et~al.}}]{Planck2015kSZ}
\bibinfo{author}{\bibnamefont{{Planck Collaboration}}},
  \bibinfo{author}{\bibfnamefont{P.~A.~R.} \bibnamefont{{Ade}}},
  \bibinfo{author}{\bibfnamefont{N.}~\bibnamefont{{Aghanim}}},
  \bibinfo{author}{\bibfnamefont{M.}~\bibnamefont{{Arnaud}}},
  \bibinfo{author}{\bibfnamefont{M.}~\bibnamefont{{Ashdown}}},
  \bibinfo{author}{\bibfnamefont{E.}~\bibnamefont{{Aubourg}}},
  \bibinfo{author}{\bibfnamefont{J.}~\bibnamefont{{Aumont}}},
  \bibinfo{author}{\bibfnamefont{C.}~\bibnamefont{{Baccigalupi}}},
  \bibinfo{author}{\bibfnamefont{A.~J.} \bibnamefont{{Banday}}},
  \bibinfo{author}{\bibfnamefont{R.~B.} \bibnamefont{{Barreiro}}},
  \bibnamefont{et~al.}, \bibinfo{journal}{\aap} \textbf{\bibinfo{volume}{586}},
  \bibinfo{eid}{A140} (\bibinfo{year}{2016}{\natexlab{c}}),
  \eprint{1504.03339}.

\bibitem[{\citenamefont{{Planck Collaboration}
  et~al.}(2014{\natexlab{a}})\citenamefont{{Planck Collaboration}, {Ade},
  {Aghanim}, {Arnaud}, {Ashdown}, {Aumont}, {Baccigalupi}, {Balbi}, {Banday},
  {Barreiro} et~al.}}]{Planck2013kSZ}
\bibinfo{author}{\bibnamefont{{Planck Collaboration}}},
  \bibinfo{author}{\bibfnamefont{P.~A.~R.} \bibnamefont{{Ade}}},
  \bibinfo{author}{\bibfnamefont{N.}~\bibnamefont{{Aghanim}}},
  \bibinfo{author}{\bibfnamefont{M.}~\bibnamefont{{Arnaud}}},
  \bibinfo{author}{\bibfnamefont{M.}~\bibnamefont{{Ashdown}}},
  \bibinfo{author}{\bibfnamefont{J.}~\bibnamefont{{Aumont}}},
  \bibinfo{author}{\bibfnamefont{C.}~\bibnamefont{{Baccigalupi}}},
  \bibinfo{author}{\bibfnamefont{A.}~\bibnamefont{{Balbi}}},
  \bibinfo{author}{\bibfnamefont{A.~J.} \bibnamefont{{Banday}}},
  \bibinfo{author}{\bibfnamefont{R.~B.} \bibnamefont{{Barreiro}}},
  \bibnamefont{et~al.}, \bibinfo{journal}{\aap} \textbf{\bibinfo{volume}{561}},
  \bibinfo{eid}{A97} (\bibinfo{year}{2014}{\natexlab{a}}), \eprint{1303.5090}.

\bibitem[{\citenamefont{{Bhattacharya}
  et~al.}(2012)\citenamefont{{Bhattacharya}, {Nagai}, {Shaw}, {Crawford}, and
  {Holder}}}]{Bhattacharya2012}
\bibinfo{author}{\bibfnamefont{S.}~\bibnamefont{{Bhattacharya}}},
  \bibinfo{author}{\bibfnamefont{D.}~\bibnamefont{{Nagai}}},
  \bibinfo{author}{\bibfnamefont{L.}~\bibnamefont{{Shaw}}},
  \bibinfo{author}{\bibfnamefont{T.}~\bibnamefont{{Crawford}}},
  \bibnamefont{and} \bibinfo{author}{\bibfnamefont{G.~P.}
  \bibnamefont{{Holder}}}, \bibinfo{journal}{\apj}
  \textbf{\bibinfo{volume}{760}}, \bibinfo{eid}{5} (\bibinfo{year}{2012}),
  \eprint{1203.6368}.

\bibitem[{\citenamefont{{Coulton} et~al.}(2018)\citenamefont{{Coulton},
  {Aiola}, {Battaglia}, {Calabrese}, {Choi}, {Devlin}, {Gallardo}, {Hill},
  {Hincks}, {Hubmayr} et~al.}}]{Coulton2018}
\bibinfo{author}{\bibfnamefont{W.~R.} \bibnamefont{{Coulton}}},
  \bibinfo{author}{\bibfnamefont{S.}~\bibnamefont{{Aiola}}},
  \bibinfo{author}{\bibfnamefont{N.}~\bibnamefont{{Battaglia}}},
  \bibinfo{author}{\bibfnamefont{E.}~\bibnamefont{{Calabrese}}},
  \bibinfo{author}{\bibfnamefont{S.~K.} \bibnamefont{{Choi}}},
  \bibinfo{author}{\bibfnamefont{M.~J.} \bibnamefont{{Devlin}}},
  \bibinfo{author}{\bibfnamefont{P.~A.} \bibnamefont{{Gallardo}}},
  \bibinfo{author}{\bibfnamefont{J.~C.} \bibnamefont{{Hill}}},
  \bibinfo{author}{\bibfnamefont{A.~D.} \bibnamefont{{Hincks}}},
  \bibinfo{author}{\bibfnamefont{J.}~\bibnamefont{{Hubmayr}}},
  \bibnamefont{et~al.}, \bibinfo{journal}{\jcap} \textbf{\bibinfo{volume}{9}},
  \bibinfo{eid}{022} (\bibinfo{year}{2018}), \eprint{1711.07879}.

\bibitem[{\citenamefont{{Crawford} et~al.}(2014)\citenamefont{{Crawford},
  {Schaffer}, {Bhattacharya}, {Aird}, {Benson}, {Bleem}, {Carlstrom}, {Chang},
  {Cho}, {Crites} et~al.}}]{Crawford2014}
\bibinfo{author}{\bibfnamefont{T.~M.} \bibnamefont{{Crawford}}},
  \bibinfo{author}{\bibfnamefont{K.~K.} \bibnamefont{{Schaffer}}},
  \bibinfo{author}{\bibfnamefont{S.}~\bibnamefont{{Bhattacharya}}},
  \bibinfo{author}{\bibfnamefont{K.~A.} \bibnamefont{{Aird}}},
  \bibinfo{author}{\bibfnamefont{B.~A.} \bibnamefont{{Benson}}},
  \bibinfo{author}{\bibfnamefont{L.~E.} \bibnamefont{{Bleem}}},
  \bibinfo{author}{\bibfnamefont{J.~E.} \bibnamefont{{Carlstrom}}},
  \bibinfo{author}{\bibfnamefont{C.~L.} \bibnamefont{{Chang}}},
  \bibinfo{author}{\bibfnamefont{H.-M.} \bibnamefont{{Cho}}},
  \bibinfo{author}{\bibfnamefont{A.~T.} \bibnamefont{{Crites}}},
  \bibnamefont{et~al.}, \bibinfo{journal}{\apj} \textbf{\bibinfo{volume}{784}},
  \bibinfo{eid}{143} (\bibinfo{year}{2014}), \eprint{1303.3535}.

\bibitem[{\citenamefont{{Hill} and {Sherwin}}(2013)}]{Hill-Sherwin2013}
\bibinfo{author}{\bibfnamefont{J.~C.} \bibnamefont{{Hill}}} \bibnamefont{and}
  \bibinfo{author}{\bibfnamefont{B.~D.} \bibnamefont{{Sherwin}}},
  \bibinfo{journal}{\prd} \textbf{\bibinfo{volume}{87}}, \bibinfo{eid}{023527}
  (\bibinfo{year}{2013}), \eprint{1205.5794}.

\bibitem[{\citenamefont{{Hill} et~al.}(2014)\citenamefont{{Hill}, {Sherwin},
  {Smith}, {Addison}, {Battaglia}, {Battistelli}, {Bond}, {Calabrese},
  {Devlin}, {Dunkley} et~al.}}]{Hill2014}
\bibinfo{author}{\bibfnamefont{J.~C.} \bibnamefont{{Hill}}},
  \bibinfo{author}{\bibfnamefont{B.~D.} \bibnamefont{{Sherwin}}},
  \bibinfo{author}{\bibfnamefont{K.~M.} \bibnamefont{{Smith}}},
  \bibinfo{author}{\bibfnamefont{G.~E.} \bibnamefont{{Addison}}},
  \bibinfo{author}{\bibfnamefont{N.}~\bibnamefont{{Battaglia}}},
  \bibinfo{author}{\bibfnamefont{E.~S.} \bibnamefont{{Battistelli}}},
  \bibinfo{author}{\bibfnamefont{J.~R.} \bibnamefont{{Bond}}},
  \bibinfo{author}{\bibfnamefont{E.}~\bibnamefont{{Calabrese}}},
  \bibinfo{author}{\bibfnamefont{M.~J.} \bibnamefont{{Devlin}}},
  \bibinfo{author}{\bibfnamefont{J.}~\bibnamefont{{Dunkley}}},
  \bibnamefont{et~al.}, \bibinfo{journal}{arXiv e-prints}
  (\bibinfo{year}{2014}), \eprint{1411.8004}.

\bibitem[{\citenamefont{{Planck Collaboration}
  et~al.}(2016{\natexlab{d}})\citenamefont{{Planck Collaboration}, {Aghanim},
  {Arnaud}, {Ashdown}, {Aumont}, {Baccigalupi}, {Banday}, {Barreiro},
  {Bartlett}, {Bartolo} et~al.}}]{Planck2015ymap}
\bibinfo{author}{\bibnamefont{{Planck Collaboration}}},
  \bibinfo{author}{\bibfnamefont{N.}~\bibnamefont{{Aghanim}}},
  \bibinfo{author}{\bibfnamefont{M.}~\bibnamefont{{Arnaud}}},
  \bibinfo{author}{\bibfnamefont{M.}~\bibnamefont{{Ashdown}}},
  \bibinfo{author}{\bibfnamefont{J.}~\bibnamefont{{Aumont}}},
  \bibinfo{author}{\bibfnamefont{C.}~\bibnamefont{{Baccigalupi}}},
  \bibinfo{author}{\bibfnamefont{A.~J.} \bibnamefont{{Banday}}},
  \bibinfo{author}{\bibfnamefont{R.~B.} \bibnamefont{{Barreiro}}},
  \bibinfo{author}{\bibfnamefont{J.~G.} \bibnamefont{{Bartlett}}},
  \bibinfo{author}{\bibfnamefont{N.}~\bibnamefont{{Bartolo}}},
  \bibnamefont{et~al.}, \bibinfo{journal}{\aap} \textbf{\bibinfo{volume}{594}},
  \bibinfo{eid}{A22} (\bibinfo{year}{2016}{\natexlab{d}}), \eprint{1502.01596}.

\bibitem[{\citenamefont{{Thiele} et~al.}(2019)\citenamefont{{Thiele}, {Hill},
  and {Smith}}}]{Thiele2019}
\bibinfo{author}{\bibfnamefont{L.}~\bibnamefont{{Thiele}}},
  \bibinfo{author}{\bibfnamefont{J.~C.} \bibnamefont{{Hill}}},
  \bibnamefont{and} \bibinfo{author}{\bibfnamefont{K.~M.}
  \bibnamefont{{Smith}}}, \bibinfo{journal}{\prd}
  \textbf{\bibinfo{volume}{99}}, \bibinfo{eid}{103511} (\bibinfo{year}{2019}).

\bibitem[{\citenamefont{{Wilson} et~al.}(2012)\citenamefont{{Wilson},
  {Sherwin}, {Hill}, {Addison}, {Battaglia}, {Bond}, {Das}, {Devlin},
  {Dunkley}, {D{\"u}nner} et~al.}}]{Wilson2012}
\bibinfo{author}{\bibfnamefont{M.~J.} \bibnamefont{{Wilson}}},
  \bibinfo{author}{\bibfnamefont{B.~D.} \bibnamefont{{Sherwin}}},
  \bibinfo{author}{\bibfnamefont{J.~C.} \bibnamefont{{Hill}}},
  \bibinfo{author}{\bibfnamefont{G.}~\bibnamefont{{Addison}}},
  \bibinfo{author}{\bibfnamefont{N.}~\bibnamefont{{Battaglia}}},
  \bibinfo{author}{\bibfnamefont{J.~R.} \bibnamefont{{Bond}}},
  \bibinfo{author}{\bibfnamefont{S.}~\bibnamefont{{Das}}},
  \bibinfo{author}{\bibfnamefont{M.~J.} \bibnamefont{{Devlin}}},
  \bibinfo{author}{\bibfnamefont{J.}~\bibnamefont{{Dunkley}}},
  \bibinfo{author}{\bibfnamefont{R.}~\bibnamefont{{D{\"u}nner}}},
  \bibnamefont{et~al.}, \bibinfo{journal}{\prd} \textbf{\bibinfo{volume}{86}},
  \bibinfo{eid}{122005} (\bibinfo{year}{2012}), \eprint{1203.6633}.

\bibitem[{\citenamefont{{Alonso} et~al.}(2018)\citenamefont{{Alonso}, {Hill},
  {Hlo{\v z}ek}, and {Spergel}}}]{Alonso2018}
\bibinfo{author}{\bibfnamefont{D.}~\bibnamefont{{Alonso}}},
  \bibinfo{author}{\bibfnamefont{J.~C.} \bibnamefont{{Hill}}},
  \bibinfo{author}{\bibfnamefont{R.}~\bibnamefont{{Hlo{\v z}ek}}},
  \bibnamefont{and} \bibinfo{author}{\bibfnamefont{D.~N.}
  \bibnamefont{{Spergel}}}, \bibinfo{journal}{\prd}
  \textbf{\bibinfo{volume}{97}}, \bibinfo{eid}{063514} (\bibinfo{year}{2018}),
  \eprint{1709.01489}.

\bibitem[{\citenamefont{{Battaglia} et~al.}(2015)\citenamefont{{Battaglia},
  {Hill}, and {Murray}}}]{BHM2015}
\bibinfo{author}{\bibfnamefont{N.}~\bibnamefont{{Battaglia}}},
  \bibinfo{author}{\bibfnamefont{J.~C.} \bibnamefont{{Hill}}},
  \bibnamefont{and} \bibinfo{author}{\bibfnamefont{N.}~\bibnamefont{{Murray}}},
  \bibinfo{journal}{\apj} \textbf{\bibinfo{volume}{812}}, \bibinfo{eid}{154}
  (\bibinfo{year}{2015}), \eprint{1412.5593}.

\bibitem[{\citenamefont{{Hill} et~al.}(2018)\citenamefont{{Hill}, {Baxter},
  {Lidz}, {Greco}, and {Jain}}}]{Hill2018}
\bibinfo{author}{\bibfnamefont{J.~C.} \bibnamefont{{Hill}}},
  \bibinfo{author}{\bibfnamefont{E.~J.} \bibnamefont{{Baxter}}},
  \bibinfo{author}{\bibfnamefont{A.}~\bibnamefont{{Lidz}}},
  \bibinfo{author}{\bibfnamefont{J.~P.} \bibnamefont{{Greco}}},
  \bibnamefont{and} \bibinfo{author}{\bibfnamefont{B.}~\bibnamefont{{Jain}}},
  \bibinfo{journal}{\prd} \textbf{\bibinfo{volume}{97}}, \bibinfo{eid}{083501}
  (\bibinfo{year}{2018}), \eprint{1706.03753}.

\bibitem[{\citenamefont{{Hill} and {Spergel}}(2014)}]{Hill-Spergel2014}
\bibinfo{author}{\bibfnamefont{J.~C.} \bibnamefont{{Hill}}} \bibnamefont{and}
  \bibinfo{author}{\bibfnamefont{D.~N.} \bibnamefont{{Spergel}}},
  \bibinfo{journal}{\jcap} \textbf{\bibinfo{volume}{2}}, \bibinfo{eid}{030}
  (\bibinfo{year}{2014}), \eprint{1312.4525}.

\bibitem[{\citenamefont{{Pandey}
  et~al.}(2019{\natexlab{a}})\citenamefont{{Pandey}, {Baxter}, and
  {Hill}}}]{Pandey2019b}
\bibinfo{author}{\bibfnamefont{S.}~\bibnamefont{{Pandey}}},
  \bibinfo{author}{\bibfnamefont{E.~J.} \bibnamefont{{Baxter}}},
  \bibnamefont{and} \bibinfo{author}{\bibfnamefont{J.~C.}
  \bibnamefont{{Hill}}}, \bibinfo{journal}{arXiv e-prints}
  (\bibinfo{year}{2019}{\natexlab{a}}), \eprint{1909.00405}.

\bibitem[{\citenamefont{{Pandey}
  et~al.}(2019{\natexlab{b}})\citenamefont{{Pandey}, {Baxter}, {Xu},
  {Orlowski-Scherer}, {Zhu}, {Lidz}, {Aguirre}, {DeRose}, {Devlin}, {Hill}
  et~al.}}]{Pandey2019a}
\bibinfo{author}{\bibfnamefont{S.}~\bibnamefont{{Pandey}}},
  \bibinfo{author}{\bibfnamefont{E.~J.} \bibnamefont{{Baxter}}},
  \bibinfo{author}{\bibfnamefont{Z.}~\bibnamefont{{Xu}}},
  \bibinfo{author}{\bibfnamefont{J.}~\bibnamefont{{Orlowski-Scherer}}},
  \bibinfo{author}{\bibfnamefont{N.}~\bibnamefont{{Zhu}}},
  \bibinfo{author}{\bibfnamefont{A.}~\bibnamefont{{Lidz}}},
  \bibinfo{author}{\bibfnamefont{J.}~\bibnamefont{{Aguirre}}},
  \bibinfo{author}{\bibfnamefont{J.}~\bibnamefont{{DeRose}}},
  \bibinfo{author}{\bibfnamefont{M.}~\bibnamefont{{Devlin}}},
  \bibinfo{author}{\bibfnamefont{J.~C.} \bibnamefont{{Hill}}},
  \bibnamefont{et~al.}, \bibinfo{journal}{\prd} \textbf{\bibinfo{volume}{100}},
  \bibinfo{eid}{063519} (\bibinfo{year}{2019}{\natexlab{b}}),
  \eprint{1904.13347}.

\bibitem[{\citenamefont{{Van Waerbeke} et~al.}(2014)\citenamefont{{Van
  Waerbeke}, {Hinshaw}, and {Murray}}}]{VanWaerbeke2014}
\bibinfo{author}{\bibfnamefont{L.}~\bibnamefont{{Van Waerbeke}}},
  \bibinfo{author}{\bibfnamefont{G.}~\bibnamefont{{Hinshaw}}},
  \bibnamefont{and} \bibinfo{author}{\bibfnamefont{N.}~\bibnamefont{{Murray}}},
  \bibinfo{journal}{\prd} \textbf{\bibinfo{volume}{89}}, \bibinfo{eid}{023508}
  (\bibinfo{year}{2014}), \eprint{1310.5721}.

\bibitem[{\citenamefont{{Vikram} et~al.}(2017)\citenamefont{{Vikram}, {Lidz},
  and {Jain}}}]{Vikram2017}
\bibinfo{author}{\bibfnamefont{V.}~\bibnamefont{{Vikram}}},
  \bibinfo{author}{\bibfnamefont{A.}~\bibnamefont{{Lidz}}}, \bibnamefont{and}
  \bibinfo{author}{\bibfnamefont{B.}~\bibnamefont{{Jain}}},
  \bibinfo{journal}{\mnras} \textbf{\bibinfo{volume}{467}},
  \bibinfo{pages}{2315} (\bibinfo{year}{2017}), \eprint{1608.04160}.

\bibitem[{\citenamefont{{Benson} et~al.}(2014)\citenamefont{{Benson}, {Ade},
  {Ahmed}, {Allen}, {Arnold}, {Austermann}, {Bender}, {Bleem}, {Carlstrom},
  {Chang} et~al.}}]{spt3g}
\bibinfo{author}{\bibfnamefont{B.~A.} \bibnamefont{{Benson}}},
  \bibinfo{author}{\bibfnamefont{P.~A.~R.} \bibnamefont{{Ade}}},
  \bibinfo{author}{\bibfnamefont{Z.}~\bibnamefont{{Ahmed}}},
  \bibinfo{author}{\bibfnamefont{S.~W.} \bibnamefont{{Allen}}},
  \bibinfo{author}{\bibfnamefont{K.}~\bibnamefont{{Arnold}}},
  \bibinfo{author}{\bibfnamefont{J.~E.} \bibnamefont{{Austermann}}},
  \bibinfo{author}{\bibfnamefont{A.~N.} \bibnamefont{{Bender}}},
  \bibinfo{author}{\bibfnamefont{L.~E.} \bibnamefont{{Bleem}}},
  \bibinfo{author}{\bibfnamefont{J.~E.} \bibnamefont{{Carlstrom}}},
  \bibinfo{author}{\bibfnamefont{C.~L.} \bibnamefont{{Chang}}},
  \bibnamefont{et~al.}, in \emph{\bibinfo{booktitle}{Millimeter, Submillimeter,
  and Far-Infrared Detectors and Instrumentation for Astronomy VII}}
  (\bibinfo{year}{2014}), vol. \bibinfo{volume}{9153} of
  \emph{\bibinfo{series}{Proceedings of SPIE}}, p. \bibinfo{pages}{91531P},
  \eprint{1407.2973}.

\bibitem[{\citenamefont{{Everett} et~al.}(2018)\citenamefont{{Everett}, {Ade},
  {Ahmed}, {Anderson}, {Austermann}, {Avva}, {Thakur}, {Bender}, {Benson},
  {Carlstrom} et~al.}}]{spt3g2}
\bibinfo{author}{\bibfnamefont{W.}~\bibnamefont{{Everett}}},
  \bibinfo{author}{\bibfnamefont{P.~A.~R.} \bibnamefont{{Ade}}},
  \bibinfo{author}{\bibfnamefont{Z.}~\bibnamefont{{Ahmed}}},
  \bibinfo{author}{\bibfnamefont{A.~J.} \bibnamefont{{Anderson}}},
  \bibinfo{author}{\bibfnamefont{J.~E.} \bibnamefont{{Austermann}}},
  \bibinfo{author}{\bibfnamefont{J.~S.} \bibnamefont{{Avva}}},
  \bibinfo{author}{\bibfnamefont{R.~B.} \bibnamefont{{Thakur}}},
  \bibinfo{author}{\bibfnamefont{A.~N.} \bibnamefont{{Bender}}},
  \bibinfo{author}{\bibfnamefont{B.~A.} \bibnamefont{{Benson}}},
  \bibinfo{author}{\bibfnamefont{J.~E.} \bibnamefont{{Carlstrom}}},
  \bibnamefont{et~al.}, \bibinfo{journal}{Journal of Low Temperature Physics}
  \textbf{\bibinfo{volume}{193}}, \bibinfo{pages}{1085} (\bibinfo{year}{2018}),
  \eprint{1902.09640}.

\bibitem[{\citenamefont{Anderson et~al.}(2018)}]{spt3g3}
\bibinfo{author}{\bibfnamefont{A.~J.} \bibnamefont{Anderson}}
  \bibnamefont{et~al.} (\bibinfo{collaboration}{SPT}), \bibinfo{journal}{J.
  Low. Temp. Phys.} \textbf{\bibinfo{volume}{193}}, \bibinfo{pages}{1057}
  (\bibinfo{year}{2018}).

\bibitem[{\citenamefont{{Henderson} et~al.}(2016)\citenamefont{{Henderson},
  {Allison}, {Austermann}, {Baildon}, {Battaglia}, {Beall}, {Becker}, {De
  Bernardis}, {Bond}, {Calabrese} et~al.}}]{Henderson2016}
\bibinfo{author}{\bibfnamefont{S.~W.} \bibnamefont{{Henderson}}},
  \bibinfo{author}{\bibfnamefont{R.}~\bibnamefont{{Allison}}},
  \bibinfo{author}{\bibfnamefont{J.}~\bibnamefont{{Austermann}}},
  \bibinfo{author}{\bibfnamefont{T.}~\bibnamefont{{Baildon}}},
  \bibinfo{author}{\bibfnamefont{N.}~\bibnamefont{{Battaglia}}},
  \bibinfo{author}{\bibfnamefont{J.~A.} \bibnamefont{{Beall}}},
  \bibinfo{author}{\bibfnamefont{D.}~\bibnamefont{{Becker}}},
  \bibinfo{author}{\bibfnamefont{F.}~\bibnamefont{{De Bernardis}}},
  \bibinfo{author}{\bibfnamefont{J.~R.} \bibnamefont{{Bond}}},
  \bibinfo{author}{\bibfnamefont{E.}~\bibnamefont{{Calabrese}}},
  \bibnamefont{et~al.}, \bibinfo{journal}{Journal of Low Temperature Physics}
  \textbf{\bibinfo{volume}{184}}, \bibinfo{pages}{772} (\bibinfo{year}{2016}),
  \eprint{1510.02809}.

\bibitem[{\citenamefont{{Ade} et~al.}(2019)\citenamefont{{Ade}, {Aguirre},
  {Ahmed}, {Aiola}, {Ali}, {Alonso}, {Alvarez}, {Arnold}, {Ashton},
  {Austermann} et~al.}}]{SO2019}
\bibinfo{author}{\bibfnamefont{P.}~\bibnamefont{{Ade}}},
  \bibinfo{author}{\bibfnamefont{J.}~\bibnamefont{{Aguirre}}},
  \bibinfo{author}{\bibfnamefont{Z.}~\bibnamefont{{Ahmed}}},
  \bibinfo{author}{\bibfnamefont{S.}~\bibnamefont{{Aiola}}},
  \bibinfo{author}{\bibfnamefont{A.}~\bibnamefont{{Ali}}},
  \bibinfo{author}{\bibfnamefont{D.}~\bibnamefont{{Alonso}}},
  \bibinfo{author}{\bibfnamefont{M.~A.} \bibnamefont{{Alvarez}}},
  \bibinfo{author}{\bibfnamefont{K.}~\bibnamefont{{Arnold}}},
  \bibinfo{author}{\bibfnamefont{P.}~\bibnamefont{{Ashton}}},
  \bibinfo{author}{\bibfnamefont{J.}~\bibnamefont{{Austermann}}},
  \bibnamefont{et~al.}, \bibinfo{journal}{\jcap} \textbf{\bibinfo{volume}{2}},
  \bibinfo{eid}{056} (\bibinfo{year}{2019}), \eprint{1808.07445}.

\bibitem[{\citenamefont{{Stacey} et~al.}(2018)\citenamefont{{Stacey},
  {Aravena}, {Basu}, {Battaglia}, {Beringue}, {Bertoldi}, {Bond}, {Breysse},
  {Bustos}, {Chapman} et~al.}}]{CCATp2018}
\bibinfo{author}{\bibfnamefont{G.~J.} \bibnamefont{{Stacey}}},
  \bibinfo{author}{\bibfnamefont{M.}~\bibnamefont{{Aravena}}},
  \bibinfo{author}{\bibfnamefont{K.}~\bibnamefont{{Basu}}},
  \bibinfo{author}{\bibfnamefont{N.}~\bibnamefont{{Battaglia}}},
  \bibinfo{author}{\bibfnamefont{B.}~\bibnamefont{{Beringue}}},
  \bibinfo{author}{\bibfnamefont{F.}~\bibnamefont{{Bertoldi}}},
  \bibinfo{author}{\bibfnamefont{J.~R.} \bibnamefont{{Bond}}},
  \bibinfo{author}{\bibfnamefont{P.}~\bibnamefont{{Breysse}}},
  \bibinfo{author}{\bibfnamefont{R.}~\bibnamefont{{Bustos}}},
  \bibinfo{author}{\bibfnamefont{S.}~\bibnamefont{{Chapman}}},
  \bibnamefont{et~al.}, in \emph{\bibinfo{booktitle}{\procspie}}
  (\bibinfo{year}{2018}), vol. \bibinfo{volume}{10700} of
  \emph{\bibinfo{series}{Society of Photo-Optical Instrumentation Engineers
  (SPIE) Conference Series}}, p. \bibinfo{pages}{107001M}, \eprint{1807.04354}.

\bibitem[{\citenamefont{{Aghanim} et~al.}(2019)\citenamefont{{Aghanim},
  {Douspis}, {Hurier}, {Crichton}, {Diego}, {Hasselfield}, {Macias-Perez},
  {Marriage}, {Pointecouteau}, {Remazeilles} et~al.}}]{PACT2019}
\bibinfo{author}{\bibfnamefont{N.}~\bibnamefont{{Aghanim}}},
  \bibinfo{author}{\bibfnamefont{M.}~\bibnamefont{{Douspis}}},
  \bibinfo{author}{\bibfnamefont{G.}~\bibnamefont{{Hurier}}},
  \bibinfo{author}{\bibfnamefont{D.}~\bibnamefont{{Crichton}}},
  \bibinfo{author}{\bibfnamefont{J.~M.} \bibnamefont{{Diego}}},
  \bibinfo{author}{\bibfnamefont{M.}~\bibnamefont{{Hasselfield}}},
  \bibinfo{author}{\bibfnamefont{J.}~\bibnamefont{{Macias-Perez}}},
  \bibinfo{author}{\bibfnamefont{T.~A.} \bibnamefont{{Marriage}}},
  \bibinfo{author}{\bibfnamefont{E.}~\bibnamefont{{Pointecouteau}}},
  \bibinfo{author}{\bibfnamefont{M.}~\bibnamefont{{Remazeilles}}},
  \bibnamefont{et~al.}, \bibinfo{journal}{\aap} \textbf{\bibinfo{volume}{632}},
  \bibinfo{eid}{A47} (\bibinfo{year}{2019}).

\bibitem[{\citenamefont{{Chown} et~al.}(2018)\citenamefont{{Chown}, {Omori},
  {Aylor}, {Benson}, {Bleem}, {Carlstrom}, {Chang}, {Cho}, {Crawford}, {Crites}
  et~al.}}]{ChownEtAl}
\bibinfo{author}{\bibfnamefont{R.}~\bibnamefont{{Chown}}},
  \bibinfo{author}{\bibfnamefont{Y.}~\bibnamefont{{Omori}}},
  \bibinfo{author}{\bibfnamefont{K.}~\bibnamefont{{Aylor}}},
  \bibinfo{author}{\bibfnamefont{B.~A.} \bibnamefont{{Benson}}},
  \bibinfo{author}{\bibfnamefont{L.~E.} \bibnamefont{{Bleem}}},
  \bibinfo{author}{\bibfnamefont{J.~E.} \bibnamefont{{Carlstrom}}},
  \bibinfo{author}{\bibfnamefont{C.~L.} \bibnamefont{{Chang}}},
  \bibinfo{author}{\bibfnamefont{H.~M.} \bibnamefont{{Cho}}},
  \bibinfo{author}{\bibfnamefont{T.~M.} \bibnamefont{{Crawford}}},
  \bibinfo{author}{\bibfnamefont{A.~T.} \bibnamefont{{Crites}}},
  \bibnamefont{et~al.}, \bibinfo{journal}{\apjs}
  \textbf{\bibinfo{volume}{239}}, \bibinfo{eid}{10} (\bibinfo{year}{2018}),
  \eprint{1803.10682}.

\bibitem[{\citenamefont{{Melin} et~al.}(2006)\citenamefont{{Melin}, {Bartlett},
  and {Delabrouille}}}]{JB2006}
\bibinfo{author}{\bibfnamefont{J.-B.} \bibnamefont{{Melin}}},
  \bibinfo{author}{\bibfnamefont{J.~G.} \bibnamefont{{Bartlett}}},
  \bibnamefont{and}
  \bibinfo{author}{\bibfnamefont{J.}~\bibnamefont{{Delabrouille}}},
  \bibinfo{journal}{\aap} \textbf{\bibinfo{volume}{459}}, \bibinfo{pages}{341}
  (\bibinfo{year}{2006}), \eprint{astro-ph/0602424}.

\bibitem[{\citenamefont{{Planck Collaboration}
  et~al.}(2016{\natexlab{e}})\citenamefont{{Planck Collaboration}, {Ade},
  {Aghanim}, {Arnaud}, {Ashdown}, {Aumont}, {Baccigalupi}, {Banday},
  {Barreiro}, {Barrena} et~al.}}]{PlanckClusters2015}
\bibinfo{author}{\bibnamefont{{Planck Collaboration}}},
  \bibinfo{author}{\bibfnamefont{P.~A.~R.} \bibnamefont{{Ade}}},
  \bibinfo{author}{\bibfnamefont{N.}~\bibnamefont{{Aghanim}}},
  \bibinfo{author}{\bibfnamefont{M.}~\bibnamefont{{Arnaud}}},
  \bibinfo{author}{\bibfnamefont{M.}~\bibnamefont{{Ashdown}}},
  \bibinfo{author}{\bibfnamefont{J.}~\bibnamefont{{Aumont}}},
  \bibinfo{author}{\bibfnamefont{C.}~\bibnamefont{{Baccigalupi}}},
  \bibinfo{author}{\bibfnamefont{A.~J.} \bibnamefont{{Banday}}},
  \bibinfo{author}{\bibfnamefont{R.~B.} \bibnamefont{{Barreiro}}},
  \bibinfo{author}{\bibfnamefont{R.}~\bibnamefont{{Barrena}}},
  \bibnamefont{et~al.}, \bibinfo{journal}{\aap} \textbf{\bibinfo{volume}{594}},
  \bibinfo{eid}{A27} (\bibinfo{year}{2016}{\natexlab{e}}), \eprint{1502.01598}.

\bibitem[{\citenamefont{{Williamson} et~al.}(2011)\citenamefont{{Williamson},
  {Benson}, {High}, {Vand erlinde}, {Ade}, {Aird}, {Andersson}, {Armstrong},
  {Ashby}, {Bautz} et~al.}}]{Williamson2011}
\bibinfo{author}{\bibfnamefont{R.}~\bibnamefont{{Williamson}}},
  \bibinfo{author}{\bibfnamefont{B.~A.} \bibnamefont{{Benson}}},
  \bibinfo{author}{\bibfnamefont{F.~W.} \bibnamefont{{High}}},
  \bibinfo{author}{\bibfnamefont{K.}~\bibnamefont{{Vand erlinde}}},
  \bibinfo{author}{\bibfnamefont{P.~A.~R.} \bibnamefont{{Ade}}},
  \bibinfo{author}{\bibfnamefont{K.~A.} \bibnamefont{{Aird}}},
  \bibinfo{author}{\bibfnamefont{K.}~\bibnamefont{{Andersson}}},
  \bibinfo{author}{\bibfnamefont{R.}~\bibnamefont{{Armstrong}}},
  \bibinfo{author}{\bibfnamefont{M.~L.~N.} \bibnamefont{{Ashby}}},
  \bibinfo{author}{\bibfnamefont{M.}~\bibnamefont{{Bautz}}},
  \bibnamefont{et~al.}, \bibinfo{journal}{\apj} \textbf{\bibinfo{volume}{738}},
  \bibinfo{eid}{139} (\bibinfo{year}{2011}), \eprint{1101.1290}.

\bibitem[{\citenamefont{{Hilton} et~al.}(2018)\citenamefont{{Hilton},
  {Hasselfield}, {Sif{\'o}n}, {Battaglia}, {Aiola}, {Bharadwaj}, {Bond},
  {Choi}, {Crichton}, {Datta} et~al.}}]{HiltonEtAl}
\bibinfo{author}{\bibfnamefont{M.}~\bibnamefont{{Hilton}}},
  \bibinfo{author}{\bibfnamefont{M.}~\bibnamefont{{Hasselfield}}},
  \bibinfo{author}{\bibfnamefont{C.}~\bibnamefont{{Sif{\'o}n}}},
  \bibinfo{author}{\bibfnamefont{N.}~\bibnamefont{{Battaglia}}},
  \bibinfo{author}{\bibfnamefont{S.}~\bibnamefont{{Aiola}}},
  \bibinfo{author}{\bibfnamefont{V.}~\bibnamefont{{Bharadwaj}}},
  \bibinfo{author}{\bibfnamefont{J.~R.} \bibnamefont{{Bond}}},
  \bibinfo{author}{\bibfnamefont{S.~K.} \bibnamefont{{Choi}}},
  \bibinfo{author}{\bibfnamefont{D.}~\bibnamefont{{Crichton}}},
  \bibinfo{author}{\bibfnamefont{R.}~\bibnamefont{{Datta}}},
  \bibnamefont{et~al.}, \bibinfo{journal}{\apjs}
  \textbf{\bibinfo{volume}{235}}, \bibinfo{eid}{20} (\bibinfo{year}{2018}),
  \eprint{1709.05600}.

\bibitem[{\citenamefont{{Marriage} et~al.}(2011)\citenamefont{{Marriage},
  {Acquaviva}, {Ade}, {Aguirre}, {Amiri}, {Appel}, {Barrientos}, {Battistelli},
  {Bond}, {Brown} et~al.}}]{2011ApJ...737...61M}
\bibinfo{author}{\bibfnamefont{T.~A.} \bibnamefont{{Marriage}}},
  \bibinfo{author}{\bibfnamefont{V.}~\bibnamefont{{Acquaviva}}},
  \bibinfo{author}{\bibfnamefont{P.~A.~R.} \bibnamefont{{Ade}}},
  \bibinfo{author}{\bibfnamefont{P.}~\bibnamefont{{Aguirre}}},
  \bibinfo{author}{\bibfnamefont{M.}~\bibnamefont{{Amiri}}},
  \bibinfo{author}{\bibfnamefont{J.~W.} \bibnamefont{{Appel}}},
  \bibinfo{author}{\bibfnamefont{L.~F.} \bibnamefont{{Barrientos}}},
  \bibinfo{author}{\bibfnamefont{E.~S.} \bibnamefont{{Battistelli}}},
  \bibinfo{author}{\bibfnamefont{J.~R.} \bibnamefont{{Bond}}},
  \bibinfo{author}{\bibfnamefont{B.}~\bibnamefont{{Brown}}},
  \bibnamefont{et~al.}, \bibinfo{journal}{\apj} \textbf{\bibinfo{volume}{737}},
  \bibinfo{eid}{61} (\bibinfo{year}{2011}), \eprint{1010.1065}.

\bibitem[{\citenamefont{{Louis} et~al.}(2017)\citenamefont{{Louis}, {Grace},
  {Hasselfield}, {Lungu}, {Maurin}, {Addison}, {Ade}, {Aiola}, {Allison},
  {Amiri} et~al.}}]{Louis2017}
\bibinfo{author}{\bibfnamefont{T.}~\bibnamefont{{Louis}}},
  \bibinfo{author}{\bibfnamefont{E.}~\bibnamefont{{Grace}}},
  \bibinfo{author}{\bibfnamefont{M.}~\bibnamefont{{Hasselfield}}},
  \bibinfo{author}{\bibfnamefont{M.}~\bibnamefont{{Lungu}}},
  \bibinfo{author}{\bibfnamefont{L.}~\bibnamefont{{Maurin}}},
  \bibinfo{author}{\bibfnamefont{G.~E.} \bibnamefont{{Addison}}},
  \bibinfo{author}{\bibfnamefont{P.~A.~R.} \bibnamefont{{Ade}}},
  \bibinfo{author}{\bibfnamefont{S.}~\bibnamefont{{Aiola}}},
  \bibinfo{author}{\bibfnamefont{R.}~\bibnamefont{{Allison}}},
  \bibinfo{author}{\bibfnamefont{M.}~\bibnamefont{{Amiri}}},
  \bibnamefont{et~al.}, \bibinfo{journal}{\jcap}
  \textbf{\bibinfo{volume}{2017}}, \bibinfo{eid}{031} (\bibinfo{year}{2017}),
  \eprint{1610.02360}.

\bibitem[{\citenamefont{{Naess} et~al.}(2014)\citenamefont{{Naess},
  {Hasselfield}, {McMahon}, {Niemack}, {Addison}, {Ade}, {Allison}, {Amiri},
  {Battaglia}, {Beall} et~al.}}]{Naess2014}
\bibinfo{author}{\bibfnamefont{S.}~\bibnamefont{{Naess}}},
  \bibinfo{author}{\bibfnamefont{M.}~\bibnamefont{{Hasselfield}}},
  \bibinfo{author}{\bibfnamefont{J.}~\bibnamefont{{McMahon}}},
  \bibinfo{author}{\bibfnamefont{M.~D.} \bibnamefont{{Niemack}}},
  \bibinfo{author}{\bibfnamefont{G.~E.} \bibnamefont{{Addison}}},
  \bibinfo{author}{\bibfnamefont{P.~A.~R.} \bibnamefont{{Ade}}},
  \bibinfo{author}{\bibfnamefont{R.}~\bibnamefont{{Allison}}},
  \bibinfo{author}{\bibfnamefont{M.}~\bibnamefont{{Amiri}}},
  \bibinfo{author}{\bibfnamefont{N.}~\bibnamefont{{Battaglia}}},
  \bibinfo{author}{\bibfnamefont{J.~A.} \bibnamefont{{Beall}}},
  \bibnamefont{et~al.}, \bibinfo{journal}{\jcap}
  \textbf{\bibinfo{volume}{2014}}, \bibinfo{eid}{007} (\bibinfo{year}{2014}),
  \eprint{1405.5524}.

\bibitem[{\citenamefont{{Thornton} et~al.}(2016)\citenamefont{{Thornton},
  {Ade}, {Aiola}, {Angil{\`e}}, {Amiri}, {Beall}, {Becker}, {Cho}, {Choi},
  {Corlies} et~al.}}]{thornton/2016}
\bibinfo{author}{\bibfnamefont{R.~J.} \bibnamefont{{Thornton}}},
  \bibinfo{author}{\bibfnamefont{P.~A.~R.} \bibnamefont{{Ade}}},
  \bibinfo{author}{\bibfnamefont{S.}~\bibnamefont{{Aiola}}},
  \bibinfo{author}{\bibfnamefont{F.~E.} \bibnamefont{{Angil{\`e}}}},
  \bibinfo{author}{\bibfnamefont{M.}~\bibnamefont{{Amiri}}},
  \bibinfo{author}{\bibfnamefont{J.~A.} \bibnamefont{{Beall}}},
  \bibinfo{author}{\bibfnamefont{D.~T.} \bibnamefont{{Becker}}},
  \bibinfo{author}{\bibfnamefont{H.-M.} \bibnamefont{{Cho}}},
  \bibinfo{author}{\bibfnamefont{S.~K.} \bibnamefont{{Choi}}},
  \bibinfo{author}{\bibfnamefont{P.}~\bibnamefont{{Corlies}}},
  \bibnamefont{et~al.}, \bibinfo{journal}{\apjs}
  \textbf{\bibinfo{volume}{227}}, \bibinfo{eid}{21} (\bibinfo{year}{2016}),
  \eprint{1605.06569}.

\bibitem[{\citenamefont{{Bennett} et~al.}(2003)\citenamefont{{Bennett}, {Hill},
  {Hinshaw}, {Nolta}, {Odegard}, {Page}, {Spergel}, {Weiland}, {Wright},
  {Halpern} et~al.}}]{Bennett2003}
\bibinfo{author}{\bibfnamefont{C.~L.} \bibnamefont{{Bennett}}},
  \bibinfo{author}{\bibfnamefont{R.~S.} \bibnamefont{{Hill}}},
  \bibinfo{author}{\bibfnamefont{G.}~\bibnamefont{{Hinshaw}}},
  \bibinfo{author}{\bibfnamefont{M.~R.} \bibnamefont{{Nolta}}},
  \bibinfo{author}{\bibfnamefont{N.}~\bibnamefont{{Odegard}}},
  \bibinfo{author}{\bibfnamefont{L.}~\bibnamefont{{Page}}},
  \bibinfo{author}{\bibfnamefont{D.~N.} \bibnamefont{{Spergel}}},
  \bibinfo{author}{\bibfnamefont{J.~L.} \bibnamefont{{Weiland}}},
  \bibinfo{author}{\bibfnamefont{E.~L.} \bibnamefont{{Wright}}},
  \bibinfo{author}{\bibfnamefont{M.}~\bibnamefont{{Halpern}}},
  \bibnamefont{et~al.}, \bibinfo{journal}{\apjs}
  \textbf{\bibinfo{volume}{148}}, \bibinfo{pages}{97} (\bibinfo{year}{2003}),
  \eprint{astro-ph/0302208}.

\bibitem[{\citenamefont{{Delabrouille}
  et~al.}(2009)\citenamefont{{Delabrouille}, {Cardoso}, {Le Jeune}, {Betoule},
  {Fay}, and {Guilloux}}}]{Delabrouille2009}
\bibinfo{author}{\bibfnamefont{J.}~\bibnamefont{{Delabrouille}}},
  \bibinfo{author}{\bibfnamefont{J.-F.} \bibnamefont{{Cardoso}}},
  \bibinfo{author}{\bibfnamefont{M.}~\bibnamefont{{Le Jeune}}},
  \bibinfo{author}{\bibfnamefont{M.}~\bibnamefont{{Betoule}}},
  \bibinfo{author}{\bibfnamefont{G.}~\bibnamefont{{Fay}}}, \bibnamefont{and}
  \bibinfo{author}{\bibfnamefont{F.}~\bibnamefont{{Guilloux}}},
  \bibinfo{journal}{\aap} \textbf{\bibinfo{volume}{493}}, \bibinfo{pages}{835}
  (\bibinfo{year}{2009}), \eprint{0807.0773}.

\bibitem[{\citenamefont{{Eriksen} et~al.}(2004)\citenamefont{{Eriksen},
  {Banday}, {G{\'o}rski}, and {Lilje}}}]{Eriksen2004}
\bibinfo{author}{\bibfnamefont{H.~K.} \bibnamefont{{Eriksen}}},
  \bibinfo{author}{\bibfnamefont{A.~J.} \bibnamefont{{Banday}}},
  \bibinfo{author}{\bibfnamefont{K.~M.} \bibnamefont{{G{\'o}rski}}},
  \bibnamefont{and} \bibinfo{author}{\bibfnamefont{P.~B.}
  \bibnamefont{{Lilje}}}, \bibinfo{journal}{\apj}
  \textbf{\bibinfo{volume}{612}}, \bibinfo{pages}{633} (\bibinfo{year}{2004}),
  \eprint{astro-ph/0403098}.

\bibitem[{\citenamefont{{Remazeilles} et~al.}(2011)\citenamefont{{Remazeilles},
  {Delabrouille}, and {Cardoso}}}]{Remazeilles2011}
\bibinfo{author}{\bibfnamefont{M.}~\bibnamefont{{Remazeilles}}},
  \bibinfo{author}{\bibfnamefont{J.}~\bibnamefont{{Delabrouille}}},
  \bibnamefont{and} \bibinfo{author}{\bibfnamefont{J.-F.}
  \bibnamefont{{Cardoso}}}, \bibinfo{journal}{\mnras}
  \textbf{\bibinfo{volume}{410}}, \bibinfo{pages}{2481} (\bibinfo{year}{2011}),
  \eprint{1006.5599}.

\bibitem[{\citenamefont{{Planck Collaboration}
  et~al.}(2018{\natexlab{c}})\citenamefont{{Planck Collaboration}, {Aghanim},
  {Akrami}, {Ashdown}, {Aumont}, {Baccigalupi}, {Ballardini}, {Banday},
  {Barreiro}, {Bartolo} et~al.}}]{PlanckCosmology}
\bibinfo{author}{\bibnamefont{{Planck Collaboration}}},
  \bibinfo{author}{\bibfnamefont{N.}~\bibnamefont{{Aghanim}}},
  \bibinfo{author}{\bibfnamefont{Y.}~\bibnamefont{{Akrami}}},
  \bibinfo{author}{\bibfnamefont{M.}~\bibnamefont{{Ashdown}}},
  \bibinfo{author}{\bibfnamefont{J.}~\bibnamefont{{Aumont}}},
  \bibinfo{author}{\bibfnamefont{C.}~\bibnamefont{{Baccigalupi}}},
  \bibinfo{author}{\bibfnamefont{M.}~\bibnamefont{{Ballardini}}},
  \bibinfo{author}{\bibfnamefont{A.~J.} \bibnamefont{{Banday}}},
  \bibinfo{author}{\bibfnamefont{R.~B.} \bibnamefont{{Barreiro}}},
  \bibinfo{author}{\bibfnamefont{N.}~\bibnamefont{{Bartolo}}},
  \bibnamefont{et~al.}, \bibinfo{journal}{arXiv e-prints}
  \bibinfo{eid}{arXiv:1807.06209} (\bibinfo{year}{2018}{\natexlab{c}}),
  \eprint{1807.06209}.

\bibitem[{\citenamefont{{The Dark Energy Survey Collaboration}}(2005)}]{DES}
\bibinfo{author}{\bibnamefont{{The Dark Energy Survey Collaboration}}},
  \bibinfo{journal}{ArXiv Astrophysics e-prints}  (\bibinfo{year}{2005}),
  \eprint{astro-ph/0510346}.

\bibitem[{\citenamefont{{de Jong} et~al.}(2013)\citenamefont{{de Jong},
  {Kuijken}, {Applegate}, {Begeman}, {Belikov}, {Blake}, {Bout}, {Boxhoorn},
  {Buddelmeijer}, {Buddendiek} et~al.}}]{KIDS}
\bibinfo{author}{\bibfnamefont{J.~T.~A.} \bibnamefont{{de Jong}}},
  \bibinfo{author}{\bibfnamefont{K.}~\bibnamefont{{Kuijken}}},
  \bibinfo{author}{\bibfnamefont{D.}~\bibnamefont{{Applegate}}},
  \bibinfo{author}{\bibfnamefont{K.}~\bibnamefont{{Begeman}}},
  \bibinfo{author}{\bibfnamefont{A.}~\bibnamefont{{Belikov}}},
  \bibinfo{author}{\bibfnamefont{C.}~\bibnamefont{{Blake}}},
  \bibinfo{author}{\bibfnamefont{J.}~\bibnamefont{{Bout}}},
  \bibinfo{author}{\bibfnamefont{D.}~\bibnamefont{{Boxhoorn}}},
  \bibinfo{author}{\bibfnamefont{H.}~\bibnamefont{{Buddelmeijer}}},
  \bibinfo{author}{\bibfnamefont{A.}~\bibnamefont{{Buddendiek}}},
  \bibnamefont{et~al.}, \bibinfo{journal}{The Messenger}
  \textbf{\bibinfo{volume}{154}}, \bibinfo{pages}{44} (\bibinfo{year}{2013}).

\bibitem[{\citenamefont{Miyazaki et~al.}(2012)\citenamefont{Miyazaki, Komiyama,
  Nakaya, Kamata, Doi, Hamana, Karoji, Furusawa, Kawanomoto, Morokuma
  et~al.}}]{HSC}
\bibinfo{author}{\bibfnamefont{S.}~\bibnamefont{Miyazaki}},
  \bibinfo{author}{\bibfnamefont{Y.}~\bibnamefont{Komiyama}},
  \bibinfo{author}{\bibfnamefont{H.}~\bibnamefont{Nakaya}},
  \bibinfo{author}{\bibfnamefont{Y.}~\bibnamefont{Kamata}},
  \bibinfo{author}{\bibfnamefont{Y.}~\bibnamefont{Doi}},
  \bibinfo{author}{\bibfnamefont{T.}~\bibnamefont{Hamana}},
  \bibinfo{author}{\bibfnamefont{H.}~\bibnamefont{Karoji}},
  \bibinfo{author}{\bibfnamefont{H.}~\bibnamefont{Furusawa}},
  \bibinfo{author}{\bibfnamefont{S.}~\bibnamefont{Kawanomoto}},
  \bibinfo{author}{\bibfnamefont{T.}~\bibnamefont{Morokuma}},
  \bibnamefont{et~al.}, in \emph{\bibinfo{booktitle}{Ground-based and Airborne
  Instrumentation for Astronomy IV}} (\bibinfo{organization}{International
  Society for Optics and Photonics}, \bibinfo{year}{2012}), vol.
  \bibinfo{volume}{8446}, p. \bibinfo{pages}{84460Z}.

\bibitem[{\citenamefont{{LSST Science Collaboration}
  et~al.}(2009)\citenamefont{{LSST Science Collaboration}, {Abell}, {Allison},
  {Anderson}, {Andrew}, {Angel}, {Armus}, {Arnett}, {Asztalos}, {Axelrod}
  et~al.}}]{LSST}
\bibinfo{author}{\bibnamefont{{LSST Science Collaboration}}},
  \bibinfo{author}{\bibfnamefont{P.~A.} \bibnamefont{{Abell}}},
  \bibinfo{author}{\bibfnamefont{J.}~\bibnamefont{{Allison}}},
  \bibinfo{author}{\bibfnamefont{S.~F.} \bibnamefont{{Anderson}}},
  \bibinfo{author}{\bibfnamefont{J.~R.} \bibnamefont{{Andrew}}},
  \bibinfo{author}{\bibfnamefont{J.~R.~P.} \bibnamefont{{Angel}}},
  \bibinfo{author}{\bibfnamefont{L.}~\bibnamefont{{Armus}}},
  \bibinfo{author}{\bibfnamefont{D.}~\bibnamefont{{Arnett}}},
  \bibinfo{author}{\bibfnamefont{S.~J.} \bibnamefont{{Asztalos}}},
  \bibinfo{author}{\bibfnamefont{T.~S.} \bibnamefont{{Axelrod}}},
  \bibnamefont{et~al.}, \bibinfo{journal}{ArXiv e-prints}
  (\bibinfo{year}{2009}), \eprint{0912.0201}.

\bibitem[{\citenamefont{{DESI Collaboration} et~al.}(2016)\citenamefont{{DESI
  Collaboration}, {Aghamousa}, {Aguilar}, {Ahlen}, {Alam}, {Allen}, {Allende
  Prieto}, {Annis}, {Bailey}, {Balland} et~al.}}]{DESI}
\bibinfo{author}{\bibnamefont{{DESI Collaboration}}},
  \bibinfo{author}{\bibfnamefont{A.}~\bibnamefont{{Aghamousa}}},
  \bibinfo{author}{\bibfnamefont{J.}~\bibnamefont{{Aguilar}}},
  \bibinfo{author}{\bibfnamefont{S.}~\bibnamefont{{Ahlen}}},
  \bibinfo{author}{\bibfnamefont{S.}~\bibnamefont{{Alam}}},
  \bibinfo{author}{\bibfnamefont{L.~E.} \bibnamefont{{Allen}}},
  \bibinfo{author}{\bibfnamefont{C.}~\bibnamefont{{Allende Prieto}}},
  \bibinfo{author}{\bibfnamefont{J.}~\bibnamefont{{Annis}}},
  \bibinfo{author}{\bibfnamefont{S.}~\bibnamefont{{Bailey}}},
  \bibinfo{author}{\bibfnamefont{C.}~\bibnamefont{{Balland}}},
  \bibnamefont{et~al.}, \bibinfo{journal}{arXiv e-prints}
  \bibinfo{eid}{arXiv:1611.00036} (\bibinfo{year}{2016}), \eprint{1611.00036}.

\bibitem[{\citenamefont{{Amiaux} et~al.}(2012)\citenamefont{{Amiaux},
  {Scaramella}, {Mellier}, {Altieri}, {Burigana}, {Da Silva}, {Gomez}, {Hoar},
  {Laureijs}, {Maiorano} et~al.}}]{EUCLID}
\bibinfo{author}{\bibfnamefont{J.}~\bibnamefont{{Amiaux}}},
  \bibinfo{author}{\bibfnamefont{R.}~\bibnamefont{{Scaramella}}},
  \bibinfo{author}{\bibfnamefont{Y.}~\bibnamefont{{Mellier}}},
  \bibinfo{author}{\bibfnamefont{B.}~\bibnamefont{{Altieri}}},
  \bibinfo{author}{\bibfnamefont{C.}~\bibnamefont{{Burigana}}},
  \bibinfo{author}{\bibfnamefont{A.}~\bibnamefont{{Da Silva}}},
  \bibinfo{author}{\bibfnamefont{P.}~\bibnamefont{{Gomez}}},
  \bibinfo{author}{\bibfnamefont{J.}~\bibnamefont{{Hoar}}},
  \bibinfo{author}{\bibfnamefont{R.}~\bibnamefont{{Laureijs}}},
  \bibinfo{author}{\bibfnamefont{E.}~\bibnamefont{{Maiorano}}},
  \bibnamefont{et~al.}, in \emph{\bibinfo{booktitle}{Space Telescopes and
  Instrumentation 2012: Optical, Infrared, and Millimeter Wave}}
  (\bibinfo{year}{2012}), vol. \bibinfo{volume}{8442} of
  \emph{\bibinfo{series}{Proceedings of SPIE}}, p. \bibinfo{pages}{84420Z},
  \eprint{1209.2228}.

\bibitem[{\citenamefont{{Alam} et~al.}(2015)\citenamefont{{Alam}, {Albareti},
  {Allende Prieto}, {Anders}, {Anderson}, {Anderton}, {Andrews}, {Armengaud},
  {Aubourg}, {Bailey} et~al.}}]{SDSS}
\bibinfo{author}{\bibfnamefont{S.}~\bibnamefont{{Alam}}},
  \bibinfo{author}{\bibfnamefont{F.~D.} \bibnamefont{{Albareti}}},
  \bibinfo{author}{\bibfnamefont{C.}~\bibnamefont{{Allende Prieto}}},
  \bibinfo{author}{\bibfnamefont{F.}~\bibnamefont{{Anders}}},
  \bibinfo{author}{\bibfnamefont{S.~F.} \bibnamefont{{Anderson}}},
  \bibinfo{author}{\bibfnamefont{T.}~\bibnamefont{{Anderton}}},
  \bibinfo{author}{\bibfnamefont{B.~H.} \bibnamefont{{Andrews}}},
  \bibinfo{author}{\bibfnamefont{E.}~\bibnamefont{{Armengaud}}},
  \bibinfo{author}{\bibfnamefont{{\'E}.}~\bibnamefont{{Aubourg}}},
  \bibinfo{author}{\bibfnamefont{S.}~\bibnamefont{{Bailey}}},
  \bibnamefont{et~al.}, \bibinfo{journal}{\apjs}
  \textbf{\bibinfo{volume}{219}}, \bibinfo{eid}{12} (\bibinfo{year}{2015}),
  \eprint{1501.00963}.

\bibitem[{\citenamefont{{Wright} et~al.}(2010)\citenamefont{{Wright},
  {Eisenhardt}, {Mainzer}, {Ressler}, {Cutri}, {Jarrett}, {Kirkpatrick},
  {Padgett}, {McMillan}, {Skrutskie} et~al.}}]{WISE2010}
\bibinfo{author}{\bibfnamefont{E.~L.} \bibnamefont{{Wright}}},
  \bibinfo{author}{\bibfnamefont{P.~R.~M.} \bibnamefont{{Eisenhardt}}},
  \bibinfo{author}{\bibfnamefont{A.~K.} \bibnamefont{{Mainzer}}},
  \bibinfo{author}{\bibfnamefont{M.~E.} \bibnamefont{{Ressler}}},
  \bibinfo{author}{\bibfnamefont{R.~M.} \bibnamefont{{Cutri}}},
  \bibinfo{author}{\bibfnamefont{T.}~\bibnamefont{{Jarrett}}},
  \bibinfo{author}{\bibfnamefont{J.~D.} \bibnamefont{{Kirkpatrick}}},
  \bibinfo{author}{\bibfnamefont{D.}~\bibnamefont{{Padgett}}},
  \bibinfo{author}{\bibfnamefont{R.~S.} \bibnamefont{{McMillan}}},
  \bibinfo{author}{\bibfnamefont{M.}~\bibnamefont{{Skrutskie}}},
  \bibnamefont{et~al.}, \bibinfo{journal}{\aj} \textbf{\bibinfo{volume}{140}},
  \bibinfo{pages}{1868} (\bibinfo{year}{2010}), \eprint{1008.0031}.

\bibitem[{\citenamefont{{Green} et~al.}(2012)\citenamefont{{Green},
  {Schechter}, {Baltay}, {Bean}, {Bennett}, {Brown}, {Conselice}, {Donahue},
  {Fan}, {Gaudi} et~al.}}]{WFIRST}
\bibinfo{author}{\bibfnamefont{J.}~\bibnamefont{{Green}}},
  \bibinfo{author}{\bibfnamefont{P.}~\bibnamefont{{Schechter}}},
  \bibinfo{author}{\bibfnamefont{C.}~\bibnamefont{{Baltay}}},
  \bibinfo{author}{\bibfnamefont{R.}~\bibnamefont{{Bean}}},
  \bibinfo{author}{\bibfnamefont{D.}~\bibnamefont{{Bennett}}},
  \bibinfo{author}{\bibfnamefont{R.}~\bibnamefont{{Brown}}},
  \bibinfo{author}{\bibfnamefont{C.}~\bibnamefont{{Conselice}}},
  \bibinfo{author}{\bibfnamefont{M.}~\bibnamefont{{Donahue}}},
  \bibinfo{author}{\bibfnamefont{X.}~\bibnamefont{{Fan}}},
  \bibinfo{author}{\bibfnamefont{B.~S.} \bibnamefont{{Gaudi}}},
  \bibnamefont{et~al.}, \bibinfo{journal}{ArXiv e-prints}
  (\bibinfo{year}{2012}), \eprint{1208.4012}.

\bibitem[{\citenamefont{{Merloni} et~al.}(2012)\citenamefont{{Merloni},
  {Predehl}, {Becker}, {B{\"o}hringer}, {Boller}, {Brunner}, {Brusa},
  {Dennerl}, {Freyberg}, {Friedrich} et~al.}}]{EROSITA}
\bibinfo{author}{\bibfnamefont{A.}~\bibnamefont{{Merloni}}},
  \bibinfo{author}{\bibfnamefont{P.}~\bibnamefont{{Predehl}}},
  \bibinfo{author}{\bibfnamefont{W.}~\bibnamefont{{Becker}}},
  \bibinfo{author}{\bibfnamefont{H.}~\bibnamefont{{B{\"o}hringer}}},
  \bibinfo{author}{\bibfnamefont{T.}~\bibnamefont{{Boller}}},
  \bibinfo{author}{\bibfnamefont{H.}~\bibnamefont{{Brunner}}},
  \bibinfo{author}{\bibfnamefont{M.}~\bibnamefont{{Brusa}}},
  \bibinfo{author}{\bibfnamefont{K.}~\bibnamefont{{Dennerl}}},
  \bibinfo{author}{\bibfnamefont{M.}~\bibnamefont{{Freyberg}}},
  \bibinfo{author}{\bibfnamefont{P.}~\bibnamefont{{Friedrich}}},
  \bibnamefont{et~al.}, \bibinfo{journal}{arXiv e-prints}
  \bibinfo{eid}{arXiv:1209.3114} (\bibinfo{year}{2012}), \eprint{1209.3114}.

\bibitem[{\citenamefont{{Voges} et~al.}(1999)\citenamefont{{Voges},
  {Aschenbach}, {Boller}, {Br{\"a}uninger}, {Briel}, {Burkert}, {Dennerl},
  {Englhauser}, {Gruber}, {Haberl} et~al.}}]{ROSAT}
\bibinfo{author}{\bibfnamefont{W.}~\bibnamefont{{Voges}}},
  \bibinfo{author}{\bibfnamefont{B.}~\bibnamefont{{Aschenbach}}},
  \bibinfo{author}{\bibfnamefont{T.}~\bibnamefont{{Boller}}},
  \bibinfo{author}{\bibfnamefont{H.}~\bibnamefont{{Br{\"a}uninger}}},
  \bibinfo{author}{\bibfnamefont{U.}~\bibnamefont{{Briel}}},
  \bibinfo{author}{\bibfnamefont{W.}~\bibnamefont{{Burkert}}},
  \bibinfo{author}{\bibfnamefont{K.}~\bibnamefont{{Dennerl}}},
  \bibinfo{author}{\bibfnamefont{J.}~\bibnamefont{{Englhauser}}},
  \bibinfo{author}{\bibfnamefont{R.}~\bibnamefont{{Gruber}}},
  \bibinfo{author}{\bibfnamefont{F.}~\bibnamefont{{Haberl}}},
  \bibnamefont{et~al.}, \bibinfo{journal}{\aap} \textbf{\bibinfo{volume}{349}},
  \bibinfo{pages}{389} (\bibinfo{year}{1999}), \eprint{astro-ph/9909315}.

\bibitem[{\citenamefont{{Battaglia} et~al.}(2019)\citenamefont{{Battaglia},
  {Hill}, {Amodeo}, {Bartlett}, {Basu}, {Erler}, {Ferraro}, {Hernquist},
  {Madhavacheril}, {McQuinn} et~al.}}]{BH2020SWP}
\bibinfo{author}{\bibfnamefont{N.}~\bibnamefont{{Battaglia}}},
  \bibinfo{author}{\bibfnamefont{J.~C.} \bibnamefont{{Hill}}},
  \bibinfo{author}{\bibfnamefont{S.}~\bibnamefont{{Amodeo}}},
  \bibinfo{author}{\bibfnamefont{J.~G.} \bibnamefont{{Bartlett}}},
  \bibinfo{author}{\bibfnamefont{K.}~\bibnamefont{{Basu}}},
  \bibinfo{author}{\bibfnamefont{J.}~\bibnamefont{{Erler}}},
  \bibinfo{author}{\bibfnamefont{S.}~\bibnamefont{{Ferraro}}},
  \bibinfo{author}{\bibfnamefont{L.}~\bibnamefont{{Hernquist}}},
  \bibinfo{author}{\bibfnamefont{M.}~\bibnamefont{{Madhavacheril}}},
  \bibinfo{author}{\bibfnamefont{M.}~\bibnamefont{{McQuinn}}},
  \bibnamefont{et~al.}, \bibinfo{journal}{\baas} \textbf{\bibinfo{volume}{51}},
  \bibinfo{eid}{297} (\bibinfo{year}{2019}), \eprint{1903.04647}.

\bibitem[{\citenamefont{Madhavacheril and Hill}(2018)}]{MMHill}
\bibinfo{author}{\bibfnamefont{M.~S.} \bibnamefont{Madhavacheril}}
  \bibnamefont{and} \bibinfo{author}{\bibfnamefont{J.~C.} \bibnamefont{Hill}},
  \bibinfo{journal}{Phys. Rev. D} \textbf{\bibinfo{volume}{98}},
  \bibinfo{pages}{023534} (\bibinfo{year}{2018}),
  \urlprefix\url{https://link.aps.org/doi/10.1103/PhysRevD.98.023534}.

\bibitem[{\citenamefont{{Fixsen}}(2009)}]{Fixsen2009}
\bibinfo{author}{\bibfnamefont{D.~J.} \bibnamefont{{Fixsen}}},
  \bibinfo{journal}{\apj} \textbf{\bibinfo{volume}{707}}, \bibinfo{pages}{916}
  (\bibinfo{year}{2009}), \eprint{0911.1955}.

\bibitem[{\citenamefont{{Addison}
  et~al.}(2012{\natexlab{a}})\citenamefont{{Addison}, {Dunkley}, and
  {Spergel}}}]{Addison2012}
\bibinfo{author}{\bibfnamefont{G.~E.} \bibnamefont{{Addison}}},
  \bibinfo{author}{\bibfnamefont{J.}~\bibnamefont{{Dunkley}}},
  \bibnamefont{and} \bibinfo{author}{\bibfnamefont{D.~N.}
  \bibnamefont{{Spergel}}}, \bibinfo{journal}{\mnras}
  \textbf{\bibinfo{volume}{427}}, \bibinfo{pages}{1741}
  (\bibinfo{year}{2012}{\natexlab{a}}), \eprint{1204.5927}.

\bibitem[{\citenamefont{{Ward} et~al.}(2018)\citenamefont{{Ward}, {Alonso},
  {Errard}, {Devlin}, and {Hasselfield}}}]{WardBandpass}
\bibinfo{author}{\bibfnamefont{J.~T.} \bibnamefont{{Ward}}},
  \bibinfo{author}{\bibfnamefont{D.}~\bibnamefont{{Alonso}}},
  \bibinfo{author}{\bibfnamefont{J.}~\bibnamefont{{Errard}}},
  \bibinfo{author}{\bibfnamefont{M.~J.} \bibnamefont{{Devlin}}},
  \bibnamefont{and}
  \bibinfo{author}{\bibfnamefont{M.}~\bibnamefont{{Hasselfield}}},
  \bibinfo{journal}{\apj} \textbf{\bibinfo{volume}{861}}, \bibinfo{eid}{82}
  (\bibinfo{year}{2018}), \eprint{1803.07630}.

\bibitem[{\citenamefont{{Remazeilles} et~al.}(2013)\citenamefont{{Remazeilles},
  {Aghanim}, and {Douspis}}}]{HeteroILC}
\bibinfo{author}{\bibfnamefont{M.}~\bibnamefont{{Remazeilles}}},
  \bibinfo{author}{\bibfnamefont{N.}~\bibnamefont{{Aghanim}}},
  \bibnamefont{and}
  \bibinfo{author}{\bibfnamefont{M.}~\bibnamefont{{Douspis}}},
  \bibinfo{journal}{\mnras} \textbf{\bibinfo{volume}{430}},
  \bibinfo{pages}{370} (\bibinfo{year}{2013}), \eprint{1207.4683}.

\bibitem[{\citenamefont{{Bobin} et~al.}(2016)\citenamefont{{Bobin}, {Sureau},
  and {Starck}}}]{Bobin2016}
\bibinfo{author}{\bibfnamefont{J.}~\bibnamefont{{Bobin}}},
  \bibinfo{author}{\bibfnamefont{F.}~\bibnamefont{{Sureau}}}, \bibnamefont{and}
  \bibinfo{author}{\bibfnamefont{J.~L.} \bibnamefont{{Starck}}},
  \bibinfo{journal}{\aap} \textbf{\bibinfo{volume}{591}}, \bibinfo{eid}{A50}
  (\bibinfo{year}{2016}), \eprint{1511.08690}.

\bibitem[{\citenamefont{{Hajian} et~al.}(2012)\citenamefont{{Hajian}, {Viero},
  {Addison}, {Aguirre}, {Appel}, {Battaglia}, {Bock}, {Bond}, {Das}, {Devlin}
  et~al.}}]{CIBDecorr1}
\bibinfo{author}{\bibfnamefont{A.}~\bibnamefont{{Hajian}}},
  \bibinfo{author}{\bibfnamefont{M.~P.} \bibnamefont{{Viero}}},
  \bibinfo{author}{\bibfnamefont{G.}~\bibnamefont{{Addison}}},
  \bibinfo{author}{\bibfnamefont{P.}~\bibnamefont{{Aguirre}}},
  \bibinfo{author}{\bibfnamefont{J.~W.} \bibnamefont{{Appel}}},
  \bibinfo{author}{\bibfnamefont{N.}~\bibnamefont{{Battaglia}}},
  \bibinfo{author}{\bibfnamefont{J.~J.} \bibnamefont{{Bock}}},
  \bibinfo{author}{\bibfnamefont{J.~R.} \bibnamefont{{Bond}}},
  \bibinfo{author}{\bibfnamefont{S.}~\bibnamefont{{Das}}},
  \bibinfo{author}{\bibfnamefont{M.~J.} \bibnamefont{{Devlin}}},
  \bibnamefont{et~al.}, \bibinfo{journal}{\apj} \textbf{\bibinfo{volume}{744}},
  \bibinfo{eid}{40} (\bibinfo{year}{2012}), \eprint{1101.1517}.

\bibitem[{\citenamefont{{Addison}
  et~al.}(2012{\natexlab{b}})\citenamefont{{Addison}, {Dunkley}, {Hajian},
  {Viero}, {Bond}, {Das}, {Devlin}, {Halpern}, {Hincks}, {Hlozek}
  et~al.}}]{CIBDecorr2}
\bibinfo{author}{\bibfnamefont{G.~E.} \bibnamefont{{Addison}}},
  \bibinfo{author}{\bibfnamefont{J.}~\bibnamefont{{Dunkley}}},
  \bibinfo{author}{\bibfnamefont{A.}~\bibnamefont{{Hajian}}},
  \bibinfo{author}{\bibfnamefont{M.}~\bibnamefont{{Viero}}},
  \bibinfo{author}{\bibfnamefont{J.~R.} \bibnamefont{{Bond}}},
  \bibinfo{author}{\bibfnamefont{S.}~\bibnamefont{{Das}}},
  \bibinfo{author}{\bibfnamefont{M.~J.} \bibnamefont{{Devlin}}},
  \bibinfo{author}{\bibfnamefont{M.}~\bibnamefont{{Halpern}}},
  \bibinfo{author}{\bibfnamefont{A.~D.} \bibnamefont{{Hincks}}},
  \bibinfo{author}{\bibfnamefont{R.}~\bibnamefont{{Hlozek}}},
  \bibnamefont{et~al.}, \bibinfo{journal}{\apj} \textbf{\bibinfo{volume}{752}},
  \bibinfo{eid}{120} (\bibinfo{year}{2012}{\natexlab{b}}), \eprint{1108.4614}.

\bibitem[{\citenamefont{{Planck Collaboration}
  et~al.}(2014{\natexlab{b}})\citenamefont{{Planck Collaboration}, {Ade},
  {Aghanim}, {Armitage-Caplan}, {Arnaud}, {Ashdown}, {Atrio-Barand ela},
  {Aumont}, {Baccigalupi}, {Banday} et~al.}}]{CIBDecorr3}
\bibinfo{author}{\bibnamefont{{Planck Collaboration}}},
  \bibinfo{author}{\bibfnamefont{P.~A.~R.} \bibnamefont{{Ade}}},
  \bibinfo{author}{\bibfnamefont{N.}~\bibnamefont{{Aghanim}}},
  \bibinfo{author}{\bibfnamefont{C.}~\bibnamefont{{Armitage-Caplan}}},
  \bibinfo{author}{\bibfnamefont{M.}~\bibnamefont{{Arnaud}}},
  \bibinfo{author}{\bibfnamefont{M.}~\bibnamefont{{Ashdown}}},
  \bibinfo{author}{\bibfnamefont{F.}~\bibnamefont{{Atrio-Barand ela}}},
  \bibinfo{author}{\bibfnamefont{J.}~\bibnamefont{{Aumont}}},
  \bibinfo{author}{\bibfnamefont{C.}~\bibnamefont{{Baccigalupi}}},
  \bibinfo{author}{\bibfnamefont{A.~J.} \bibnamefont{{Banday}}},
  \bibnamefont{et~al.}, \bibinfo{journal}{\aap} \textbf{\bibinfo{volume}{571}},
  \bibinfo{eid}{A30} (\bibinfo{year}{2014}{\natexlab{b}}), \eprint{1309.0382}.

\bibitem[{\citenamefont{{Viero} et~al.}(2013)\citenamefont{{Viero}, {Wang},
  {Zemcov}, {Addison}, {Amblard}, {Arumugam}, {Aussel}, {B{\'e}thermin},
  {Bock}, {Boselli} et~al.}}]{CIBDecorr4}
\bibinfo{author}{\bibfnamefont{M.~P.} \bibnamefont{{Viero}}},
  \bibinfo{author}{\bibfnamefont{L.}~\bibnamefont{{Wang}}},
  \bibinfo{author}{\bibfnamefont{M.}~\bibnamefont{{Zemcov}}},
  \bibinfo{author}{\bibfnamefont{G.}~\bibnamefont{{Addison}}},
  \bibinfo{author}{\bibfnamefont{A.}~\bibnamefont{{Amblard}}},
  \bibinfo{author}{\bibfnamefont{V.}~\bibnamefont{{Arumugam}}},
  \bibinfo{author}{\bibfnamefont{H.}~\bibnamefont{{Aussel}}},
  \bibinfo{author}{\bibfnamefont{M.}~\bibnamefont{{B{\'e}thermin}}},
  \bibinfo{author}{\bibfnamefont{J.}~\bibnamefont{{Bock}}},
  \bibinfo{author}{\bibfnamefont{A.}~\bibnamefont{{Boselli}}},
  \bibnamefont{et~al.}, \bibinfo{journal}{\apj} \textbf{\bibinfo{volume}{772}},
  \bibinfo{eid}{77} (\bibinfo{year}{2013}), \eprint{1208.5049}.

\bibitem[{\citenamefont{{Viero} et~al.}(2019)\citenamefont{{Viero},
  {Reichardt}, {Benson}, {Bleem}, {Bock}, {Carlstrom}, {Chang}, {Cho},
  {Crawford}, {Crites} et~al.}}]{CIBDecorr5}
\bibinfo{author}{\bibfnamefont{M.~P.} \bibnamefont{{Viero}}},
  \bibinfo{author}{\bibfnamefont{C.~L.} \bibnamefont{{Reichardt}}},
  \bibinfo{author}{\bibfnamefont{B.~A.} \bibnamefont{{Benson}}},
  \bibinfo{author}{\bibfnamefont{L.~E.} \bibnamefont{{Bleem}}},
  \bibinfo{author}{\bibfnamefont{J.}~\bibnamefont{{Bock}}},
  \bibinfo{author}{\bibfnamefont{J.~E.} \bibnamefont{{Carlstrom}}},
  \bibinfo{author}{\bibfnamefont{C.~L.} \bibnamefont{{Chang}}},
  \bibinfo{author}{\bibfnamefont{H.~M.} \bibnamefont{{Cho}}},
  \bibinfo{author}{\bibfnamefont{T.~M.} \bibnamefont{{Crawford}}},
  \bibinfo{author}{\bibfnamefont{A.~T.} \bibnamefont{{Crites}}},
  \bibnamefont{et~al.}, \bibinfo{journal}{\apj} \textbf{\bibinfo{volume}{881}},
  \bibinfo{eid}{96} (\bibinfo{year}{2019}), \eprint{1810.10643}.

\bibitem[{\citenamefont{{Sunyaev} and {Zeldovich}}(1970)}]{SZ1970}
\bibinfo{author}{\bibfnamefont{R.~A.} \bibnamefont{{Sunyaev}}}
  \bibnamefont{and} \bibinfo{author}{\bibfnamefont{Y.~B.}
  \bibnamefont{{Zeldovich}}}, \bibinfo{journal}{\apss}
  \textbf{\bibinfo{volume}{7}}, \bibinfo{pages}{3} (\bibinfo{year}{1970}).

\bibitem[{\citenamefont{{Zeldovich} and {Sunyaev}}(1969)}]{ZS1969}
\bibinfo{author}{\bibfnamefont{Y.~B.} \bibnamefont{{Zeldovich}}}
  \bibnamefont{and} \bibinfo{author}{\bibfnamefont{R.~A.}
  \bibnamefont{{Sunyaev}}}, \bibinfo{journal}{\apss}
  \textbf{\bibinfo{volume}{4}}, \bibinfo{pages}{301} (\bibinfo{year}{1969}).

\bibitem[{\citenamefont{{Nozawa} et~al.}(2006)\citenamefont{{Nozawa}, {Itoh},
  {Suda}, and {Ohhata}}}]{Nozawa2006}
\bibinfo{author}{\bibfnamefont{S.}~\bibnamefont{{Nozawa}}},
  \bibinfo{author}{\bibfnamefont{N.}~\bibnamefont{{Itoh}}},
  \bibinfo{author}{\bibfnamefont{Y.}~\bibnamefont{{Suda}}}, \bibnamefont{and}
  \bibinfo{author}{\bibfnamefont{Y.}~\bibnamefont{{Ohhata}}},
  \bibinfo{journal}{Nuovo Cimento B Serie} \textbf{\bibinfo{volume}{121}},
  \bibinfo{pages}{487} (\bibinfo{year}{2006}), \eprint{astro-ph/0507466}.

\bibitem[{\citenamefont{{Chluba} et~al.}(2017)\citenamefont{{Chluba}, {Hill},
  and {Abitbol}}}]{Chluba2017}
\bibinfo{author}{\bibfnamefont{J.}~\bibnamefont{{Chluba}}},
  \bibinfo{author}{\bibfnamefont{J.~C.} \bibnamefont{{Hill}}},
  \bibnamefont{and} \bibinfo{author}{\bibfnamefont{M.~H.}
  \bibnamefont{{Abitbol}}}, \bibinfo{journal}{\mnras}
  \textbf{\bibinfo{volume}{472}}, \bibinfo{pages}{1195} (\bibinfo{year}{2017}),
  \eprint{1701.00274}.

\bibitem[{\citenamefont{{Remazeilles} and {Chluba}}(2019)}]{Remazeilles2019}
\bibinfo{author}{\bibfnamefont{M.}~\bibnamefont{{Remazeilles}}}
  \bibnamefont{and} \bibinfo{author}{\bibfnamefont{J.}~\bibnamefont{{Chluba}}},
  \bibinfo{journal}{arXiv e-prints} \bibinfo{eid}{arXiv:1907.00916}
  (\bibinfo{year}{2019}), \eprint{1907.00916}.

\bibitem[{\citenamefont{{Planck Collaboration}
  et~al.}(2014{\natexlab{c}})\citenamefont{{Planck Collaboration}, {Ade},
  {Aghanim}, {Armitage-Caplan}, {Arnaud}, {Ashdown}, {Atrio-Barand ela},
  {Aumont}, {Baccigalupi}, {Banday} et~al.}}]{Planck2013CIB}
\bibinfo{author}{\bibnamefont{{Planck Collaboration}}},
  \bibinfo{author}{\bibfnamefont{P.~A.~R.} \bibnamefont{{Ade}}},
  \bibinfo{author}{\bibfnamefont{N.}~\bibnamefont{{Aghanim}}},
  \bibinfo{author}{\bibfnamefont{C.}~\bibnamefont{{Armitage-Caplan}}},
  \bibinfo{author}{\bibfnamefont{M.}~\bibnamefont{{Arnaud}}},
  \bibinfo{author}{\bibfnamefont{M.}~\bibnamefont{{Ashdown}}},
  \bibinfo{author}{\bibfnamefont{F.}~\bibnamefont{{Atrio-Barand ela}}},
  \bibinfo{author}{\bibfnamefont{J.}~\bibnamefont{{Aumont}}},
  \bibinfo{author}{\bibfnamefont{C.}~\bibnamefont{{Baccigalupi}}},
  \bibinfo{author}{\bibfnamefont{A.~J.} \bibnamefont{{Banday}}},
  \bibnamefont{et~al.}, \bibinfo{journal}{\aap} \textbf{\bibinfo{volume}{571}},
  \bibinfo{eid}{A30} (\bibinfo{year}{2014}{\natexlab{c}}), \eprint{1309.0382}.

\bibitem[{\citenamefont{{Planck Collaboration}
  et~al.}(2016{\natexlab{f}})\citenamefont{{Planck Collaboration}, {Ade},
  {Aghanim}, {Ashdown}, {Aumont}, {Baccigalupi}, {Band ay}, {Barreiro},
  {Bartolo}, {Battaner} et~al.}}]{LFIMaps}
\bibinfo{author}{\bibnamefont{{Planck Collaboration}}},
  \bibinfo{author}{\bibfnamefont{P.~A.~R.} \bibnamefont{{Ade}}},
  \bibinfo{author}{\bibfnamefont{N.}~\bibnamefont{{Aghanim}}},
  \bibinfo{author}{\bibfnamefont{M.}~\bibnamefont{{Ashdown}}},
  \bibinfo{author}{\bibfnamefont{J.}~\bibnamefont{{Aumont}}},
  \bibinfo{author}{\bibfnamefont{C.}~\bibnamefont{{Baccigalupi}}},
  \bibinfo{author}{\bibfnamefont{A.~J.} \bibnamefont{{Band ay}}},
  \bibinfo{author}{\bibfnamefont{R.~B.} \bibnamefont{{Barreiro}}},
  \bibinfo{author}{\bibfnamefont{N.}~\bibnamefont{{Bartolo}}},
  \bibinfo{author}{\bibfnamefont{E.}~\bibnamefont{{Battaner}}},
  \bibnamefont{et~al.}, \bibinfo{journal}{\aap} \textbf{\bibinfo{volume}{594}},
  \bibinfo{eid}{A6} (\bibinfo{year}{2016}{\natexlab{f}}), \eprint{1502.01585}.

\bibitem[{\citenamefont{{Planck Collaboration}
  et~al.}(2016{\natexlab{g}})\citenamefont{{Planck Collaboration}, {Adam},
  {Ade}, {Aghanim}, {Arnaud}, {Ashdown}, {Aumont}, {Baccigalupi}, {Banday},
  {Barreiro} et~al.}}]{Planck2015HFIcal}
\bibinfo{author}{\bibnamefont{{Planck Collaboration}}},
  \bibinfo{author}{\bibfnamefont{R.}~\bibnamefont{{Adam}}},
  \bibinfo{author}{\bibfnamefont{P.~A.~R.} \bibnamefont{{Ade}}},
  \bibinfo{author}{\bibfnamefont{N.}~\bibnamefont{{Aghanim}}},
  \bibinfo{author}{\bibfnamefont{M.}~\bibnamefont{{Arnaud}}},
  \bibinfo{author}{\bibfnamefont{M.}~\bibnamefont{{Ashdown}}},
  \bibinfo{author}{\bibfnamefont{J.}~\bibnamefont{{Aumont}}},
  \bibinfo{author}{\bibfnamefont{C.}~\bibnamefont{{Baccigalupi}}},
  \bibinfo{author}{\bibfnamefont{A.~J.} \bibnamefont{{Banday}}},
  \bibinfo{author}{\bibfnamefont{R.~B.} \bibnamefont{{Barreiro}}},
  \bibnamefont{et~al.}, \bibinfo{journal}{\aap} \textbf{\bibinfo{volume}{594}},
  \bibinfo{eid}{A8} (\bibinfo{year}{2016}{\natexlab{g}}), \eprint{1502.01587}.

\bibitem[{\citenamefont{{Choi} et~al.}(2020)\citenamefont{{Choi},
  {Hasselfield}, {Ho}, {Koopman}, {Lungu}, {Abitbol}, {Addison}, {Ade},
  {Aiola}, {Alonso} et~al.}}]{ChoiEtAl}
\bibinfo{author}{\bibfnamefont{S.~K.} \bibnamefont{{Choi}}},
  \bibinfo{author}{\bibfnamefont{M.}~\bibnamefont{{Hasselfield}}},
  \bibinfo{author}{\bibfnamefont{S.-P.~P.} \bibnamefont{{Ho}}},
  \bibinfo{author}{\bibfnamefont{B.}~\bibnamefont{{Koopman}}},
  \bibinfo{author}{\bibfnamefont{M.}~\bibnamefont{{Lungu}}},
  \bibinfo{author}{\bibfnamefont{M.~H.} \bibnamefont{{Abitbol}}},
  \bibinfo{author}{\bibfnamefont{G.~E.} \bibnamefont{{Addison}}},
  \bibinfo{author}{\bibfnamefont{P.~A.~R.} \bibnamefont{{Ade}}},
  \bibinfo{author}{\bibfnamefont{S.}~\bibnamefont{{Aiola}}},
  \bibinfo{author}{\bibfnamefont{D.}~\bibnamefont{{Alonso}}},
  \bibnamefont{et~al.}, \bibinfo{journal}{arXiv e-prints}
  \bibinfo{eid}{arXiv:2007.07289} (\bibinfo{year}{2020}), \eprint{2007.07289}.

\bibitem[{\citenamefont{{Aiola} et~al.}(2020)\citenamefont{{Aiola},
  {Calabrese}, {Maurin}, {Naess}, {Schmitt}, {Abitbol}, {Addison}, {Ade},
  {Alonso}, {Amiri} et~al.}}]{AiolaEtAl}
\bibinfo{author}{\bibfnamefont{S.}~\bibnamefont{{Aiola}}},
  \bibinfo{author}{\bibfnamefont{E.}~\bibnamefont{{Calabrese}}},
  \bibinfo{author}{\bibfnamefont{L.}~\bibnamefont{{Maurin}}},
  \bibinfo{author}{\bibfnamefont{S.}~\bibnamefont{{Naess}}},
  \bibinfo{author}{\bibfnamefont{B.~L.} \bibnamefont{{Schmitt}}},
  \bibinfo{author}{\bibfnamefont{M.~H.} \bibnamefont{{Abitbol}}},
  \bibinfo{author}{\bibfnamefont{G.~E.} \bibnamefont{{Addison}}},
  \bibinfo{author}{\bibfnamefont{P.~A.~R.} \bibnamefont{{Ade}}},
  \bibinfo{author}{\bibfnamefont{D.}~\bibnamefont{{Alonso}}},
  \bibinfo{author}{\bibfnamefont{M.}~\bibnamefont{{Amiri}}},
  \bibnamefont{et~al.}, \bibinfo{journal}{arXiv e-prints}
  \bibinfo{eid}{arXiv:2007.07288} (\bibinfo{year}{2020}), \eprint{2007.07288}.

\bibitem[{\citenamefont{{Planck Collaboration}
  et~al.}(2014{\natexlab{d}})\citenamefont{{Planck Collaboration}, {Aghanim},
  {Armitage-Caplan}, {Arnaud}, {Ashdown}, {Atrio-Barandela}, {Aumont},
  {Baccigalupi}, {Banday}, {Barreiro} et~al.}}]{Planck2013LFIdataproc}
\bibinfo{author}{\bibnamefont{{Planck Collaboration}}},
  \bibinfo{author}{\bibfnamefont{N.}~\bibnamefont{{Aghanim}}},
  \bibinfo{author}{\bibfnamefont{C.}~\bibnamefont{{Armitage-Caplan}}},
  \bibinfo{author}{\bibfnamefont{M.}~\bibnamefont{{Arnaud}}},
  \bibinfo{author}{\bibfnamefont{M.}~\bibnamefont{{Ashdown}}},
  \bibinfo{author}{\bibfnamefont{F.}~\bibnamefont{{Atrio-Barandela}}},
  \bibinfo{author}{\bibfnamefont{J.}~\bibnamefont{{Aumont}}},
  \bibinfo{author}{\bibfnamefont{C.}~\bibnamefont{{Baccigalupi}}},
  \bibinfo{author}{\bibfnamefont{A.~J.} \bibnamefont{{Banday}}},
  \bibinfo{author}{\bibfnamefont{R.~B.} \bibnamefont{{Barreiro}}},
  \bibnamefont{et~al.}, \bibinfo{journal}{\aap} \textbf{\bibinfo{volume}{571}},
  \bibinfo{eid}{A2} (\bibinfo{year}{2014}{\natexlab{d}}), \eprint{1303.5063}.

\bibitem[{\citenamefont{{Planck Collaboration}
  et~al.}(2014{\natexlab{e}})\citenamefont{{Planck Collaboration}, {Ade},
  {Aghanim}, {Armitage-Caplan}, {Arnaud}, {Ashdown}, {Atrio-Barandela},
  {Aumont}, {Baccigalupi}, {Banday} et~al.}}]{Planck2013HFIspectralresp}
\bibinfo{author}{\bibnamefont{{Planck Collaboration}}},
  \bibinfo{author}{\bibfnamefont{P.~A.~R.} \bibnamefont{{Ade}}},
  \bibinfo{author}{\bibfnamefont{N.}~\bibnamefont{{Aghanim}}},
  \bibinfo{author}{\bibfnamefont{C.}~\bibnamefont{{Armitage-Caplan}}},
  \bibinfo{author}{\bibfnamefont{M.}~\bibnamefont{{Arnaud}}},
  \bibinfo{author}{\bibfnamefont{M.}~\bibnamefont{{Ashdown}}},
  \bibinfo{author}{\bibfnamefont{F.}~\bibnamefont{{Atrio-Barandela}}},
  \bibinfo{author}{\bibfnamefont{J.}~\bibnamefont{{Aumont}}},
  \bibinfo{author}{\bibfnamefont{C.}~\bibnamefont{{Baccigalupi}}},
  \bibinfo{author}{\bibfnamefont{A.~J.} \bibnamefont{{Banday}}},
  \bibnamefont{et~al.}, \bibinfo{journal}{\aap} \textbf{\bibinfo{volume}{571}},
  \bibinfo{eid}{A9} (\bibinfo{year}{2014}{\natexlab{e}}), \eprint{1303.5070}.

\bibitem[{\citenamefont{{Planck Collaboration}
  et~al.}(2016{\natexlab{h}})\citenamefont{{Planck Collaboration}, {Ade},
  {Aghanim}, {Ashdown}, {Aumont}, {Baccigalupi}, {Band ay}, {Barreiro},
  {Bartolo}, {Battaner} et~al.}}]{Planck2015LFIbeams}
\bibinfo{author}{\bibnamefont{{Planck Collaboration}}},
  \bibinfo{author}{\bibfnamefont{P.~A.~R.} \bibnamefont{{Ade}}},
  \bibinfo{author}{\bibfnamefont{N.}~\bibnamefont{{Aghanim}}},
  \bibinfo{author}{\bibfnamefont{M.}~\bibnamefont{{Ashdown}}},
  \bibinfo{author}{\bibfnamefont{J.}~\bibnamefont{{Aumont}}},
  \bibinfo{author}{\bibfnamefont{C.}~\bibnamefont{{Baccigalupi}}},
  \bibinfo{author}{\bibfnamefont{A.~J.} \bibnamefont{{Band ay}}},
  \bibinfo{author}{\bibfnamefont{R.~B.} \bibnamefont{{Barreiro}}},
  \bibinfo{author}{\bibfnamefont{N.}~\bibnamefont{{Bartolo}}},
  \bibinfo{author}{\bibfnamefont{E.}~\bibnamefont{{Battaner}}},
  \bibnamefont{et~al.}, \bibinfo{journal}{\aap} \textbf{\bibinfo{volume}{594}},
  \bibinfo{eid}{A4} (\bibinfo{year}{2016}{\natexlab{h}}), \eprint{1502.01584}.

\bibitem[{\citenamefont{{Planck Collaboration}
  et~al.}(2016{\natexlab{i}})\citenamefont{{Planck Collaboration}, {Adam},
  {Ade}, {Aghanim}, {Arnaud}, {Ashdown}, {Aumont}, {Baccigalupi}, {Banday},
  {Barreiro} et~al.}}]{Planck2015HFIbeams}
\bibinfo{author}{\bibnamefont{{Planck Collaboration}}},
  \bibinfo{author}{\bibfnamefont{R.}~\bibnamefont{{Adam}}},
  \bibinfo{author}{\bibfnamefont{P.~A.~R.} \bibnamefont{{Ade}}},
  \bibinfo{author}{\bibfnamefont{N.}~\bibnamefont{{Aghanim}}},
  \bibinfo{author}{\bibfnamefont{M.}~\bibnamefont{{Arnaud}}},
  \bibinfo{author}{\bibfnamefont{M.}~\bibnamefont{{Ashdown}}},
  \bibinfo{author}{\bibfnamefont{J.}~\bibnamefont{{Aumont}}},
  \bibinfo{author}{\bibfnamefont{C.}~\bibnamefont{{Baccigalupi}}},
  \bibinfo{author}{\bibfnamefont{A.~J.} \bibnamefont{{Banday}}},
  \bibinfo{author}{\bibfnamefont{R.~B.} \bibnamefont{{Barreiro}}},
  \bibnamefont{et~al.}, \bibinfo{journal}{\aap} \textbf{\bibinfo{volume}{594}},
  \bibinfo{eid}{A7} (\bibinfo{year}{2016}{\natexlab{i}}), \eprint{1502.01586}.

\bibitem[{\citenamefont{{G{\'o}rski}
  et~al.}(2005{\natexlab{a}})\citenamefont{{G{\'o}rski}, {Hivon}, {Banday},
  {Wandelt}, {Hansen}, {Reinecke}, and {Bartelmann}}}]{HEALPix}
\bibinfo{author}{\bibfnamefont{K.~M.} \bibnamefont{{G{\'o}rski}}},
  \bibinfo{author}{\bibfnamefont{E.}~\bibnamefont{{Hivon}}},
  \bibinfo{author}{\bibfnamefont{A.~J.} \bibnamefont{{Banday}}},
  \bibinfo{author}{\bibfnamefont{B.~D.} \bibnamefont{{Wandelt}}},
  \bibinfo{author}{\bibfnamefont{F.~K.} \bibnamefont{{Hansen}}},
  \bibinfo{author}{\bibfnamefont{M.}~\bibnamefont{{Reinecke}}},
  \bibnamefont{and}
  \bibinfo{author}{\bibfnamefont{M.}~\bibnamefont{{Bartelmann}}},
  \bibinfo{journal}{\apj} \textbf{\bibinfo{volume}{622}}, \bibinfo{pages}{759}
  (\bibinfo{year}{2005}{\natexlab{a}}), \eprint{astro-ph/0409513}.

\bibitem[{\citenamefont{{Calabretta} and {Greisen}}(2002)}]{WCS1}
\bibinfo{author}{\bibfnamefont{M.~R.} \bibnamefont{{Calabretta}}}
  \bibnamefont{and} \bibinfo{author}{\bibfnamefont{E.~W.}
  \bibnamefont{{Greisen}}}, \bibinfo{journal}{\aap}
  \textbf{\bibinfo{volume}{395}}, \bibinfo{pages}{1077} (\bibinfo{year}{2002}),
  \eprint{astro-ph/0207413}.

\bibitem[{\citenamefont{{Planck Collaboration}
  et~al.}(2016{\natexlab{j}})\citenamefont{{Planck Collaboration}, {Ade},
  {Aghanim}, {Arg{\"u}eso}, {Arnaud}, {Ashdown}, {Aumont}, {Baccigalupi},
  {Banday}, {Barreiro} et~al.}}]{Planck2015PCCS}
\bibinfo{author}{\bibnamefont{{Planck Collaboration}}},
  \bibinfo{author}{\bibfnamefont{P.~A.~R.} \bibnamefont{{Ade}}},
  \bibinfo{author}{\bibfnamefont{N.}~\bibnamefont{{Aghanim}}},
  \bibinfo{author}{\bibfnamefont{F.}~\bibnamefont{{Arg{\"u}eso}}},
  \bibinfo{author}{\bibfnamefont{M.}~\bibnamefont{{Arnaud}}},
  \bibinfo{author}{\bibfnamefont{M.}~\bibnamefont{{Ashdown}}},
  \bibinfo{author}{\bibfnamefont{J.}~\bibnamefont{{Aumont}}},
  \bibinfo{author}{\bibfnamefont{C.}~\bibnamefont{{Baccigalupi}}},
  \bibinfo{author}{\bibfnamefont{A.~J.} \bibnamefont{{Banday}}},
  \bibinfo{author}{\bibfnamefont{R.~B.} \bibnamefont{{Barreiro}}},
  \bibnamefont{et~al.}, \bibinfo{journal}{\aap} \textbf{\bibinfo{volume}{594}},
  \bibinfo{eid}{A26} (\bibinfo{year}{2016}{\natexlab{j}}), \eprint{1507.02058}.

\bibitem[{\citenamefont{{Naess}}(2019)}]{NaessSources}
\bibinfo{author}{\bibfnamefont{S.~K.} \bibnamefont{{Naess}}},
  \bibinfo{journal}{arXiv e-prints} \bibinfo{eid}{arXiv:1906.08030}
  (\bibinfo{year}{2019}), \eprint{1906.08030}.

\bibitem[{\citenamefont{Bucher and Louis}(2012)}]{Bucher:2012}
\bibinfo{author}{\bibfnamefont{M.}~\bibnamefont{Bucher}} \bibnamefont{and}
  \bibinfo{author}{\bibfnamefont{T.}~\bibnamefont{Louis}},
  \bibinfo{journal}{Monthly Notices of the Royal Astronomical Society}
  \textbf{\bibinfo{volume}{424}}, \bibinfo{pages}{1694} (\bibinfo{year}{2012}),
  \urlprefix\url{http://dx.doi.org/10.1111/j.1365-2966.2012.21138.x}.

\bibitem[{\citenamefont{{Lewis} et~al.}(2000)\citenamefont{{Lewis},
  {Challinor}, and {Lasenby}}}]{CAMB}
\bibinfo{author}{\bibfnamefont{A.}~\bibnamefont{{Lewis}}},
  \bibinfo{author}{\bibfnamefont{A.}~\bibnamefont{{Challinor}}},
  \bibnamefont{and}
  \bibinfo{author}{\bibfnamefont{A.}~\bibnamefont{{Lasenby}}},
  \bibinfo{journal}{\apj} \textbf{\bibinfo{volume}{538}}, \bibinfo{pages}{473}
  (\bibinfo{year}{2000}), \eprint{astro-ph/9911177}.

\bibitem[{\citenamefont{{Darwish} et~al.}(2020)\citenamefont{{Darwish},
  {Madhavacheril}, {Sherwin}, {Aiola}, {Battaglia}, {Beall}, {Becker}, {Bond},
  {Calabrese}, {Choi} et~al.}}]{DarwishEtAl}
\bibinfo{author}{\bibfnamefont{O.}~\bibnamefont{{Darwish}}},
  \bibinfo{author}{\bibfnamefont{M.~S.} \bibnamefont{{Madhavacheril}}},
  \bibinfo{author}{\bibfnamefont{B.}~\bibnamefont{{Sherwin}}},
  \bibinfo{author}{\bibfnamefont{S.}~\bibnamefont{{Aiola}}},
  \bibinfo{author}{\bibfnamefont{N.}~\bibnamefont{{Battaglia}}},
  \bibinfo{author}{\bibfnamefont{J.~A.} \bibnamefont{{Beall}}},
  \bibinfo{author}{\bibfnamefont{D.~T.} \bibnamefont{{Becker}}},
  \bibinfo{author}{\bibfnamefont{J.~R.} \bibnamefont{{Bond}}},
  \bibinfo{author}{\bibfnamefont{E.}~\bibnamefont{{Calabrese}}},
  \bibinfo{author}{\bibfnamefont{S.}~\bibnamefont{{Choi}}},
  \bibnamefont{et~al.}, \bibinfo{journal}{arXiv e-prints}
  \bibinfo{eid}{arXiv:2004.01139} (\bibinfo{year}{2020}), \eprint{2004.01139}.

\bibitem[{\citenamefont{{P{\^a}ris} et~al.}(2018)\citenamefont{{P{\^a}ris},
  {Petitjean}, {Aubourg}, {Myers}, {Streblyanska}, {Lyke}, {Anderson},
  {Armengaud}, {Bautista}, {Blanton} et~al.}}]{SDSSQuasars}
\bibinfo{author}{\bibfnamefont{I.}~\bibnamefont{{P{\^a}ris}}},
  \bibinfo{author}{\bibfnamefont{P.}~\bibnamefont{{Petitjean}}},
  \bibinfo{author}{\bibfnamefont{{\'E}.}~\bibnamefont{{Aubourg}}},
  \bibinfo{author}{\bibfnamefont{A.~D.} \bibnamefont{{Myers}}},
  \bibinfo{author}{\bibfnamefont{A.}~\bibnamefont{{Streblyanska}}},
  \bibinfo{author}{\bibfnamefont{B.~W.} \bibnamefont{{Lyke}}},
  \bibinfo{author}{\bibfnamefont{S.~F.} \bibnamefont{{Anderson}}},
  \bibinfo{author}{\bibfnamefont{{\'E}.}~\bibnamefont{{Armengaud}}},
  \bibinfo{author}{\bibfnamefont{J.}~\bibnamefont{{Bautista}}},
  \bibinfo{author}{\bibfnamefont{M.~R.} \bibnamefont{{Blanton}}},
  \bibnamefont{et~al.}, \bibinfo{journal}{\aap} \textbf{\bibinfo{volume}{613}},
  \bibinfo{eid}{A51} (\bibinfo{year}{2018}), \eprint{1712.05029}.

\bibitem[{\citenamefont{{Crichton} et~al.}(2016)\citenamefont{{Crichton},
  {Gralla}, {Hall}, {Marriage}, {Zakamska}, {Battaglia}, {Bond}, {Devlin},
  {Hill}, {Hilton} et~al.}}]{CrichtonQuasars}
\bibinfo{author}{\bibfnamefont{D.}~\bibnamefont{{Crichton}}},
  \bibinfo{author}{\bibfnamefont{M.~B.} \bibnamefont{{Gralla}}},
  \bibinfo{author}{\bibfnamefont{K.}~\bibnamefont{{Hall}}},
  \bibinfo{author}{\bibfnamefont{T.~A.} \bibnamefont{{Marriage}}},
  \bibinfo{author}{\bibfnamefont{N.~L.} \bibnamefont{{Zakamska}}},
  \bibinfo{author}{\bibfnamefont{N.}~\bibnamefont{{Battaglia}}},
  \bibinfo{author}{\bibfnamefont{J.~R.} \bibnamefont{{Bond}}},
  \bibinfo{author}{\bibfnamefont{M.~J.} \bibnamefont{{Devlin}}},
  \bibinfo{author}{\bibfnamefont{J.~C.} \bibnamefont{{Hill}}},
  \bibinfo{author}{\bibfnamefont{M.}~\bibnamefont{{Hilton}}},
  \bibnamefont{et~al.}, \bibinfo{journal}{\mnras}
  \textbf{\bibinfo{volume}{458}}, \bibinfo{pages}{1478} (\bibinfo{year}{2016}),
  \eprint{1510.05656}.

\bibitem[{\citenamefont{{Hall} et~al.}(2019)\citenamefont{{Hall}, {Zakamska},
  {Addison}, {Battaglia}, {Crichton}, {Devlin}, {Dunkley}, {Gralla}, {Hill},
  {Hilton} et~al.}}]{HallQuasars}
\bibinfo{author}{\bibfnamefont{K.~R.} \bibnamefont{{Hall}}},
  \bibinfo{author}{\bibfnamefont{N.~L.} \bibnamefont{{Zakamska}}},
  \bibinfo{author}{\bibfnamefont{G.~E.} \bibnamefont{{Addison}}},
  \bibinfo{author}{\bibfnamefont{N.}~\bibnamefont{{Battaglia}}},
  \bibinfo{author}{\bibfnamefont{D.}~\bibnamefont{{Crichton}}},
  \bibinfo{author}{\bibfnamefont{M.}~\bibnamefont{{Devlin}}},
  \bibinfo{author}{\bibfnamefont{J.}~\bibnamefont{{Dunkley}}},
  \bibinfo{author}{\bibfnamefont{M.}~\bibnamefont{{Gralla}}},
  \bibinfo{author}{\bibfnamefont{J.~C.} \bibnamefont{{Hill}}},
  \bibinfo{author}{\bibfnamefont{M.}~\bibnamefont{{Hilton}}},
  \bibnamefont{et~al.}, \bibinfo{journal}{\mnras}  (\bibinfo{year}{2019}).

\bibitem[{\citenamefont{{Battaglia} et~al.}(2010)\citenamefont{{Battaglia},
  {Bond}, {Pfrommer}, {Sievers}, and {Sijacki}}}]{Battaglia2010}
\bibinfo{author}{\bibfnamefont{N.}~\bibnamefont{{Battaglia}}},
  \bibinfo{author}{\bibfnamefont{J.~R.} \bibnamefont{{Bond}}},
  \bibinfo{author}{\bibfnamefont{C.}~\bibnamefont{{Pfrommer}}},
  \bibinfo{author}{\bibfnamefont{J.~L.} \bibnamefont{{Sievers}}},
  \bibnamefont{and}
  \bibinfo{author}{\bibfnamefont{D.}~\bibnamefont{{Sijacki}}},
  \bibinfo{journal}{\apj} \textbf{\bibinfo{volume}{725}}, \bibinfo{pages}{91}
  (\bibinfo{year}{2010}), \eprint{1003.4256}.

\bibitem[{\citenamefont{{Battaglia} et~al.}(2012)\citenamefont{{Battaglia},
  {Bond}, {Pfrommer}, and {Sievers}}}]{Battaglia2012tSZPS}
\bibinfo{author}{\bibfnamefont{N.}~\bibnamefont{{Battaglia}}},
  \bibinfo{author}{\bibfnamefont{J.~R.} \bibnamefont{{Bond}}},
  \bibinfo{author}{\bibfnamefont{C.}~\bibnamefont{{Pfrommer}}},
  \bibnamefont{and} \bibinfo{author}{\bibfnamefont{J.~L.}
  \bibnamefont{{Sievers}}}, \bibinfo{journal}{\apj}
  \textbf{\bibinfo{volume}{758}}, \bibinfo{eid}{75} (\bibinfo{year}{2012}),
  \eprint{1109.3711}.

\bibitem[{\citenamefont{{Osborne} et~al.}(2014)\citenamefont{{Osborne},
  {Hanson}, and {Dor{\'e}}}}]{Osborne}
\bibinfo{author}{\bibfnamefont{S.~J.} \bibnamefont{{Osborne}}},
  \bibinfo{author}{\bibfnamefont{D.}~\bibnamefont{{Hanson}}}, \bibnamefont{and}
  \bibinfo{author}{\bibfnamefont{O.}~\bibnamefont{{Dor{\'e}}}},
  \bibinfo{journal}{\jcap} \textbf{\bibinfo{volume}{2014}}, \bibinfo{eid}{024}
  (\bibinfo{year}{2014}), \eprint{1310.7547}.

\bibitem[{\citenamefont{{van Engelen} et~al.}(2014)\citenamefont{{van Engelen},
  {Bhattacharya}, {Sehgal}, {Holder}, {Zahn}, and {Nagai}}}]{vanEngelen2014}
\bibinfo{author}{\bibfnamefont{A.}~\bibnamefont{{van Engelen}}},
  \bibinfo{author}{\bibfnamefont{S.}~\bibnamefont{{Bhattacharya}}},
  \bibinfo{author}{\bibfnamefont{N.}~\bibnamefont{{Sehgal}}},
  \bibinfo{author}{\bibfnamefont{G.~P.} \bibnamefont{{Holder}}},
  \bibinfo{author}{\bibfnamefont{O.}~\bibnamefont{{Zahn}}}, \bibnamefont{and}
  \bibinfo{author}{\bibfnamefont{D.}~\bibnamefont{{Nagai}}},
  \bibinfo{journal}{\apj} \textbf{\bibinfo{volume}{786}}, \bibinfo{eid}{13}
  (\bibinfo{year}{2014}), \eprint{1310.7023}.

\bibitem[{\citenamefont{{Omori} et~al.}(2019)\citenamefont{{Omori},
  {Giannantonio}, {Porredon}, {Baxter}, {Chang}, {Crocce}, {Fosalba},
  {Alarcon}, {Banik}, {Blazek} et~al.}}]{Omori2019}
\bibinfo{author}{\bibfnamefont{Y.}~\bibnamefont{{Omori}}},
  \bibinfo{author}{\bibfnamefont{T.}~\bibnamefont{{Giannantonio}}},
  \bibinfo{author}{\bibfnamefont{A.}~\bibnamefont{{Porredon}}},
  \bibinfo{author}{\bibfnamefont{E.~J.} \bibnamefont{{Baxter}}},
  \bibinfo{author}{\bibfnamefont{C.}~\bibnamefont{{Chang}}},
  \bibinfo{author}{\bibfnamefont{M.}~\bibnamefont{{Crocce}}},
  \bibinfo{author}{\bibfnamefont{P.}~\bibnamefont{{Fosalba}}},
  \bibinfo{author}{\bibfnamefont{A.}~\bibnamefont{{Alarcon}}},
  \bibinfo{author}{\bibfnamefont{N.}~\bibnamefont{{Banik}}},
  \bibinfo{author}{\bibfnamefont{J.}~\bibnamefont{{Blazek}}},
  \bibnamefont{et~al.}, \bibinfo{journal}{\prd} \textbf{\bibinfo{volume}{100}},
  \bibinfo{eid}{043501} (\bibinfo{year}{2019}), \eprint{1810.02342}.

\bibitem[{\citenamefont{{Baxter} et~al.}(2019)\citenamefont{{Baxter}, {Omori},
  {Chang}, {Giannantonio}, {Kirk}, {Krause}, {Blazek}, {Bleem}, {Choi},
  {Crawford} et~al.}}]{Baxter2019}
\bibinfo{author}{\bibfnamefont{E.~J.} \bibnamefont{{Baxter}}},
  \bibinfo{author}{\bibfnamefont{Y.}~\bibnamefont{{Omori}}},
  \bibinfo{author}{\bibfnamefont{C.}~\bibnamefont{{Chang}}},
  \bibinfo{author}{\bibfnamefont{T.}~\bibnamefont{{Giannantonio}}},
  \bibinfo{author}{\bibfnamefont{D.}~\bibnamefont{{Kirk}}},
  \bibinfo{author}{\bibfnamefont{E.}~\bibnamefont{{Krause}}},
  \bibinfo{author}{\bibfnamefont{J.}~\bibnamefont{{Blazek}}},
  \bibinfo{author}{\bibfnamefont{L.}~\bibnamefont{{Bleem}}},
  \bibinfo{author}{\bibfnamefont{A.}~\bibnamefont{{Choi}}},
  \bibinfo{author}{\bibfnamefont{T.~M.} \bibnamefont{{Crawford}}},
  \bibnamefont{et~al.}, \bibinfo{journal}{\prd} \textbf{\bibinfo{volume}{99}},
  \bibinfo{eid}{023508} (\bibinfo{year}{2019}), \eprint{1802.05257}.

\bibitem[{\citenamefont{{Prince} et~al.}(2018)\citenamefont{{Prince},
  {Moodley}, {Ridl}, and {Bucher}}}]{Prince2018}
\bibinfo{author}{\bibfnamefont{H.}~\bibnamefont{{Prince}}},
  \bibinfo{author}{\bibfnamefont{K.}~\bibnamefont{{Moodley}}},
  \bibinfo{author}{\bibfnamefont{J.}~\bibnamefont{{Ridl}}}, \bibnamefont{and}
  \bibinfo{author}{\bibfnamefont{M.}~\bibnamefont{{Bucher}}},
  \bibinfo{journal}{\jcap} \textbf{\bibinfo{volume}{2018}}, \bibinfo{eid}{034}
  (\bibinfo{year}{2018}), \eprint{1709.02227}.

\bibitem[{\citenamefont{{Schaan} and {Ferraro}}(2019)}]{SchaanFerraro}
\bibinfo{author}{\bibfnamefont{E.}~\bibnamefont{{Schaan}}} \bibnamefont{and}
  \bibinfo{author}{\bibfnamefont{S.}~\bibnamefont{{Ferraro}}},
  \bibinfo{journal}{\prl} \textbf{\bibinfo{volume}{122}}, \bibinfo{eid}{181301}
  (\bibinfo{year}{2019}), \eprint{1804.06403}.

\bibitem[{\citenamefont{{Planck Collaboration}
  et~al.}(2016{\natexlab{k}})\citenamefont{{Planck Collaboration}, {Ade},
  {Aghanim}, {Arnaud}, {Arroja}, {Ashdown}, {Aumont}, {Baccigalupi},
  {Ballardini}, {Banday} et~al.}}]{Planck2015NG}
\bibinfo{author}{\bibnamefont{{Planck Collaboration}}},
  \bibinfo{author}{\bibfnamefont{P.~A.~R.} \bibnamefont{{Ade}}},
  \bibinfo{author}{\bibfnamefont{N.}~\bibnamefont{{Aghanim}}},
  \bibinfo{author}{\bibfnamefont{M.}~\bibnamefont{{Arnaud}}},
  \bibinfo{author}{\bibfnamefont{F.}~\bibnamefont{{Arroja}}},
  \bibinfo{author}{\bibfnamefont{M.}~\bibnamefont{{Ashdown}}},
  \bibinfo{author}{\bibfnamefont{J.}~\bibnamefont{{Aumont}}},
  \bibinfo{author}{\bibfnamefont{C.}~\bibnamefont{{Baccigalupi}}},
  \bibinfo{author}{\bibfnamefont{M.}~\bibnamefont{{Ballardini}}},
  \bibinfo{author}{\bibfnamefont{A.~J.} \bibnamefont{{Banday}}},
  \bibnamefont{et~al.}, \bibinfo{journal}{\aap} \textbf{\bibinfo{volume}{594}},
  \bibinfo{eid}{A17} (\bibinfo{year}{2016}{\natexlab{k}}), \eprint{1502.01592}.

\bibitem[{\citenamefont{{Planck Collaboration}
  et~al.}(2019{\natexlab{b}})\citenamefont{{Planck Collaboration}, {Akrami},
  {Arroja}, {Ashdown}, {Aumont}, {Baccigalupi}, {Ballardini}, {Banday},
  {Barreiro}, {Bartolo} et~al.}}]{Planck2018NG}
\bibinfo{author}{\bibnamefont{{Planck Collaboration}}},
  \bibinfo{author}{\bibfnamefont{Y.}~\bibnamefont{{Akrami}}},
  \bibinfo{author}{\bibfnamefont{F.}~\bibnamefont{{Arroja}}},
  \bibinfo{author}{\bibfnamefont{M.}~\bibnamefont{{Ashdown}}},
  \bibinfo{author}{\bibfnamefont{J.}~\bibnamefont{{Aumont}}},
  \bibinfo{author}{\bibfnamefont{C.}~\bibnamefont{{Baccigalupi}}},
  \bibinfo{author}{\bibfnamefont{M.}~\bibnamefont{{Ballardini}}},
  \bibinfo{author}{\bibfnamefont{A.~J.} \bibnamefont{{Banday}}},
  \bibinfo{author}{\bibfnamefont{R.~B.} \bibnamefont{{Barreiro}}},
  \bibinfo{author}{\bibfnamefont{N.}~\bibnamefont{{Bartolo}}},
  \bibnamefont{et~al.}, \bibinfo{journal}{arXiv e-prints}
  \bibinfo{eid}{arXiv:1905.05697} (\bibinfo{year}{2019}{\natexlab{b}}),
  \eprint{1905.05697}.

\bibitem[{\citenamefont{{Hill}}(2018)}]{Hill2018NG}
\bibinfo{author}{\bibfnamefont{J.~C.} \bibnamefont{{Hill}}},
  \bibinfo{journal}{\prd} \textbf{\bibinfo{volume}{98}}, \bibinfo{eid}{083542}
  (\bibinfo{year}{2018}), \eprint{1807.07324}.

\bibitem[{\citenamefont{{Jung} et~al.}(2018)\citenamefont{{Jung}, {Racine}, and
  {van Tent}}}]{Jung2018}
\bibinfo{author}{\bibfnamefont{G.}~\bibnamefont{{Jung}}},
  \bibinfo{author}{\bibfnamefont{B.}~\bibnamefont{{Racine}}}, \bibnamefont{and}
  \bibinfo{author}{\bibfnamefont{B.}~\bibnamefont{{van Tent}}},
  \bibinfo{journal}{\jcap} \textbf{\bibinfo{volume}{2018}}, \bibinfo{eid}{047}
  (\bibinfo{year}{2018}), \eprint{1810.01727}.

\bibitem[{\citenamefont{{Dor{\'e}} et~al.}(2004)\citenamefont{{Dor{\'e}},
  {Hennawi}, and {Spergel}}}]{Dore2004}
\bibinfo{author}{\bibfnamefont{O.}~\bibnamefont{{Dor{\'e}}}},
  \bibinfo{author}{\bibfnamefont{J.~F.} \bibnamefont{{Hennawi}}},
  \bibnamefont{and} \bibinfo{author}{\bibfnamefont{D.~N.}
  \bibnamefont{{Spergel}}}, \bibinfo{journal}{\apj}
  \textbf{\bibinfo{volume}{606}}, \bibinfo{pages}{46} (\bibinfo{year}{2004}),
  \eprint{astro-ph/0309337}.

\bibitem[{\citenamefont{{Smith} et~al.}(2018)\citenamefont{{Smith},
  {Madhavacheril}, {M{\"u}nchmeyer}, {Ferraro}, {Giri}, and
  {Johnson}}}]{Smith2018}
\bibinfo{author}{\bibfnamefont{K.~M.} \bibnamefont{{Smith}}},
  \bibinfo{author}{\bibfnamefont{M.~S.} \bibnamefont{{Madhavacheril}}},
  \bibinfo{author}{\bibfnamefont{M.}~\bibnamefont{{M{\"u}nchmeyer}}},
  \bibinfo{author}{\bibfnamefont{S.}~\bibnamefont{{Ferraro}}},
  \bibinfo{author}{\bibfnamefont{U.}~\bibnamefont{{Giri}}}, \bibnamefont{and}
  \bibinfo{author}{\bibfnamefont{M.~C.} \bibnamefont{{Johnson}}},
  \bibinfo{journal}{arXiv e-prints} \bibinfo{eid}{arXiv:1810.13423}
  (\bibinfo{year}{2018}), \eprint{1810.13423}.

\bibitem[{\citenamefont{{Ho} et~al.}(2009)\citenamefont{{Ho}, {Dedeo}, and
  {Spergel}}}]{HoTemplate}
\bibinfo{author}{\bibfnamefont{S.}~\bibnamefont{{Ho}}},
  \bibinfo{author}{\bibfnamefont{S.}~\bibnamefont{{Dedeo}}}, \bibnamefont{and}
  \bibinfo{author}{\bibfnamefont{D.}~\bibnamefont{{Spergel}}},
  \bibinfo{journal}{arXiv e-prints} \bibinfo{eid}{arXiv:0903.2845}
  (\bibinfo{year}{2009}), \eprint{0903.2845}.

\bibitem[{\citenamefont{{Schaan} et~al.}(2016)\citenamefont{{Schaan},
  {Ferraro}, {Vargas-Maga{\~n}a}, {Smith}, {Ho}, {Aiola}, {Battaglia}, {Bond},
  {De Bernardis}, {Calabrese} et~al.}}]{SchaanTemplate}
\bibinfo{author}{\bibfnamefont{E.}~\bibnamefont{{Schaan}}},
  \bibinfo{author}{\bibfnamefont{S.}~\bibnamefont{{Ferraro}}},
  \bibinfo{author}{\bibfnamefont{M.}~\bibnamefont{{Vargas-Maga{\~n}a}}},
  \bibinfo{author}{\bibfnamefont{K.~M.} \bibnamefont{{Smith}}},
  \bibinfo{author}{\bibfnamefont{S.}~\bibnamefont{{Ho}}},
  \bibinfo{author}{\bibfnamefont{S.}~\bibnamefont{{Aiola}}},
  \bibinfo{author}{\bibfnamefont{N.}~\bibnamefont{{Battaglia}}},
  \bibinfo{author}{\bibfnamefont{J.~R.} \bibnamefont{{Bond}}},
  \bibinfo{author}{\bibfnamefont{F.}~\bibnamefont{{De Bernardis}}},
  \bibinfo{author}{\bibfnamefont{E.}~\bibnamefont{{Calabrese}}},
  \bibnamefont{et~al.}, \bibinfo{journal}{\prd} \textbf{\bibinfo{volume}{93}},
  \bibinfo{eid}{082002} (\bibinfo{year}{2016}), \eprint{1510.06442}.

\bibitem[{\citenamefont{{Hand} et~al.}(2012)\citenamefont{{Hand}, {Addison},
  {Aubourg}, {Battaglia}, {Battistelli}, {Bizyaev}, {Bond}, {Brewington},
  {Brinkmann}, {Brown} et~al.}}]{HandPairwise}
\bibinfo{author}{\bibfnamefont{N.}~\bibnamefont{{Hand}}},
  \bibinfo{author}{\bibfnamefont{G.~E.} \bibnamefont{{Addison}}},
  \bibinfo{author}{\bibfnamefont{E.}~\bibnamefont{{Aubourg}}},
  \bibinfo{author}{\bibfnamefont{N.}~\bibnamefont{{Battaglia}}},
  \bibinfo{author}{\bibfnamefont{E.~S.} \bibnamefont{{Battistelli}}},
  \bibinfo{author}{\bibfnamefont{D.}~\bibnamefont{{Bizyaev}}},
  \bibinfo{author}{\bibfnamefont{J.~R.} \bibnamefont{{Bond}}},
  \bibinfo{author}{\bibfnamefont{H.}~\bibnamefont{{Brewington}}},
  \bibinfo{author}{\bibfnamefont{J.}~\bibnamefont{{Brinkmann}}},
  \bibinfo{author}{\bibfnamefont{B.~R.} \bibnamefont{{Brown}}},
  \bibnamefont{et~al.}, \bibinfo{journal}{Physical Review Letters}
  \textbf{\bibinfo{volume}{109}}, \bibinfo{eid}{041101} (\bibinfo{year}{2012}),
  \eprint{1203.4219}.

\bibitem[{\citenamefont{{De Bernardis} et~al.}(2017)\citenamefont{{De
  Bernardis}, {Aiola}, {Vavagiakis}, {Battaglia}, {Niemack}, {Beall}, {Becker},
  {Bond}, {Calabrese}, {Cho} et~al.}}]{deBernardisPairwise}
\bibinfo{author}{\bibfnamefont{F.}~\bibnamefont{{De Bernardis}}},
  \bibinfo{author}{\bibfnamefont{S.}~\bibnamefont{{Aiola}}},
  \bibinfo{author}{\bibfnamefont{E.~M.} \bibnamefont{{Vavagiakis}}},
  \bibinfo{author}{\bibfnamefont{N.}~\bibnamefont{{Battaglia}}},
  \bibinfo{author}{\bibfnamefont{M.~D.} \bibnamefont{{Niemack}}},
  \bibinfo{author}{\bibfnamefont{J.}~\bibnamefont{{Beall}}},
  \bibinfo{author}{\bibfnamefont{D.~T.} \bibnamefont{{Becker}}},
  \bibinfo{author}{\bibfnamefont{J.~R.} \bibnamefont{{Bond}}},
  \bibinfo{author}{\bibfnamefont{E.}~\bibnamefont{{Calabrese}}},
  \bibinfo{author}{\bibfnamefont{H.}~\bibnamefont{{Cho}}},
  \bibnamefont{et~al.}, \bibinfo{journal}{\jcap} \textbf{\bibinfo{volume}{3}},
  \bibinfo{eid}{008} (\bibinfo{year}{2017}), \eprint{1607.02139}.

\bibitem[{\citenamefont{{Planck Collaboration}
  et~al.}(2013)\citenamefont{{Planck Collaboration}, {Ade}, {Aghanim},
  {Arnaud}, {Ashdown}, {Atrio-Barandela}, {Aumont}, {Baccigalupi}, {Balbi},
  {Banday} et~al.}}]{Planck2013stack}
\bibinfo{author}{\bibnamefont{{Planck Collaboration}}},
  \bibinfo{author}{\bibfnamefont{P.~A.~R.} \bibnamefont{{Ade}}},
  \bibinfo{author}{\bibfnamefont{N.}~\bibnamefont{{Aghanim}}},
  \bibinfo{author}{\bibfnamefont{M.}~\bibnamefont{{Arnaud}}},
  \bibinfo{author}{\bibfnamefont{M.}~\bibnamefont{{Ashdown}}},
  \bibinfo{author}{\bibfnamefont{F.}~\bibnamefont{{Atrio-Barandela}}},
  \bibinfo{author}{\bibfnamefont{J.}~\bibnamefont{{Aumont}}},
  \bibinfo{author}{\bibfnamefont{C.}~\bibnamefont{{Baccigalupi}}},
  \bibinfo{author}{\bibfnamefont{A.}~\bibnamefont{{Balbi}}},
  \bibinfo{author}{\bibfnamefont{A.~J.} \bibnamefont{{Banday}}},
  \bibnamefont{et~al.}, \bibinfo{journal}{\aap} \textbf{\bibinfo{volume}{557}},
  \bibinfo{eid}{A52} (\bibinfo{year}{2013}), \eprint{1212.4131}.

\bibitem[{\citenamefont{{Greco} et~al.}(2015)\citenamefont{{Greco}, {Hill},
  {Spergel}, and {Battaglia}}}]{Greco2015}
\bibinfo{author}{\bibfnamefont{J.~P.} \bibnamefont{{Greco}}},
  \bibinfo{author}{\bibfnamefont{J.~C.} \bibnamefont{{Hill}}},
  \bibinfo{author}{\bibfnamefont{D.~N.} \bibnamefont{{Spergel}}},
  \bibnamefont{and}
  \bibinfo{author}{\bibfnamefont{N.}~\bibnamefont{{Battaglia}}},
  \bibinfo{journal}{\apj} \textbf{\bibinfo{volume}{808}}, \bibinfo{eid}{151}
  (\bibinfo{year}{2015}), \eprint{1409.6747}.

\bibitem[{\citenamefont{{Le Brun} et~al.}(2015)\citenamefont{{Le Brun},
  {McCarthy}, and {Melin}}}]{LeBrun2015}
\bibinfo{author}{\bibfnamefont{A.~M.~C.} \bibnamefont{{Le Brun}}},
  \bibinfo{author}{\bibfnamefont{I.~G.} \bibnamefont{{McCarthy}}},
  \bibnamefont{and} \bibinfo{author}{\bibfnamefont{J.-B.}
  \bibnamefont{{Melin}}}, \bibinfo{journal}{\mnras}
  \textbf{\bibinfo{volume}{451}}, \bibinfo{pages}{3868} (\bibinfo{year}{2015}),
  \eprint{1501.05666}.

\bibitem[{\citenamefont{{Abazajian} et~al.}(2019)\citenamefont{{Abazajian},
  {Addison}, {Adshead}, {Ahmed}, {Allen}, {Alonso}, {Alvarez}, {Anderson},
  {Arnold}, {Baccigalupi} et~al.}}]{CMBS4DSR}
\bibinfo{author}{\bibfnamefont{K.}~\bibnamefont{{Abazajian}}},
  \bibinfo{author}{\bibfnamefont{G.}~\bibnamefont{{Addison}}},
  \bibinfo{author}{\bibfnamefont{P.}~\bibnamefont{{Adshead}}},
  \bibinfo{author}{\bibfnamefont{Z.}~\bibnamefont{{Ahmed}}},
  \bibinfo{author}{\bibfnamefont{S.~W.} \bibnamefont{{Allen}}},
  \bibinfo{author}{\bibfnamefont{D.}~\bibnamefont{{Alonso}}},
  \bibinfo{author}{\bibfnamefont{M.}~\bibnamefont{{Alvarez}}},
  \bibinfo{author}{\bibfnamefont{A.}~\bibnamefont{{Anderson}}},
  \bibinfo{author}{\bibfnamefont{K.~S.} \bibnamefont{{Arnold}}},
  \bibinfo{author}{\bibfnamefont{C.}~\bibnamefont{{Baccigalupi}}},
  \bibnamefont{et~al.}, \bibinfo{journal}{arXiv e-prints}
  \bibinfo{eid}{arXiv:1907.04473} (\bibinfo{year}{2019}), \eprint{1907.04473}.

\bibitem[{\citenamefont{{Koopman} et~al.}(2018)\citenamefont{{Koopman},
  {Cothard}, {Choi}, {Crowley}, {Duff}, {Henderson}, {Ho}, {Hubmayr},
  {Gallardo}, {Nati} et~al.}}]{Koopman2018}
\bibinfo{author}{\bibfnamefont{B.~J.} \bibnamefont{{Koopman}}},
  \bibinfo{author}{\bibfnamefont{N.~F.} \bibnamefont{{Cothard}}},
  \bibinfo{author}{\bibfnamefont{S.~K.} \bibnamefont{{Choi}}},
  \bibinfo{author}{\bibfnamefont{K.~T.} \bibnamefont{{Crowley}}},
  \bibinfo{author}{\bibfnamefont{S.~M.} \bibnamefont{{Duff}}},
  \bibinfo{author}{\bibfnamefont{S.~W.} \bibnamefont{{Henderson}}},
  \bibinfo{author}{\bibfnamefont{S.~P.} \bibnamefont{{Ho}}},
  \bibinfo{author}{\bibfnamefont{J.}~\bibnamefont{{Hubmayr}}},
  \bibinfo{author}{\bibfnamefont{P.~A.} \bibnamefont{{Gallardo}}},
  \bibinfo{author}{\bibfnamefont{F.}~\bibnamefont{{Nati}}},
  \bibnamefont{et~al.}, \bibinfo{journal}{Journal of Low Temperature Physics}
  \textbf{\bibinfo{volume}{193}}, \bibinfo{pages}{1103} (\bibinfo{year}{2018}),
  \eprint{1711.02594}.

\bibitem[{\citenamefont{{Simon} et~al.}(2018)\citenamefont{{Simon}, {Beall},
  {Cothard}, {Duff}, {Gallardo}, {Ho}, {Hubmayr}, {Koopman}, {McMahon}, {Nati}
  et~al.}}]{Simon2018}
\bibinfo{author}{\bibfnamefont{S.~M.} \bibnamefont{{Simon}}},
  \bibinfo{author}{\bibfnamefont{J.~A.} \bibnamefont{{Beall}}},
  \bibinfo{author}{\bibfnamefont{N.~F.} \bibnamefont{{Cothard}}},
  \bibinfo{author}{\bibfnamefont{S.~M.} \bibnamefont{{Duff}}},
  \bibinfo{author}{\bibfnamefont{P.~A.} \bibnamefont{{Gallardo}}},
  \bibinfo{author}{\bibfnamefont{S.~P.} \bibnamefont{{Ho}}},
  \bibinfo{author}{\bibfnamefont{J.}~\bibnamefont{{Hubmayr}}},
  \bibinfo{author}{\bibfnamefont{B.~J.} \bibnamefont{{Koopman}}},
  \bibinfo{author}{\bibfnamefont{J.~J.} \bibnamefont{{McMahon}}},
  \bibinfo{author}{\bibfnamefont{F.}~\bibnamefont{{Nati}}},
  \bibnamefont{et~al.}, \bibinfo{journal}{Journal of Low Temperature Physics}
  \textbf{\bibinfo{volume}{193}}, \bibinfo{pages}{1041} (\bibinfo{year}{2018}).

\bibitem[{\citenamefont{Zonca et~al.}(2019)\citenamefont{Zonca, Singer, Lenz,
  Reinecke, Rosset, Hivon, and Gorski}}]{Healpix1}
\bibinfo{author}{\bibfnamefont{A.}~\bibnamefont{Zonca}},
  \bibinfo{author}{\bibfnamefont{L.}~\bibnamefont{Singer}},
  \bibinfo{author}{\bibfnamefont{D.}~\bibnamefont{Lenz}},
  \bibinfo{author}{\bibfnamefont{M.}~\bibnamefont{Reinecke}},
  \bibinfo{author}{\bibfnamefont{C.}~\bibnamefont{Rosset}},
  \bibinfo{author}{\bibfnamefont{E.}~\bibnamefont{Hivon}}, \bibnamefont{and}
  \bibinfo{author}{\bibfnamefont{K.}~\bibnamefont{Gorski}},
  \bibinfo{journal}{Journal of Open Source Software}
  \textbf{\bibinfo{volume}{4}}, \bibinfo{pages}{1298} (\bibinfo{year}{2019}),
  \urlprefix\url{https://doi.org/10.21105/joss.01298}.

\bibitem[{\citenamefont{{G{\'o}rski}
  et~al.}(2005{\natexlab{b}})\citenamefont{{G{\'o}rski}, {Hivon}, {Banday},
  {Wandelt}, {Hansen}, {Reinecke}, and {Bartelmann}}}]{Healpix2}
\bibinfo{author}{\bibfnamefont{K.~M.} \bibnamefont{{G{\'o}rski}}},
  \bibinfo{author}{\bibfnamefont{E.}~\bibnamefont{{Hivon}}},
  \bibinfo{author}{\bibfnamefont{A.~J.} \bibnamefont{{Banday}}},
  \bibinfo{author}{\bibfnamefont{B.~D.} \bibnamefont{{Wandelt}}},
  \bibinfo{author}{\bibfnamefont{F.~K.} \bibnamefont{{Hansen}}},
  \bibinfo{author}{\bibfnamefont{M.}~\bibnamefont{{Reinecke}}},
  \bibnamefont{and}
  \bibinfo{author}{\bibfnamefont{M.}~\bibnamefont{{Bartelmann}}},
  \bibinfo{journal}{\apj} \textbf{\bibinfo{volume}{622}}, \bibinfo{pages}{759}
  (\bibinfo{year}{2005}{\natexlab{b}}), \eprint{arXiv:astro-ph/0409513}.

\bibitem[{\citenamefont{{Astropy Collaboration}
  et~al.}(2013)\citenamefont{{Astropy Collaboration}, {Robitaille}, {Tollerud},
  {Greenfield}, {Droettboom}, {Bray}, {Aldcroft}, {Davis}, {Ginsburg},
  {Price-Whelan} et~al.}}]{astropy:2013}
\bibinfo{author}{\bibnamefont{{Astropy Collaboration}}},
  \bibinfo{author}{\bibfnamefont{T.~P.} \bibnamefont{{Robitaille}}},
  \bibinfo{author}{\bibfnamefont{E.~J.} \bibnamefont{{Tollerud}}},
  \bibinfo{author}{\bibfnamefont{P.}~\bibnamefont{{Greenfield}}},
  \bibinfo{author}{\bibfnamefont{M.}~\bibnamefont{{Droettboom}}},
  \bibinfo{author}{\bibfnamefont{E.}~\bibnamefont{{Bray}}},
  \bibinfo{author}{\bibfnamefont{T.}~\bibnamefont{{Aldcroft}}},
  \bibinfo{author}{\bibfnamefont{M.}~\bibnamefont{{Davis}}},
  \bibinfo{author}{\bibfnamefont{A.}~\bibnamefont{{Ginsburg}}},
  \bibinfo{author}{\bibfnamefont{A.~M.} \bibnamefont{{Price-Whelan}}},
  \bibnamefont{et~al.}, \bibinfo{journal}{\aap} \textbf{\bibinfo{volume}{558}},
  \bibinfo{eid}{A33} (\bibinfo{year}{2013}), \eprint{1307.6212}.

\bibitem[{\citenamefont{{Price-Whelan}
  et~al.}(2018)\citenamefont{{Price-Whelan}, {Sip{\H{o}}cz}, {G{\"u}nther},
  {Lim}, {Crawford}, {Conseil}, {Shupe}, {Craig}, {Dencheva}, {Ginsburg}
  et~al.}}]{astropy:2018}
\bibinfo{author}{\bibfnamefont{A.~M.} \bibnamefont{{Price-Whelan}}},
  \bibinfo{author}{\bibfnamefont{B.~M.} \bibnamefont{{Sip{\H{o}}cz}}},
  \bibinfo{author}{\bibfnamefont{H.~M.} \bibnamefont{{G{\"u}nther}}},
  \bibinfo{author}{\bibfnamefont{P.~L.} \bibnamefont{{Lim}}},
  \bibinfo{author}{\bibfnamefont{S.~M.} \bibnamefont{{Crawford}}},
  \bibinfo{author}{\bibfnamefont{S.}~\bibnamefont{{Conseil}}},
  \bibinfo{author}{\bibfnamefont{D.~L.} \bibnamefont{{Shupe}}},
  \bibinfo{author}{\bibfnamefont{M.~W.} \bibnamefont{{Craig}}},
  \bibinfo{author}{\bibfnamefont{N.}~\bibnamefont{{Dencheva}}},
  \bibinfo{author}{\bibfnamefont{A.}~\bibnamefont{{Ginsburg}}},
  \bibnamefont{et~al.}, \bibinfo{journal}{\aj} \textbf{\bibinfo{volume}{156}},
  \bibinfo{eid}{123} (\bibinfo{year}{2018}).

\bibitem[{\citenamefont{Hunter}(2007)}]{Hunter:2007}
\bibinfo{author}{\bibfnamefont{J.~D.} \bibnamefont{Hunter}},
  \bibinfo{journal}{Computing in Science \& Engineering}
  \textbf{\bibinfo{volume}{9}}, \bibinfo{pages}{90} (\bibinfo{year}{2007}).

\bibitem[{\citenamefont{{Datta} et~al.}(2019)\citenamefont{{Datta}, {Aiola},
  {Choi}, {Devlin}, {Dunkley}, {D{\"u}nner}, {Gallardo}, {Gralla}, {Halpern},
  {Hasselfield} et~al.}}]{Datta2019}
\bibinfo{author}{\bibfnamefont{R.}~\bibnamefont{{Datta}}},
  \bibinfo{author}{\bibfnamefont{S.}~\bibnamefont{{Aiola}}},
  \bibinfo{author}{\bibfnamefont{S.~K.} \bibnamefont{{Choi}}},
  \bibinfo{author}{\bibfnamefont{M.}~\bibnamefont{{Devlin}}},
  \bibinfo{author}{\bibfnamefont{J.}~\bibnamefont{{Dunkley}}},
  \bibinfo{author}{\bibfnamefont{R.}~\bibnamefont{{D{\"u}nner}}},
  \bibinfo{author}{\bibfnamefont{P.~A.} \bibnamefont{{Gallardo}}},
  \bibinfo{author}{\bibfnamefont{M.}~\bibnamefont{{Gralla}}},
  \bibinfo{author}{\bibfnamefont{M.}~\bibnamefont{{Halpern}}},
  \bibinfo{author}{\bibfnamefont{M.}~\bibnamefont{{Hasselfield}}},
  \bibnamefont{et~al.}, \bibinfo{journal}{\mnras}
  \textbf{\bibinfo{volume}{486}}, \bibinfo{pages}{5239} (\bibinfo{year}{2019}),
  \eprint{1811.01854}.

\bibitem[{\citenamefont{ALMA}(2009)}]{AATM}
\bibinfo{author}{\bibnamefont{ALMA}}, \emph{\bibinfo{title}{Aatm version 0.09}}
  (\bibinfo{year}{2009}),
  \urlprefix\url{http://www.mrao.cam.ac.uk/~bn204/mk2/alma/aatm-absorption.html}.

\end{thebibliography}

\end{document}